\documentclass[letterpaper]{aa}
\usepackage{graphicx}
\usepackage{txfonts}
\usepackage{natbib}
\usepackage{amssymb}
\newcommand{\lya}{Ly$\alpha$}

\newcommand{\ergscm}{erg\,s$^{-1}$\,cm$^{-2}$}
\newcommand{\kms}{km\,s$^{-1}$}
\newcommand{\msun}{{\em M}$_{\sun}$}

\newcommand{\oii}{\ion{O}{ii}}
\newcommand{\oiii}{\ion{O}{iii}}
\newcommand{\neiii}{\ion{Ne}{iii}}
\newcommand{\civ}{\ion{C}{iv}}
\newcommand{\nv}{\ion{N}{v}}
\newcommand{\hii}{\ion{H}{ii}}
\newcommand{\heii}{\ion{He}{ii}}

\begin{document}
   \title{Protoclusters associated with $z > 2$ radio galaxies
   \thanks{Based on observations carried out at the European
   Southern Observatory, Paranal, Chile, programs 66.A-0597,
   LP167.A-0409, 68.B-0295 and 70.A-0589. Also based on data obtained
   at the W.M.\ Keck Observatory, which is operated as a scientific
   partnership among the California Institute of Technology, the
   University of California and the National Aeronautics and Space
   Administration. The Observatory was made possible by the generous
   financial support of the W.M.\ Keck Foundation.}}

   \subtitle{I.\ Characteristics of high redshift protoclusters}

   \author{B.\ P.\ Venemans \inst{1,2} \and H.\ J.\ A.\ R\"ottgering
           \inst{1} \and G.\ K.\ Miley \inst{1} \and W.\ J.\ M.\ van
           Breugel \inst{3,8} \and C.\ De Breuck \inst{4} \and J.\ D.\
           Kurk \inst{5} \and L.\ Pentericci \inst{6} \and S.\ A.\
           Stanford \inst{3,9} \and R.\ A.\
           Overzier \inst{1} \and S.\ Croft \inst{3,8} \and H.\ Ford \inst{7}}

   \offprints{B.\ P.\ Venemans, \email{venemans@ast.cam.ac.uk}}

   \institute{Sterrewacht Leiden, P.O.\ Box 9513, 2300 RA, Leiden, 
              The Netherlands               
         \and
              Institute of Astronomy, University of Cambridge,  
             Madingley Road, Cambridge, CB3 0HA, United Kingdom
         \and
              Lawrence Livermore National Laboratory, 
              P.O. Box 808, Livermore CA, 94550, USA
         \and
              European Southern Observatory, Karl Schwarzschild
              Stra{\ss}e 2, D-85748 Garching, Germany
         \and
              INAF, Osservatorio Astrofisico di Arcetri, Largo Enrico
              Fermi 5, 50125, Firenze, Italy
         \and
              Osservatorio Astronomico di Roma, Via di Frascati 33,
              00040 Monte Porzio Catone, Italy
         \and
              Dept. of Physics \& Astronomy, The Johns Hopkins University,
              3400 North Charles Street, Baltimore MD, 21218-2686, USA
         \and
              University of California, Merced, P.O.\ Box 2039, Merced, CA
              95344, USA
         \and
              University of California, Davis, 1 Shields Ave, Davis, CA 95616,
              USA
             }

   \date{Received 29 July 2005 / accepted 5 October 2006}

   \abstract{We present the results of a large program conducted with
     the Very Large Telescope and augmented by observations with the
     Keck telescope to search for forming clusters of galaxies near
     powerful radio galaxies at $2.0 < z < 5.2$. Besides MRC 1138--262
     at $z = 2.16$, the radio galaxy observed in our pilot program, we
     obtained narrow- and broad-band images of eight radio galaxies
     and their surroundings. The imaging was used to select candidate
     \lya\ emitting galaxies in $\sim3\times3$ Mpc$^2$ areas near the
     radio galaxies. A total of 300 candidate emitters were found with
     a rest-frame \lya\ equivalent width of {\em EW}$_0 > 15$ \AA\ and
     significance $\Sigma \equiv$ {\em EW}$_0/\Delta${\em EW}$_0 >
     3$. Follow-up spectroscopy was performed on 152 candidates in
     seven of the radio galaxy fields. Of these, 139 were confirmed to be
     \lya\ emitters, four were low redshift interlopers and nine were
     non-detections. With the adopted criteria the success rate is
     $139/152 = 91$\%. In addition, 14 objects with {\em EW}$_0 < 15$
     and/or $\Sigma < 3$ were confirmed to be \lya\ emitters. Combined
     with the 15 \lya\ emitters near MRC 1138--262, we have determined
     \lya\ redshifts for 168 objects near eight radio galaxies.

     At least six of our eight fields are overdense in \lya\ emitters
     by a factor 3--5 as compared to the field density of \lya\
     emitters at similar redshifts, although the statistics in our
     highest redshift field ($z = 5.2$) are poor. Also, the emitters
     show significant clustering in velocity space. In the overdense
     fields, the width of the velocity distributions of the emitters
     is a factor 2--5 smaller than the width of the narrow-band
     filters. Taken together, we conclude that we have discovered six
     forming clusters of galaxies (protoclusters).  We estimate that
     roughly 75\% of powerful ($L_\mathrm{2.7\,GHz} > 10^{33}$
     erg\,s$^{-1}$\,Hz$^{-1}$\,sr$^{-1}$) high redshift radio galaxies
     reside in a protocluster. The protoclusters have sizes of at least
     $1.75$ Mpc, which is consistent with the structure
     sizes found by other groups. By using the volume occupied by the
     overdensities and assuming a bias parameter of $b=3-6$, we
     estimate that the protoclusters have masses in the range $2-9
     \times 10^{14}$ \msun. These protoclusters are likely to be
     progenitors of present-day (massive) clusters of galaxies. For
     the first time, we have been able to estimate the velocity
     dispersion of cluster progenitors from $z\sim5$ to $\sim2$. The
     velocity dispersion of the emitters increases with cosmic time,
     in agreement with the dark matter velocity dispersion in
     numerical simulations of forming massive clusters. 

      ---
     cosmology: large scale structure of Universe }

   \maketitle
%

\section{Introduction}
\label{introduction}

Clusters of galaxies are the largest and most massive gravitationally bound
structures in the Universe. They are interesting objects to study for
many reasons. 

First, clusters contain large numbers of galaxies at specific redshifts,
making them excellent laboratories with which to investigate the
formation and evolution of galaxies. For example, the analysis of
galaxies in $z \sim 1$ clusters showed that the stars in massive,
early-type galaxies formed at $z > 2$
\citep[e.g.,][]{ell97,sta98,bla03a,dok03,hol05}. Investigating the galaxy
population of (forming) clusters at $z > 2$ could provide knowledge of
the formation process of such massive galaxies
\citep[e.g.,][]{egg62,lar74}. 
Also because clusters are the most extreme overdense regions in the
Universe, they allow an efficient investigation of the interaction
between galaxies and their environment
\citep[e.g.,][]{miles04,tan04,zee04,got05,nak05,tra05}. 

A second reason to study clusters is that they can place constraints
on cosmology. The number density of massive clusters is a strong
function of the fundamental cosmological parameters $\Omega_M$ and
$\sigma_8$, and the evolution of cluster abundances with redshift
depends primarily on $\Omega_M$ \citep[e.g.,][]{eke96}. The number
density of rich clusters at $z > 0.5$ has already been successfully
used to constrain the values of cosmological parameters
\citep[e.g.,][]{bah97,bah98,ett03}.

Several studies of massive clusters with redshifts up to $z = 1.4$
have found little evolution in the cluster properties
\citep[e.g.,][]{toz03,has04,mau04,ros04,mul05}. Despite the large
lookback times, clusters at $z\sim1$ appear to be very similar to
local clusters. For example, the $z=1.3$ cluster RDCS 1252.9-2927 has
thermodynamical properties and metallicity that are very similar to
those of lower redshift clusters \citep{ros04}. To study when and how
clusters and their galaxies formed, a sample of clusters at $z \gg 1$
is needed.

Unfortunately, conventional methods for finding distant clusters
become impractical at $z>1$. Searches for extended X-ray sources are
difficult because the surface brightness of the X-ray emission fades
as $(1+z)^4$. Although large optical surveys have been successful in
finding galaxy clusters at $z \lesssim 1$ by searching for
concentrations of red galaxies \citep[e.g.,][]{gla02}, the detection
of $z > 1$ clusters with the same method requires sensitive, wide
field near-infrared cameras which are not yet available. In the
future, surveys exploiting the Sunyaev-Zeldovich (SZ) effect
\citep[e.g.,][]{carl02} will be able to detect clusters of galaxies at
$z \gg 1$. However, at this moment the sensitivity of SZ surveys is
not sufficient to detect any of the known clusters at $z > 1$
\citep{carl02,ros04}. In recent years, several distant forming
clusters (protoclusters) have been serendipitously discovered in field
surveys \citep[e.g.,][]{ste98,ste05,shi03,ouc05}. For example,
\citet{ste98} found a large scale structure of Lyman Break galaxies
(LBGs) at $z\sim3.1$ in one of their fields. This single discovery
demonstrated the power of such structures as it could be derived that
LBGs must be very biased tracers of mass.

A different approach to find (forming) clusters is to search for a
galaxy concentration near a presumed tracer of high density
regions. There is considerable evidence that powerful high redshift
radio galaxies might be such tracers, as they are forming massive
galaxies in dense environments \citep[e.g.,][ see e.g.\ Carilli et
al.\ 2001 for a review]
{car97,dey97,ath98,pen98,pap00,pen00b,arc01,jar01,deb02,deb03a,
deb03b,stev03,zir03,reu04}. \nocite{car01} Targeted searches for
companion galaxies near powerful radio sources at $z > 1$ have long
yielded promising results
\citep[e.g.,][]{lef96,pas96,kee99,san99,san02,bes00,hal01,nak01,bes03,wol03,bar04}.

We therefore started a large program with the Very Large Telescope
(VLT) to systematically search for galaxy overdensities near radio
galaxies in the redshift range $2 < z < 5.2$. This program was
initiated after a successful pilot project in which the environment of
the radio galaxy MRC 1138--262 at $z=2.16$ was investigated. Deep
narrow-band images of the radio galaxy and the surrounding field were
obtained to search for an excess of \lya\ emitting galaxies
\citep{kur00}. The imaging and follow-up spectroscopy resulted in
$\sim 40$ candidate \lya\ emitters \citep{kur00,kur04b} of which 15
were confirmed at $z = 2.16 \pm 0.02$ \citep{pen00a,cro05}. The
presence of a forming cluster associated with MRC 1138--262 was firmly
established by the subsequent discovery of significant populations of
(spectroscopically-confirmed) H$\alpha$ emitters
\citep{kur04a,kur04b}, QSOs \citep{pen02,cro05} and extremely red
objects \citep{kur04b}. The VLT large program was augmented by
observations with the Keck telescope, 
to take advantage of the excellent UV throughput of LRIS-B
for spectroscopy and imaging below the Lyman break of some of the
$z\sim3$ objects.

In this paper, we present the results of our program to search for
forming clusters (protoclusters) near radio galaxies up to
$z=5.2$. Early results of the program, the discovery of galaxy
overdensities at $z = 4.1$ and $z = 5.2$, have been presented in
\citet{ven02} and in \citet{ven04}. A detailed analysis of the field
towards the radio galaxy MRC 0316--257 at $z=3.13$ is given in
\citet[][ hereafter V05]{ven05}. In V05 the data reduction steps,
selection procedure of candidate \lya\ emitters and the assessment of
the data quality are described in detail. Here we will follow the same
steps as V05 for the data reduction and analysis of the other
fields. While in this paper we will focus on the environment of radio
galaxies, in the second paper of this series we will describe the
properties of the individual \lya\ emitting galaxies discovered in our
program.

The structure of this paper is as follows: in Sect.\
\ref{sample} we present the targets of our program, and in Sect.\
\ref{ima}--\ref{spec} we give the details of the imaging observations,
the selection of candidate \lya\ emitters and follow-up
spectroscopy. In Sect.\ \ref{results}, the results of the imaging and
spectroscopy are described for each individual field. A summary of the
results and the evidence for the presence of protoclusters near the
radio galaxies are given in Sect.\ \ref{resres}. The properties
of the protoclusters are presented in Sect.\ \ref{protoclusters},
followed by a discussion of our results in Sect.\ \ref{discussion}
and a summary in Sect.\ \ref{summary}.

In this article, we adopt a $\Lambda$-dominated cosmology with H$_0 =
70$ \kms\,Mpc$^{-1}$, $\Omega_{M} = 0.3$, and $\Omega_{\Lambda} = 0.7$.
Magnitudes are given in the AB system \citep{oke74}. 

\section{Observations}
\label{obs}

\subsection{Sample selection}
\label{sample}

\begin{table}
\caption{\label{fields} Details of the radio galaxies observed in
our program.}
\begin{center}
\begin{tabular}{lccll}
\hline
\hline
Name & $\alpha_\mathrm{J2000}$ & $\delta_\mathrm{J2000}$ & ~$z$ &
$L_\mathrm{2.7\,GHz}^a$ \\
\hline
BRL 1602--174 & 16 05 01.7 & $-$17 34 18.4 & 2.04 & $2.0 \times
10^{34}$ \\
MRC 2048--272 & 20 51 03.5 & $-$27 03 04.1 & 2.06 & $6.3 \times
10^{33}$ \\
MRC 1138--262 & 11 40 48.2 & $-$26 29 09.5 & 2.16 & $1.3 \times
10^{34}$ \\
MRC 0052--241 & 00 54 29.8 & $-$23 51 31.1 & 2.86 & $8.6 \times
10^{33}$ \\
MRC 0943--242 & 09 45 32.7 & $-$24 28 49.7 & 2.92 & $7.2 \times
10^{33}$ \\
MRC 0316--257 & 03 18 12.0 & $-$25 35 10.8 & 3.13 & $1.4 \times
10^{34}$ \\
TN J2009--3040 & 20 09 48.1 & $-$30 40 07.4 & 3.16 & $2.8 \times
10^{33}$ \\
TN J1338--1942 & 13 38 26.1 & $-$19 42 30.8 & 4.11 & $9.6 \times
10^{33}$ \\
TN J0924--2201 & 09 24 19.9 & $-$22 01 42.0 & 5.20 & $1.5 \times
10^{34}$ \\
\hline
\end{tabular}
\end{center}
$^a$ Radio luminosity at a rest-frame frequency of 2.7 GHz in
erg\,s$^{-1}$\,Hz$^{-1}$\,sr$^{-1}$ \\ 
\end{table}

The targets for our program were selected from a list of approximately
150 radio galaxies with known redshifts of $z > 2$. 
Our targets were chosen to have large radio luminosities
($L_\mathrm{2.7\,GHz} > 10^{33}$ erg\,s$^{-1}$\,Hz$^{-1}$\,sr$^{-1}$).
Because our goal was to search for companion \lya\ emitting galaxies
near the radio sources, the radio galaxies had to have redshifts
optimum for imaging with the narrow-band filters that were available
at the VLT. Also, the radio sources need to lie in the southern
hemisphere ($\delta \lesssim 0^\circ$) to allow deep imaging of the
source with the VLT. Applying these criteria to the list of $\sim150$
$z > 2$ radio sources reduced the number of possible targets for our
program to 16 of which nine are at $z \sim 2.1$ (this includes the
target of our pilot project, MRC 1138--262 at $z=2.16$), three are at
$z \sim 2.9$ and four are at $z \sim 3.1$. Given the relatively small
number of radio sources at $z > 2.9$ that could be observed with the
VLT, we did not apply other criteria. To extend the redshift range, we
purchased narrow-band filters that were centred on the wavelength of
the \lya\ line at a redshift of $z = 4.1$ and $z = 5.2$. The following
nine radio sources were chosen as targets for our program: BRL
1602--174 (hereafter 1602), MRC 2048--272 (2048) and MRC 1138--262
(1138, all three at $z = 2.1$), MRC 0052--241 (0052) and MRC 0943--242
(0943, both at $z = 2.9$), MRC 0316--257 (0316) and TN J2009--3040
(2009, both at $z = 3.1$), TN J1338--1942 (1338, $z = 4.1$) and TN
J0924--2201 (0924, $z = 5.2$). Because these targets were selected
only on the basis of their radio luminosity and position on the sky,
we regard our sample as representative of luminous radio sources. The
position, redshift and radio power of the targets are given in Table
\ref{fields}. Individual radio galaxies are briefly described in
Sect.\ \ref{results}.

\subsection{Imaging observations}
\label{ima}

\begin{table*}
\caption{\label{imgobs} Overview of the imaging observations of the
  radio galaxy fields. }
\begin{tabular}{llllllllrc}
\hline
\hline
Date & Telescope & Instrument & Field & Filter & $\lambda_c^{~a}$
(\AA) & $\Delta \lambda^b$ (\AA) & Seeing & $t_\mathrm{exp}^c$ & Depth$^d$ \\
\hline
2001 Mar 24 \& 25 & VLT UT2 & FORS2 & 1338-1$^e$ & FILT\_621\_5 & 6199 &
59 & 0\farcs6 & 33\,300 & 28.2 \\ 
2001 Mar 24 \& 25 & VLT UT2 & FORS2 & 1338-1$^e$ & Special $R$ & 6550 &
1650 & 0\farcs6 & ~6\,300 & 28.9 \\ 
2001 Mar 24--26 & VLT UT2 & FORS2 & 0943 & HeII/6500 & 4781 & 68 &
0\farcs7 & 22\,500 & 28.6 \\ 
2001 Mar 25 \& 26 & VLT UT2 & FORS2 & 0943 & Bessel $B$ & 4290 & 880
& 0\farcs9 & ~4\,500 & 28.7 \\ 
2001 Mar 25 \& 26 & VLT UT2 & FORS2 & 1602 & OII & 3717 & 73 &
0\farcs8 & 15\,000 & 26.6 \\ 
2001 Mar 26 & VLT UT2 & FORS2 & 1602 & Bessel $B$ & 4290 & 880 &
0\farcs65 & ~2\,700 & 27.6 \\ 
2001 May 21--23 & VLT UT2 & FORS2 & 2048 & OII & 3717 & 73 & 0\farcs95 &
25\,200 & {--}$^f$ \\ 
2001 May 22 \& 23 & VLT UT2 & FORS2 & 2048 & Bessel $B$ & 4290 & 880
& 1\farcs05 & ~3\,600 & {--}$^f$ \\ 
2001 Sep 20--22 & VLT UT4 & FORS2 & 2048 & OII & 3717 & 73 &
0\farcs95 & 25\,200 & 27.9 \\ 
2001 Sep 20 \& 21 & VLT UT4 & FORS2 & 2048 & Bessel $B$ & 4290 & 880
& 1\farcs05 & ~3\,000 & 28.8 \\ 
2001 Sep 20 \& 21 & VLT UT4 & FORS2 & 0316 & OIII/3000 & 5045 & 59 &
0\farcs7 & 23\,400 & 28.4 \\ 
2001 Sep 20 \& 21 & VLT UT4 & FORS2 & 0316 & Bessel $V$ & 5540 & 1115
& 0\farcs7 & ~4\,860 & 28.9 \\ 
2001 Sep 22 & VLT UT4 & FORS2 & 0052 & HeII & 4684 & 66 & 0\farcs75 &
~5\,400 & {--}$^f$ \\ 
2001 Oct 20 & VLT UT4 & FORS2 & 0052 & HeII & 4684 & 66 & 0\farcs75 &
18\,000 & 28.3 \\ 
2001 Oct 20 & VLT UT4 & FORS2 & 0052 & Bessel $B$ & 4290 & 880 &
0\farcs8 & ~4\,800 & 29.0 \\ 
2002 Mar 8 & VLT UT4 & FORS2 & 0924 & FILT\_753\_8 & 7528 & 89 &
0\farcs8 & 14\,400 & {--}$^f$ \\ 
2002 Mar 8 & VLT UT4 & FORS2 & 0924 & Bessel $I$ & 7680 & 1380 &
0\farcs8 & ~2\,700 & {--}$^f$ \\ 
2002 Apr 17--19 & VLT UT4 & FORS2 & 0924 & FILT\_753\_8 & 7528 & 89 &
0\farcs8 & 28\,800 & 28.1 \\ 
2002 Apr 17--19 & VLT UT4 & FORS2 & 0924 & Bessel $I$ & 7680 & 1380 &
0\farcs8 & ~8\,640 & 28.5 \\ 
2002 Apr 17--19 & VLT UT4 & FORS2 & 1338-2$^e$ & FILT\_621\_5 & 6199 & 59
& 0\farcs75 & 25\,200 & 28.2 \\  
2002 Apr 17--19 & VLT UT4 & FORS2 & 1338-2$^e$ & Special $R$ & 6550 &
1650 & 0\farcs75 & ~4\,500 & 28.9 \\
2002 Apr 17--19 & VLT UT4 & FORS2 & 2009 & OIII/3000 & 5045 & 59 &
0\farcs9 & 21\,600 & {--}$^f$ \\
2002 Apr 17--19 & VLT UT4 & FORS2 & 2009 & Bessel $V$ & 5540 & 1115 &
0\farcs85 & ~4\,800 & {--}$^f$ \\
2002 Apr 19 & VLT UT4 & FORS2 & 0924 & Bessel $V$ & 5540 & 1115 &
1\farcs05 & ~3\,600 & 28.6 \\
2002 Sep 6--8 & VLT UT4 & FORS2 & 0316 & Bessel $I$ & 7680 & 1380 &
0\farcs7 & ~4\,680 & 28.7 \\
2002 Sep 8 & VLT UT4 & FORS2 & 2009 & OIII/3000 & 5045 & 59 &
0\farcs9 & ~7\,200 & 28.0 \\
2002 Sep 8 & VLT UT4 & FORS2 & 2009 & Bessel $V$ & 5540 & 1115 &
0\farcs85 & ~2\,400 & 28.6 \\
2002 Sep 8 & VLT UT4 & FORS2 & 0052 & Bessel $V$ & 5540 & 1115 &
0\farcs75 & ~5\,400 & 29.2 \\
2002 Sep 8 & VLT UT4 & FORS2 & 0052 & Bessel $I$ & 7680 & 1380 &
0\farcs55 & ~4\,800 & 28.5 \\
2003 Jan 31 & Keck I & LRIS-B & 0316 & $u'$ & 3550 & 600 & 
1\farcs25 & ~4\,050 & {--}$^f$ \\ 
2003 Feb 1 \& 4 & Keck I & LRIS-B & 0316 & $u'$ & 3550 & 600 &
1\farcs25 & ~9\,000 & 29.8 \\
2003 Feb 4 & Keck I & LRIS-B & 0943 & $u'$ & 3550 & 600 & 
1\farcs2 & ~9\,600 & {--}$^f$ \\ 
2003 Feb 4 & Keck I & LRIS-R & 0943 & $V$ & 5473 & 948 & 
1\farcs25 & ~5\,600 & 29.4 \\ 
2004 Jan 19 & Keck I & LRIS-B & 0943 & $u'$ & 3550 & 600 & 
1\farcs2 & ~7\,000 & 29.8 \\  
2004 Jan 19 & Keck I & LRIS-R & 0943 & $I$ & 8331 & 3131 & 
0\farcs90 & ~6\,000 & 28.4 \\ 
\hline
\end{tabular} \\
$^a$ Central wavelength of the filter in \AA. \\
$^b$ Full width at half maximum (FWHM) of the filter in \AA. \\
$^c$ Total exposure time in seconds. \\
$^d$ 1 $\sigma$ depth of the resulting image in mag per
$\square$\arcsec.  \\ 
$^e$ The field towards TN J1338--1942 was observed at two different
pointings, see Sect.\ \ref{1338res}. \\
$^f$ For images obtained during two different observing sessions,
only the total depth is listed in the final entry. \\
\end{table*}

To search for structures of \lya\ emitting galaxies near the radio
galaxies, the fields surrounding the radio galaxies were observed in a
narrow-band filter and at least one broad-band filter. The narrow-band
filters were chosen to encompass the \lya\ line at the redshift of the
radio galaxy, and the broad-band filters were selected to measure the
UV continuum redward of the \lya\ line.

All the narrow-band imaging and most of the broad-band imaging
obtained in the large program were performed with the FOcal Reducer/
low dispersion Spectrograph 2 \citep[FORS2;][]{app92} in imaging
mode. Before 2002 April, the detector in FORS2 was a SiTE CCD with
2048$\times$2048 pixel$^2$ and a pixel scale of
0\farcs2\,pixel$^{-1}$. In 2002 April, the SiTE CCD was replaced by
two MIT CCDs each with 2048$\times$2048 pixel$^2$ with a scale of
0\farcs125\,pixel$^{-1}$. The pixels were binned by $2\times2$, which
decreases the readout time by a factor of 2 and gives a pixels scale
of 0\farcs25\,pixel$^{-1}$. The field of view, which is restricted by
the geometry of the Multi-Object Spectroscopy unit, is
6\farcm8$\times$6\farcm8.

Additional (broad-band) imaging was obtained using the Low Resolution
Imaging Spectrometer \citep[LRIS,][]{oke95} on the Keck I
telescope. LRIS has two arms: a blue channel which is optimised for
observations in the blue part of the optical spectrum \citep[LRIS-B,
see][ for more information]{mcc98b,ste04} and a red channel for
observing in the red (LRIS-R, Oke et al.\ 1995). The red arm is
equipped with a Tektronix CCD with 2048$\times$2048 pixel$^2$. The
pixel scale is 0\farcs21\,pixel$^{-1}$, resulting in a field of view
of 7\farcm3$\times$7\farcm3. The blue channel has two 2048$\times$4096
Marconi CCDs with a pixel scale of 0\farcs135\,pixel$^{-1}$. There is
a small gap between the CCDs of roughly 13\farcs5. The field of view
is $\sim$8\arcmin$\times$8\arcmin\footnote{The specifications for the
blue channel detectors given here apply for data taken after 2002
June.}.

Observations were split into separate exposures of typically
1200--1800 s in the narrow-band and 240--800 s in the
broad-band. Individual exposures were shifted 10\arcsec--15\arcsec\
with respect to each other to facilitate the identification of cosmic rays and
the removal of residual flat-field errors. The data were reduced using
standard routines within the reduction software package
IRAF\footnote{IRAF is distributed by the National Optical Astronomy
Observatories, which are operated by the Association of Universities
for Research in Astronomy, Inc., under cooperative agreement with the
National Science Foundation.}. These routines included bias
subtraction using either bias frames or the overscan region of the
CCD, flat fielding with twilight sky flats and illumination correction
using the unregistered science frames. 

All science images were registered on the ICRF astrometric frame of
reference \citep{ma98}, using the USNO-A2.0 catalogue
\citep{mon98a,mon98b}. The relative positions of objects in the fields
are accurate to 0\farcs1--0\farcs2. The absolute accuracy is
dominated by the uncertainty in the USNO-A2.0 catalogue of 0\farcs25
\citep{deu99}.

The photometric calibration was performed using several photometric
and spectrophotometric standard stars from the catalogues of Stone \&
Baldwin (1983), Baldwin \& Stone (1984), Oke (1990) and Landolt
(1992). \nocite{sto83,bal84,oke90,lan92} The magnitude zero-points
derived from these standard stars are consistent with each other within
2--3\%. Zero-points in the Vega system are converted to the AB system
using the transformations of \citet{bes79} and \citet{smi02}. The
zero-points were corrected for galactic extinction as estimated by
\citet{sch98}.

Table \ref{imgobs} summarizes the imaging observations of our Large
Program targets and gives the properties of the narrow- and broad-band
filters that were used. Also, the image quality (seeing) and depth (in
1 $\sigma$ limiting magnitudes per square arcsecond) are given in
Table \ref{imgobs}. 

\subsection{Candidate selection}
\label{canselect}

\begin{table}
\caption{\label{imgres} Depth, sensitivity and size of the imaging
  observations of the radio galaxies fields, and the number of
  candidates selected in each field.}
\begin{center}
\begin{tabular}{llllll}
\hline
\hline
Field & $m_{\mathrm{lim}}^a$ & $F_{5\sigma}^b$ & $L_{5\sigma}^c$ &
N$^d$ & Area$^e$ \\   
{} & {} &  erg\,s$^{-1}$\,cm$^{-2}$ & erg\,s$^{-1}$ & {} & arcmin$^2$ \\
\hline
1602 & 24.4 & $9.7 \times 10^{-17}$ & $3.4 \times 10^{42}$ & ~2~ & ~~42.3 \\ 
2048 & 25.4 & $3.3 \times 10^{-17}$ & $1.2 \times 10^{42}$ & 10~ & ~~43.6 \\ 
1138 & 25.2$^f$ & $3.5 \times 10^{-17}$$^f$ & $1.4 \times 10^{42}$$^f$
& 37$^f$ & ~~46.6$^f$ \\  
0052 & 26.1 & $1.1 \times 10^{-17}$ & $8.4 \times 10^{41}$ & 57~ & ~~44.9 \\ 
0943 & 26.1 & $7.4 \times 10^{-18}$ & $6.2 \times 10^{41}$ & 65~ & ~~46.5 \\ 
0316 & 26.3 & $6.9 \times 10^{-18}$ & $6.9 \times 10^{41}$ & 77~ & ~~45.8 \\  
2009 & 25.7 & $1.3 \times 10^{-17}$ & $1.3 \times 10^{42}$ & 21~ & ~~46.7 \\ 
1338-1 & 26.0 & $4.7 \times 10^{-18}$ & $8.9 \times 10^{41}$ & 31$^g$
& ~~40.1 \\  
1338-2 & 26.2 & $5.8 \times 10^{-18}$ & $1.1 \times 10^{42}$ & 33$^g$
& ~~48.9 \\  
0924 & 25.5 & $7.0 \times 10^{-18}$ & $2.3 \times 10^{42}$ & 14~ & ~~46.8 \\ 
\hline
\end{tabular}
\end{center}
$^a$ Magnitude at which 50\% of artificial and real point sources that were
added to the narrow-band image, were recovered. \\
$^b$ Line flux of an emitter with no continuum
that is detected at the 5$\sigma$ level in an aperture with a diameter
twice that of the seeing disc. \\
$^c$ Line luminosity of an emitter with no continuum that
is detected at the 5$\sigma$ level in an aperture with a diameter
twice that of the seeing disc. \\
$^d$ Number of candidate \lya\ emitters that fulfils the selection
criteria {\em EW}$_0 > 15$ \AA\ and {\em EW}$_0$/$\Delta${\em EW}$_0
> 3$. \\ 
$^e$ Imaging area useful for selection of \lya\ emitters. \\
$^f$ Values taken from \citet{kur04b}. \\
$^g$ The 1338-1 and 1338-2 fields overlap and have an area of 9.3
arcmin$^2$ in common. The total number of unique candidate emitters
in the two fields is 54 in an area of 79.7 arcmin$^2$. \\
\end{table}

Objects in the images were detected using the program SExtractor
\citep{ber96}. The narrow-band images were taken to detect objects,
and aperture photometry was subsequently performed on both the
narrow-band and the broad-band images. 

To assess the completeness of the source detection, artificial and real
point sources were added to the narrow-band image and recovered. The
completeness limit was defined as the narrow-band magnitude at which
50\% of the added sources was recovered (see V05 for details).

Detected objects were required to have a signal-to-noise of $>5$ in
the narrow-band image. The colors of the detected objects were
measured in circular apertures, while the ``total'' flux was measured
in an elliptical aperture.  A correction was made to the ``total''
flux to account for the flux outside the elliptical aperture. More
details on the object detection, completeness assessment, aperture
sizes for the photometry and ``total'' flux correction can be found in
V05.

Following \citet{ven02} and V05, we selected objects with a rest-frame
equivalent width {\em EW}$_0 > 15$ \AA\ and a significance $\Sigma$
$\equiv$ {\em EW}$_0$/$\Delta${\em EW}$_0$ $> 3$ as good candidate
\lya\ emitters. In V05 a detailed description is presented on how {\em
EW}$_0$ and $\Delta${\em EW}$_0$ are computed from the available
photometry. For two fields that are imaged in at least two broad-band
filters, the 0316 and 0052 fields, the UV continuum slope $\beta$
($f_\lambda \propto \lambda^\beta$) of candidate emitters was also
computed. For the other fields, a ``flat'' continuum slope $\beta =
-2$ was used to select the candidates, which is close to the median
$\beta$ of confirmed \lya\ emitters in the 0316 field ($\beta =
-1.76$, V05).  In each field the candidate emission line galaxies were
visually inspected and spurious sources (like spikes of bright,
saturated stars) were removed from the catalogues. The resulting lists
should have a very low fraction of contaminants, such as low redshift
interlopers ($\sim 5$\% as esimated by V05).

In Table \ref{imgres} the 50\% completeness limit, number of candidate
emitters and area of each observed field is listed. We also
calculated the $5 \sigma$ limiting line flux and line luminosity of an
emitter with a negligible continuum. The number of
candidates in the 1138 field is taken from \citet{kur04b}.

\subsection{Spectroscopic observations}
\label{spec}

\begin{table*}
\caption{\label{mosobs} Overview of the spectroscopic observations of
  the radio galaxy fields.}
\begin{center}
\begin{tabular}{llllllclr}
\hline
\hline
Date & Telescope & Instrument & Field & Mode$^a$ & Grism name &
Dispersion$^b$ & Resolution$^c$ & t$_\mathrm{exp}^d~~$ \\
\hline
2001 May 20 \& 22 & VLT UT2 & FORS2 & 1338-1 & MXU, mask A &
GRIS\_600RI & 1.32 & 260\,$\times$\,1\farcs0 & 31\,500 \\
2001 May 21 \& 22 & VLT UT2 & FORS2 & 1338-1 & MXU, mask B &
GRIS\_600RI & 1.32 & 265\,$\times$\,1\farcs0 & 35\,100 \\
2001 Sep 22 & VLT UT4 & FORS2 & 0316 & MOS & GRIS\_1400V & 0.50 &
130\,$\times$\,1\farcs5 & 12\,600 \\
2001 Oct 18 \& 19 & VLT UT4 & FORS2 & 2048 & MXU & GRIS\_600B &
2.40 & 410\,$\times$\,1\farcs0 & 16\,200 \\
2001 Oct 18 & VLT UT4 & FORS2 & 0316 & MXU, mask A & GRIS\_1400V
& 1.00 & 140\,$\times$\,1\farcs0 & 10\,800 \\
2001 Oct 18--20 & VLT UT4 & FORS2 & 0316 & MXU, mask B &
GRIS\_1400V & 1.00 & 140\,$\times$\,1\farcs0 & 29\,100 \\
2001 Nov 15 \& 16 & VLT UT3 & FORS1 & 0316 & PMOS & GRIS\_300V
& 2.64 & 630\,$\times$\,0\farcs8 & 19\,800 \\
2002 Jan 14 & Keck I & LRIS & 0943 & MSS, mask A & 600/4000 &
1.01 & 310\,$\times$\,0\farcs85 & 10\,800 \\
2002 Jan 15 & Keck I & LRIS & 0943 & MSS, mask B & 600/4000 &
1.01 & 325\,$\times$\,0\farcs9 & 9\,000 \\
2002 Sep 6 \& 7 & VLT UT4 & FORS2 & 2009 & MXU & GRIS\_1400V &
0.62 & 110\,$\times$\,0\farcs75 & 19\,800 \\   
2002 Sep 6 & VLT UT4 & FORS2 & 0052 & MXU, mask A & GRIS\_1400V
& 0.62 & 100\,$\times$\,0\farcs65 & 8\,400 \\
2002 Sep 6 \& 7 & VLT UT4 & FORS2 & 0052 & MXU, mask B &
GRIS\_1400V & 0.62 & 100\,$\times$\,0\farcs65 & 17\,550 \\
2002 Sep 7 & VLT UT4 & FORS2 & 0052 & MXU, mask C & GRIS\_1400V
& 0.62 & ~95\,$\times$\,0\farcs6 & 10\,800 \\
2002 Sep 8 & VLT UT4 & FORS2 & 0052 & MOS & GRIS\_1400V &
0.62 & ~95\,$\times$\,0\farcs6 & 9\,600 \\
2003 Jan 31 & Keck I & LRIS & 0943 & MSS, mask C & 600/4000 &
0.61 & 205\,$\times$\,0\farcs7 & 7\,200 \\
2003 Feb 1 & Keck I & LRIS & 0943 & MSS, mask D & 600/4000 &
0.61 & 215\,$\times$\,0\farcs75 & 7\,200 \\
2003 Feb 1 \& 4 & Keck I & LRIS & 1338-2 & MSS & 400/8500 &
1.85 & 320\,$\times$\,0\farcs75 & 9\,000 \\
2003 Mar 3 \& 4 & VLT UT4 & FORS2 & 0924 & MXU & GRIS\_600RI &
1.66 & 225\,$\times$\,0\farcs85 & 20\,676 \\
2003 Mar 4 & VLT UT4 & FORS2 & 1338-2 & MXU & GRIS\_1200R & 0.76 &
125\,$\times$\,0\farcs85 & 18\,600 \\
\hline
\end{tabular}
\end{center}
$^a$ Explanation of the different observing modes: \\
{~~} MOS: Multi-object spectroscopy mode of FORS2, performed with 19
movable slitlets with lengths of 20\arcsec--22\arcsec. \\
~~~~ MXU: Multi-object spectroscopy mode of FORS2 with a user-prepared mask. 
\\
~~~~ PMOS: Spectropolarimetry mode of FORS1 using 9 movable slitlets of
20\arcsec. \\
~~~~ MSS: Multi-slit spectroscopy mode of LRIS, which uses a custom laser-cut
mask. \\ 
$^b$ Dispersion in \AA\,pixel$^{-1}$. \\
$^c$ The resolution is given for both the dispersion and spatial
axis. The units are [\kms]\,$\times$\,[\arcsec]. \\
$^d$ Total exposure time in seconds. \\
\end{table*}

To confirm whether the candidate emission line objects are located at
the redshift of the radio galaxy, spectra were taken of candidate
emitters. Priority was given to the most luminous
candidates. In Table \ref{mosobs} a summary is given of the
spectroscopic observations of candidate emitters in the radio galaxy
fields.

Most of the spectroscopic observations were performed using
user-defined masks (the multi-object spectroscopy (MXU) mode of FORS2
and multi-slit spectroscopy (MSS) modes of LRIS in Table \ref{mosobs}). This
mode allowed us to observe between 20 and 40 objects per
slitmask. The slits typically had a length of 10\arcsec--12\arcsec\ and a
width of 1\farcs0--1\farcs4. The grisms used were selected to have
the highest throughput at the wavelength of the \lya\ line of the
radio galaxy and a resolution that matches the width of the \lya\
lines ($\sim300$ \kms, V05) to maximize the confirmation rate. 

Individual exposures were typically 1800--2700 s, which ensured that
the spectra were limited by the sky noise. Between the exposures the
pointing of the telescope was shifted 2\arcsec--5\arcsec\ along the
slits to enable more accurate sky subtraction. For the flux
calibration long slit exposures of (at least) one of the following
spectroscopic standard stars EG 274, Feige 67, Feige 110, GD 108, LTT
377, LTT 1020, LTT 1788, LTT 6248 and LTT 7987
\citep{sto83,bal84,oke90} were used. The flux calibration is accurate
to about $\sim 5$\%. This does not take into account the uncertainties
due to slit losses. Because we calculated the total flux of the
emission lines using the imaging photometry, we did
not attempt to correct the spectra for flux falling outside the slit.

The details of the reduction of the spectroscopic data is given in
V05. In the next section, we will describe the results in the
individual fields.

\section{Results}
\label{results}

In this section we describe the results of the imaging, candidate
selections and follow-up spectroscopy of the \lya\ emitters in each of
the nine radio galaxy fields, followed by a description of the diffuse
\lya\ halos of the radio galaxies. For the details of the results in
the 1138, 0316 and 0924 fields we refer to Kurk et al.\ (2000),
Pentericci et al.\ (2000), Venemans et al.\ (2004), Kurk et al.\
(2004b), V05 and Croft et al.\ (2005). Below a brief summary of the
results in these fields is given.
\nocite{kur00,pen00a,kur04b,ven04,cro05}

Redshift and \lya\ line properties were measured by fitting a Gaussian
function to the emission line. If absorption features were present, a
combination of a Gaussian and a Voigt absorption profile was
fitted. The properties of the confirmed \lya\ emitters in the various
fields are summarized in Tables \ref{2048table}--\ref{1338table}. In
the Tables, the objects are ordered on increasing right ascension.

\subsection{BRL 1602--174, $z = 2.04$}
\label{1602res}

The radio source BRL 1602--174 was optically
identified with a galaxy with m$_R = 21.4$ object along the radio
axis. A spectrum of this galaxy yielded a redshift of
$2.043 \pm 0.002$ based on four emission lines \citep{bes99}. This
places the \lya\ line of the radio galaxy in the lowest wavelength
narrow-band filter available for FORS2, the \oii\ filter. The
field was imaged for 250 minutes in the narrow-band and for 45 minutes
in the $B$-band. Due to the high extinction towards this field
\citep[$A_B\approx1$,][]{sch98} and the low quantum efficiency at
wavelengths $< 4000$ \AA\ of the FORS2 detector, only two candidate
\lya\ emitters were found. No spectra were taken in this field.  

\subsection{MRC 2048--272, $z = 2.06$}
\label{2048res}

\begin{figure*}
\includegraphics[width=17cm]{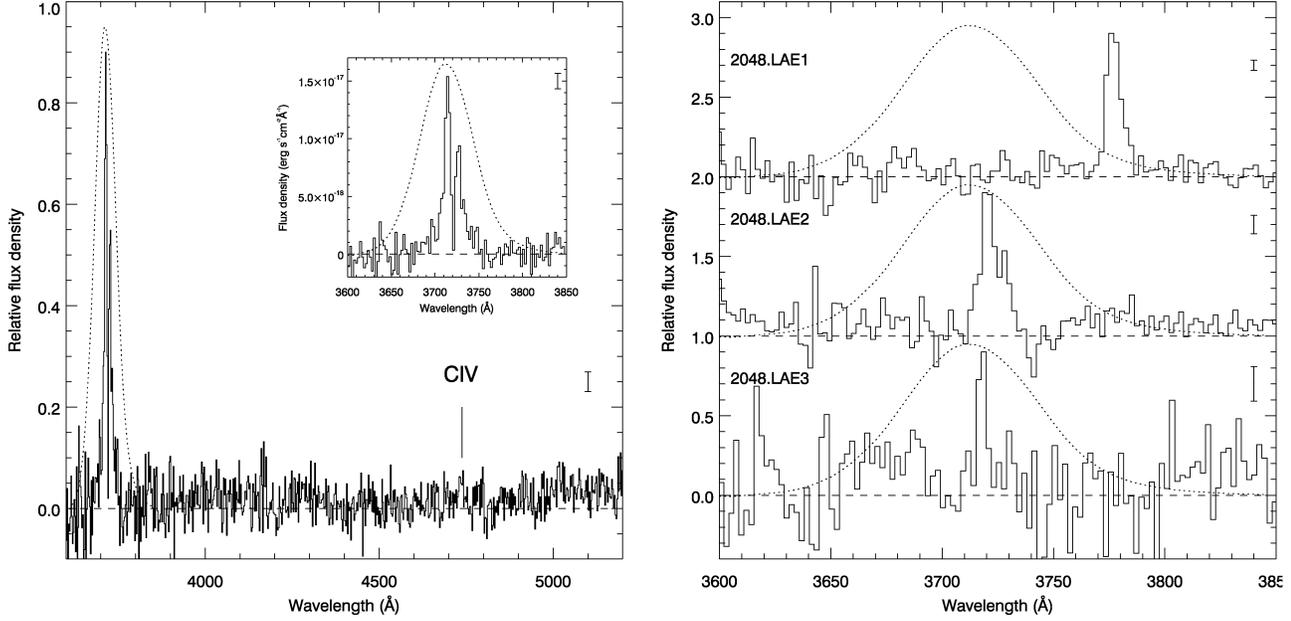}
\caption{\label{2048spec} Spectra of the radio galaxy MRC 2048--272
  ({\em left}) and the three confirmed \lya\ emitters near the radio
  galaxy ({\em right}). Inserted in the left plot is a close-up of the
  \lya\ line of the radio galaxy. The dotted curve represents the
  transmission of the narrow-band filter that was used to select the
  candidate \lya\ emitters. The uncertainty in the flux density is
  indicated by the error bar in the right corner of each spectrum.}
\end{figure*}

MRC 2048--272 was listed in the 408 MHz Molonglo Reference Catalogue
\citep{lar81} and identified with a $z = 2.06$ object by
\citet{mcc96}. High resolution imaging in the infrared with the {\em
HST} revealed three separate components within 3\arcsec\
\citep{pen01}, of which the central object was identified as the radio
galaxy. The surface brightness of the central object could be well fit by a
de Vaucouleurs profile, indicating that a dynamically relaxed stellar
population is in place in this radio galaxy \citep{pen01}. \\

\centerline{\em Imaging observations}

\noindent
We observed the field in 2001 May under moderate seeing conditions
($\sim$1\arcsec) for 7 hours in the \oii\ narrow-band filter
and for 1 hour in the $B$-band. Because of the low efficiency of the
detector at $\lambda < 4000$ \AA\ and the high extinction towards the
field of $A_B\approx0.4$ \citep{sch98}, the field was imaged again in 2001
September. The combined 14 hours of narrow-band observations of this
field have a depth comparable to that of the field surrounding MRC
1138--262 \citep[][ Table \ref{imgres}]{kur00}. In total 10 candidate
\lya\ emitters were found in this field. The number of contaminants is
expected to be low in this field, because the only strong line that
falls in the filter is [\oii] $\lambda 3727$ at a redshift of
$z < 0.007$. \\

\centerline{\em Spectroscopic observations}

\noindent
Due to geometrical constraints only three of the candidate emitters
could be observed at the same time. Additional targets were included
on the slitmask, including objects with a low equivalent width {\em
EW}$_0 < 15$ \AA. Two of the three candidates and the radio galaxy
show a line in a 4 hr spectrum (see Table \ref{2048table} and Fig.\
\ref{2048spec}). The third emitter has most likely an emission line
that is too faint ($< 3\times10^{-17}$ \ergscm) to be confirmed in a 4
hr spectroscopic observation with FORS2 at blue wavelengths
($\lambda\sim3750$ \AA). The two confirmed emitters have a redshift
very close to that of the radio galaxy with relative velocities of 100
and 10 \kms. A third \lya\ emitter was found among the candidate
emitters with a lower equivalent width. This galaxy is a bright \lya\
emitter located $\sim 4600$ \kms\ away from the radio galaxy, with the
emission line at the edge of the narrow-band filter (Fig.\
\ref{2048spec}). \\

\centerline{\em Volume density}

\noindent
\citet{kur04b} found that the 1138 field is overdense in \lya\
emitters by a factor of $4\pm2$ compared to the field (volume) density
of \lya\ emitters as derived by \citet{sti01}. Comparing the volume
density of \lya\ emitters near MRC 2048--272 to that of emitters near
MRC 1138--262 \citep{kur04b}, the density in the 2048 field is a
factor 3.4$^{+1.8}_{-1.2}$ smaller. The errors are based on Poisson
statistics in the small numbers regime \citep{geh86}. Using this
factor, the volume density of emitters near 2048 is
$1.2^{+0.8}_{-0.7}$ times the field density of emitters. A direct
comparison with \citet{sti01} gives a similar density
($n_{2048}/n_{\mathrm{field}} = 0.7^{+1.8}_{-0.6}$). The small
difference between the two density estimates (at the $0.3\sigma$
level) is due the small numbers of galaxies involved in the
comparison. The density in the 2048 field is consistent with no
overdensity of emitters near the radio galaxy.

\subsection{MRC 1138--262, $z = 2.16$}
\label{1138res}

This radio galaxy was the target of our pilot project (see also Sect.\
\ref{introduction}). Narrow- and broad-band imaging with FORS1 on the
VLT resulted in the detection of 37 candidate \lya\ emitters
\citep{kur00,kur04b}. Subsequent spectroscopy confirmed 15 \lya\
emitters to be near the radio galaxy at $z=2.16$ \citep{pen00a}. The
density of \lya\ emitters in this field is roughly a factor 4 higher
as compared to field studies \citep{pen00a,kur04b}. This field is also
overdense in X-ray sources \citep{pen02} and extremely red objects
\citep[EROs,][]{kur04b}. In total, four of the X-ray sources are
confirmed to be members of the protocluster \citep{pen02,cro05}. Also,
nine H$\alpha$ emitters were spectroscopically confirmed to be
associated with the radio galaxy \citep{kur04a,kur04b}, increasing the
number of confirmed protocluster members to $>$25.

\subsection{MRC 0052--241, $z = 2.86$}
\label{0052res}

\begin{figure}
\includegraphics[width=8.5cm]{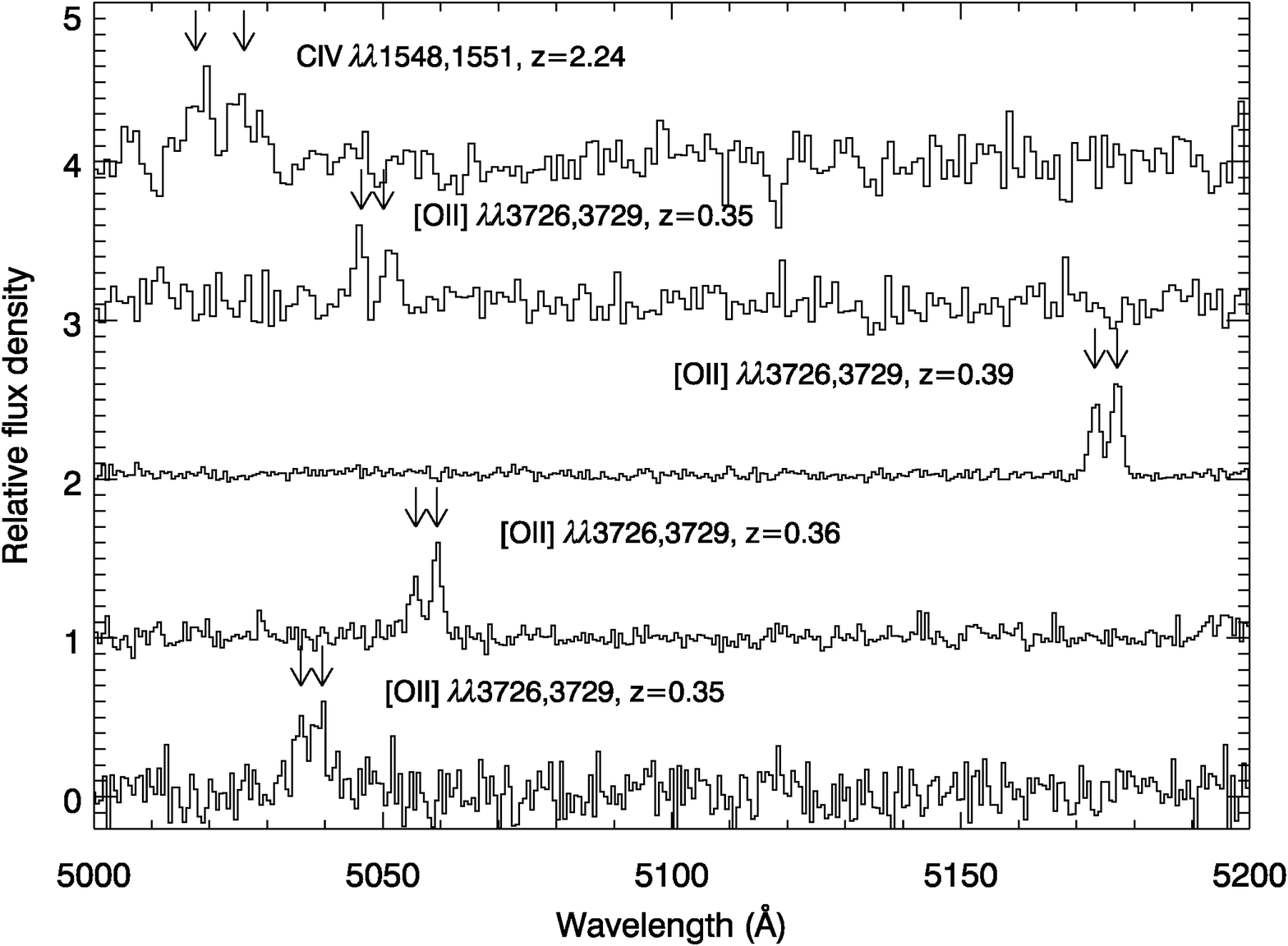}
\caption{\label{oiicivem} Spectra of low redshift emission line
  galaxies which could contaminate our sample. These spectra were
  observed through the 1400V grism that was used for the spectroscopy
  observations of the 0052, 0316 and 2009 fields. Due to the resolution of
  $R = 2100$, interloping galaxies at low redshifts can easily be
  distinguished from \lya\ emitters at $z\sim3$.}
\end{figure}

\begin{figure*}
\includegraphics[width=17cm]{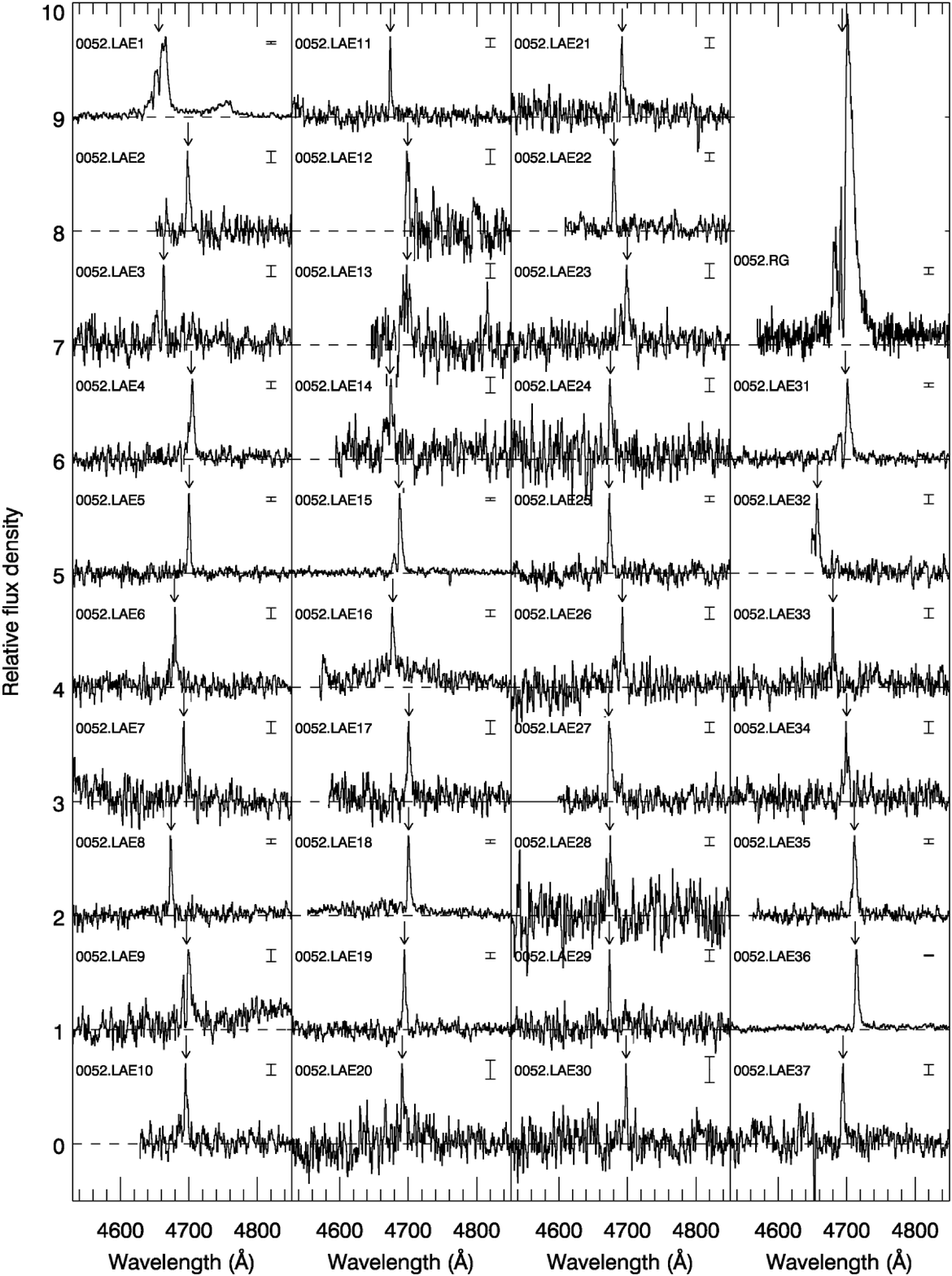}
\caption{\label{0052spectra} Spectra of the confirmed emitters near
  the radio galaxy MRC 0052--241. The spectrum of the radio galaxy is
  shown in the top-right panel of the Figure. The error bar in the
  right corner of each spectrum indicates the average error in the
  flux density of each pixel. All spectra are boxcar averaged over
  three pixels and normalized to the peak of the \lya\ line at a
  relative flux density of 0.7. The offset between the spectra is
  1.0. The arrow indicates the center of the emission line of each
  emitter. The redshift, flux and width of the emission lines can be
  found in Table \ref{0052table}.}
\end{figure*}

\begin{figure*}
\includegraphics[width=17cm]{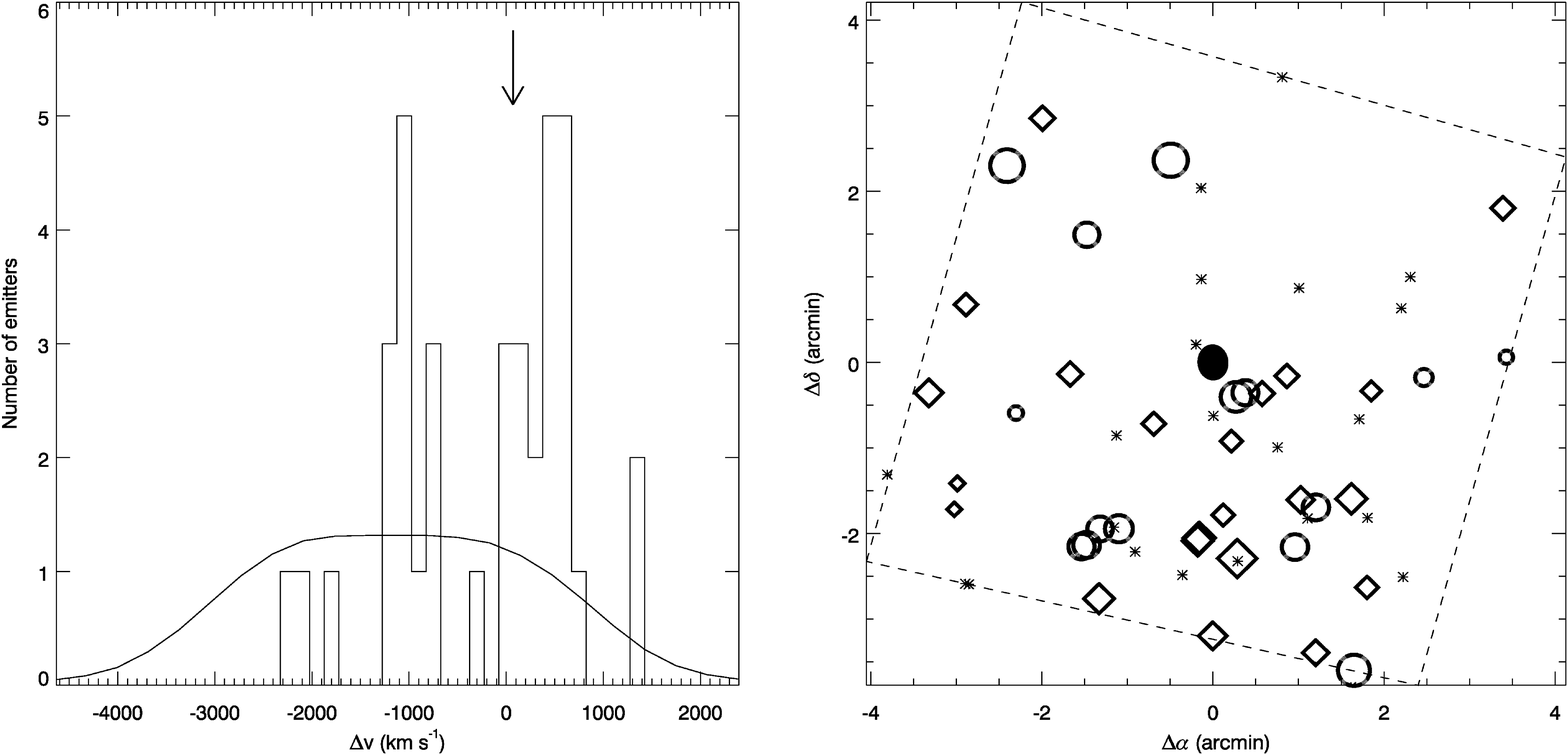}
\caption{\label{0052velskydist} {\em Left:} Velocity distribution of
  the confirmed emitters near MRC 0052--241. The median redshift
  ($z=2.8600$) of the emitters is taken as zero-point. The bin size is
  150 \kms. The relative velocity of the radio galaxy (70 \kms, see
  Table \ref{0052table}) is indicated by an arrow. The solid line
  represents the selection function of the narrow-band filter,
  normalized to the total number of confirmed emitters. {\em Right:}
  Spatial distribution of the confirmed (circles and diamonds) and
  remaining candidate (stars) \lya\ emitters. The dashed quadrangle
  represents the outline of the imaging area. The radio galaxy is
  denoted by a filled ellipse, that depicts the approximate shape and
  position angle of the radio galaxy \lya\ halo (see Table
  \ref{halotable}). The circles represent the emitters with a redshift
  smaller than the median redshift ($z < 2.8600$) and the diamonds
  those with $z > 2.8600$. The size of the symbols are scaled
  according to the velocity offset from the median, with smaller
  symbols standing for emitters with a redshift farther from the
  median.}
\end{figure*}

The optical counterpart of the radio source 0052--241 from the
Molonglo Reference Catalogue \citep{lar81} was found by \citet{mcc96}
to be a $m_R = 23.2$ object. A spectrum of this object showed strong
\lya\ emission at a redshift of $z=2.86$ \citep{mcc96}. \\

\centerline{\em Imaging and spectroscopic observations}

\noindent
At $z=2.86$ the \lya\ line is shifted into the FORS2 narrow-band
\heii\ filter. The field was imaged for 390 min in this narrow-band
and for 80, 90 and 80 min in the $B$-band, $V$-band and $I$-band
respectively (Table \ref{imgobs}). Analysis of these data resulted in
a list of 57 candidate \lya\ emitters with {\em EW}$_0 > 15$ \AA\ and
{\em EW}$_0$/$\Delta${\em EW}$_0 > 3$ (Sect.\ \ref{canselect}).

Follow-up spectroscopy of candidate emitters was carried out in 2002
September at the VLT. 36 candidates were observed in four masks with
individual exposure times between 140 and 292.5 min under
good seeing conditions (0\farcs6--0\farcs65, see Table
\ref{mosobs}). The resolution of the grism that we used (the 1400V
grism) is $R = 2100$, which is high enough to resolve the [\oii]
$\lambda\lambda 3726,3729$ and \civ\ $\lambda\lambda 1548,1551$
doublets (see Fig.\ \ref{oiicivem} for a few examples). Objects with
these emission lines in their spectrum are the main contaminants in
searches for $z\sim3$ \lya\ emitters \citep[e.g.,][]{fyn03}. The high
resolution of the 1400V grism allows us to discriminate high redshift
\lya\ emitters from low redshift interlopers. \\

\centerline{\em Results} 

\noindent
Of the 36 objects observed 35 were confirmed to be \lya\ emitters at a
redshift $z \sim 2.86$. The 36th candidate was most likely too faint
to be detected. This candidate had the smallest line flux of the
objects in the mask. In addition to the 36 good candidates, four
objects from a list with 20 candidate line emitters with $8$ \AA\ $<$
{\em EW}$_0 < 15$ \AA\ were observed. Two of them were confirmed to be
\lya\ emitters at $z\sim2.86$. One of these, emitter 0052.LAE36, has a
\lya\ line that falls at the red edge of the filter. Using the
measured redshift ($z=2.876$ instead of $z=2.86$) the calculated
equivalent width is {\em EW}$_0 \sim 24$ \AA. The spectra of the 37
confirmed \lya\ emitters and that of the radio galaxy are shown in
Fig.\ \ref{0052spectra}. \\

\centerline{\em Volume density}

\begin{figure*}
\includegraphics[width=17cm]{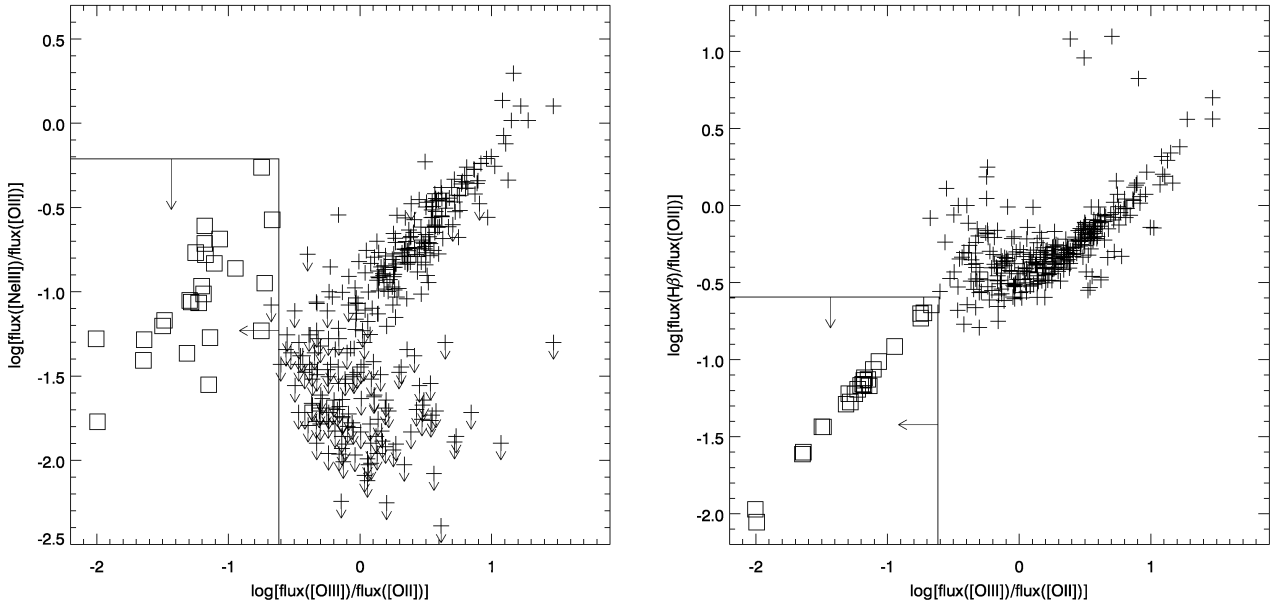}
\caption{\label{0943nooii} Limits on the observed flux ratios for the
  single emission line sources near MRC 0943--242 (squares). All
  limits are $2 \sigma$ upper limits. Line ratios of low redshift
  [\oii] emitters from \citet{ter91} are plotted as pluses. The
  measured limits on the line ratios are consistent with the
  identification of the emission lines with \lya\ (see text).}
\end{figure*}

\noindent
The density of \lya\ emitters can be compared directly to the field
density of emitters. \citet{fyn03} used the same narrow-band filter at
the VLT to observe a field which contains a damped \lya\ absorber. The
depth they reach in their images is approximately 0.7 mag deeper
compared to our narrow-band image, making their observations suitable
for comparison. \citet{fyn03} find 14 emitters (of which 13 are
spectroscopically confirmed) with a \lya\ flux $> 8 \times 10^{-18}$
\ergscm\ and a rest-frame equivalent width $> 15$ \AA. To the same
limit and in the same field of view, we find 48 (candidate)
emitters. This implies that the overdensity in the 0052 field is
$3.4^{+1.5}_{-1.0}$. 

Due to cosmic variance, the uncertainty in the number density of \lya\
emitters in a small field of view like that of the VLT images is
generally higher than the uncertainty derived assuming Poisson
statistics \citep[e.g.,][]{som04}. To overcome this, a comparison can
be made with the density of \lya\ emitters at $z \sim 3.1$ in a 0.13
deg$^2$ field as found by \citet{cia02}. This comparison gives a
volume density of \lya\ emitters in the 0052 field of
$2.5^{+1.8}_{-1.1}$ times the field density. The large (Poisson)
errors are due to the small number statistics. More recently,
\citet{hay04} measured the blank field space density of \lya\ emitters
at $z \sim 3.1$ using the Suprime-Cam on the Subaru telescope. In a
0.17 deg$^2$ field, they find 55 candidate \lya\ emitters with an
observed equivalent width $> 154$ \AA\ down to a narrow-band magnitude
of 25.3. Applying the same equivalent width and magnitude criteria to
our field and taking into account the difference in luminosity
distance gives 11 sources. The resulting density of emitters near MRC
0052--241 is a factor $3.1^{+1.4}_{-1.0}$ higher than the field
density. The weighted average of the three estimates of the
overdensity is $3.0^{+0.9}_{-0.6}$.

It is interesting to compare the density of \lya\ emitters in the 0052
field with the density of emitters near the radio galaxy MRC 0316--257
at $z = 3.13$ (V05). The depth and sensitivity of the imaging of the
two fields are very similar (see Tables \ref{imgobs} and
\ref{imgres}). There are 52 candidate emitters in the 0052 field with
a narrow-band magnitude brighter than 26.1, against 59 in the 0316
field. Taking into account the difference in the volume probed by the
narrow-band images, the ratio of the number density is
$n_{0052}/n_{0316} = 0.8^{+0.2}_{-0.2}$. Because the 0316 field is
overdense in \lya\ emitters by a factor $3.3^{+0.5}_{-0.4}$ (V05),
this means that the 0052 field has an overdensity of \lya\ emitting
galaxies of $2.7^{+0.8}_{-0.6}$, which is consistent within the errors
with the other estimates. \\

\centerline{\em Velocity and spatial distributions} 

\noindent
The velocity histogram and the spatial distribution of the candidate
and confirmed emitters are shown in Fig.\ \ref{0052velskydist}. The
redshifts of the confirmed emitters are not uniformly distributed
throughout the narrow-band filter (solid curve in Fig.\
\ref{0052velskydist}). Instead, the majority of the emitters (31 out
of the 37) appear to be clustered in two groups with velocity
dispersions of 180 and 230 \kms.  Interestingly, a similar redshift
distribution was found near the radio galaxy MRC 1138--262
\citep{pen00a}. The velocity of the radio galaxy lies $\sim$70 \kms\
from the median redshift of the emitters and falls inside the larger
of the two groups. The combination of the observed overdensity of
\lya\ emitters with the clumpy redshift distribution provides evidence
that the \lya\ emitters near MRC 0052--241 are part of a forming
cluster of galaxies at $z = 2.86$. The properties of this structure
will be discussed in Sect.\ \ref{protoclusters}.

In contrast to the velocity distribution, the emitters do not have a
preferred location on the sky. The structure of emitters appears not
to be bounded by the image, indicating that the overdensity seen in
this field extends beyond the edges of our field of view.

\subsection{MRC 0943--242, $z = 2.92$}
\label{0943res}

\begin{figure*}
\includegraphics[width=17cm]{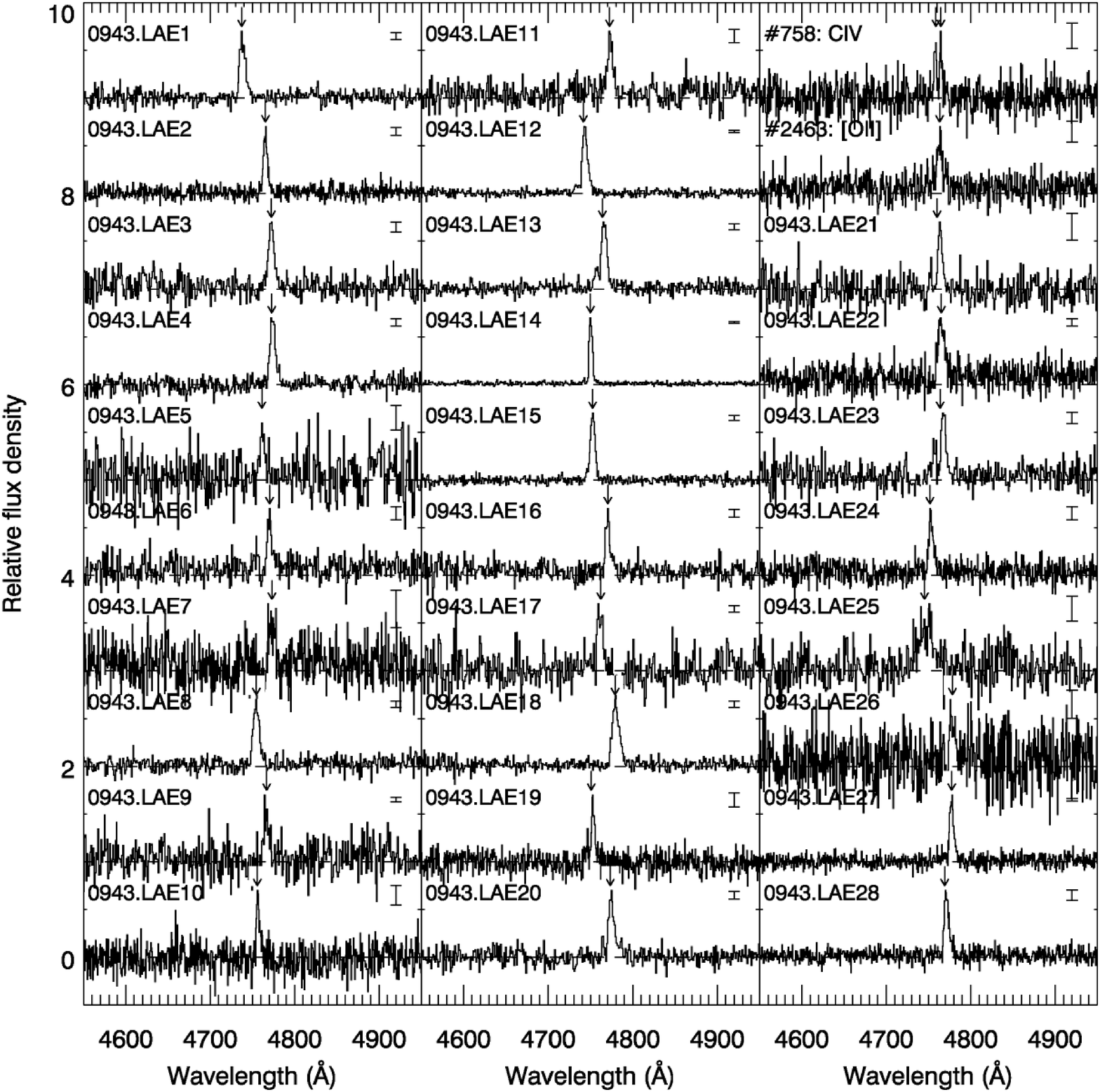}
\caption{\label{0943spectra} Spectra of the confirmed emitters and low
  redshift interlopers near the radio galaxy MRC 0943--242. The
  average uncertainty in the flux density over the wavelength range
  $4550$ \AA\ $< \lambda < 4950$ \AA\ is plotted as an error bar in
  the right corner of each spectrum. All spectra are normalized to the
  peak of the \lya\ line at a relative flux density of 0.7. The offset
  between the spectra is 1.0. The arrow indicates the center of the
  emission line of each object. The redshift, flux and width of the
  emission lines can be found in Table \ref{0943table}.}
\end{figure*}

\begin{figure*}
\includegraphics[width=17cm]{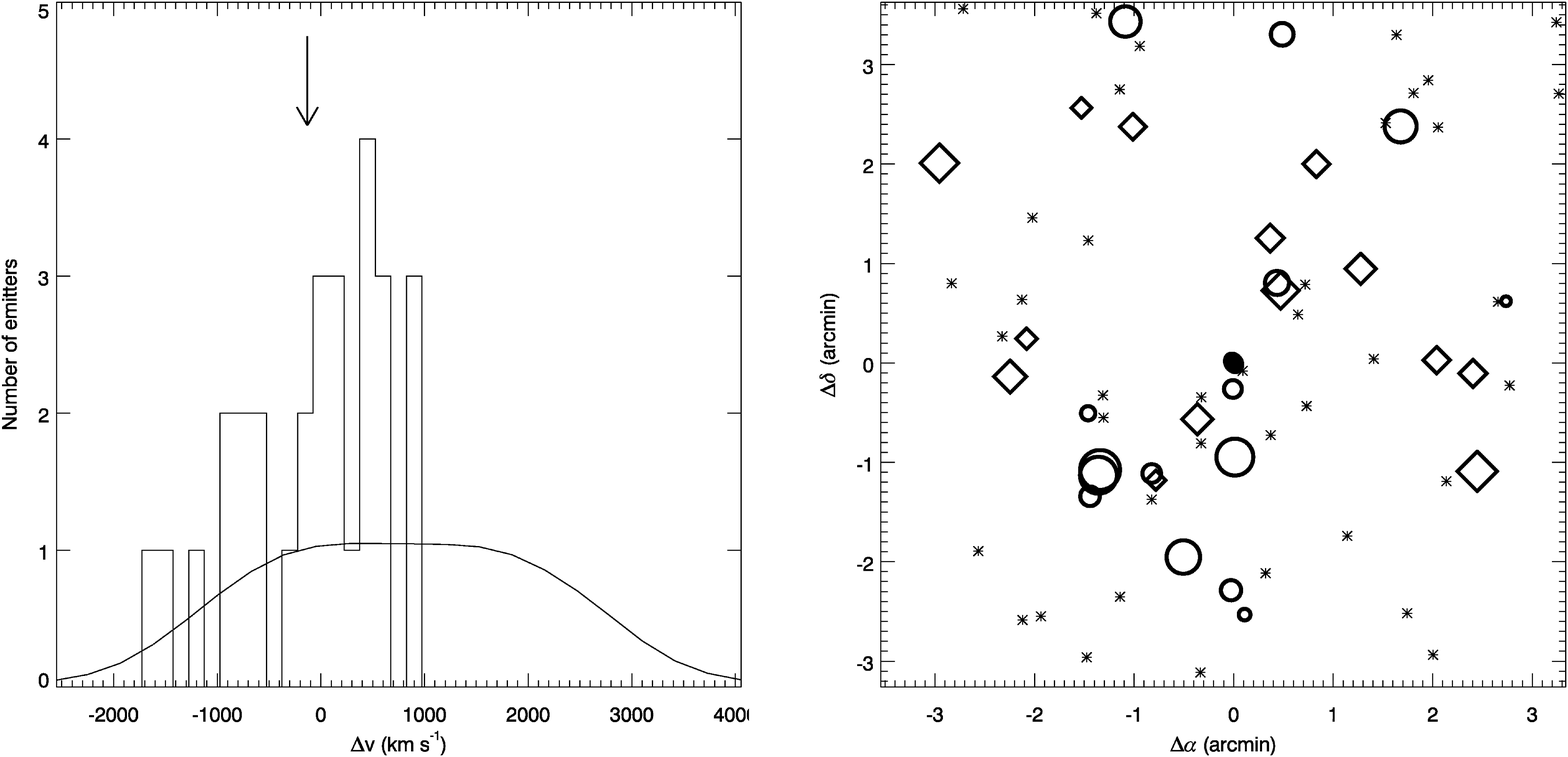}
\caption{\label{0943velskydist} Same as Fig.\ \ref{0052velskydist},
  but for the velocity distribution ({\em left}, the median
  redshift of the emitters ($z=2.9201$) is used as zeropoint) and
  spatial distribution ({\em right}) of the emitters near MRC
  0943--242.}
\end{figure*}

This powerful radio source with a flux density of 1.1 Jy at
408 MHz has a redshift of $z=2.923$ \citep{rot95,rot97}. The radio
galaxy is surrounded by a metal-enriched, low surface brightness gas
halo extending for at least 8\arcsec\ \citep[67 kpc,][]{vil03}. The
estimated dynamical mass of the gas halo is $7-44 \times 10^{11}$
\msun\ \citep{vil03,jar03}. This could be an indication that we are
witnessing the formation of a massive galaxy.
\\

\centerline{\em Imaging and spectroscopic observations}

\noindent
The field surrounding the radio galaxy was observed with the VLT in
2001 March. Images were taken in the narrow-band for 375 min and 75
min in the $B$-band (Table \ref{imgobs}). A part of the field (1.1
arcmin$^2$) was not usable to search for \lya\ emitters due to the
presence of a bright star ($R$ magnitude of 8.7) located 2\farcm5 to
the north-west of the radio galaxy. A total of 65 emission line
candidates with {\em EW}$_0 > 15$ \AA\ and $\Sigma > 3$ is found in
the images. 

Additional images of the field were taken in 2003 February and 2004
January with the LRIS instrument on the Keck I telescope. Observations
were done in the $u'$-band (for 277 min), $V$-band (93 min) and
$I$-band (100 min). The main purpose of these images is to search
for Lyman-break galaxies in the field. In this paper we use the images
only to separate possible \lya\ emitters at $z \sim 2.9$ from low
redshift line emitters.

Follow-up spectroscopy was carried out during two observing sessions
with LRIS on Keck I. In total, four user-defined slitmasks were
observed with exposure times between 120 and 180 min (Table
\ref{mosobs}). In the four masks, 30 of the 65 candidate \lya\
emitters were observed. Of these 30 objects, 26 showed an emission
line at the wavelength expected for \lya\ at $z = 2.9$, the remaining
four did not show a line (or continuum) in their spectrum. These four
unconfirmed emitters were among the five faintest emitters observed. Most
likely, the emission lines of the four objects are too faint to be
detected.
In addition to the 26 confirmed emission line galaxies, four objects
which had a predicted \lya\ equivalent width $< 15$ \AA\ also showed
an emission line in their spectrum. The spectra of these 30 emission
line objects are shown in Fig.\ \ref{0943spectra}.\\

\centerline{\em Line identification} 

\noindent
The next question is whether these 30 line emitters are due to \lya\
at $z \simeq 2.9$. In contrast to the observations of the 0052 field,
the resolution of the spectra was not high enough to resolve the
[\oii] doublet.

For three galaxies the redshift could be securely derived. One of
these objects (\# 758 in Fig.\ \ref{0943spectra}) is identified as a
\civ\ $\lambda \lambda 1548,1551$ emitter at $z = 2.07$ as its
emission line is resolved into two lines separated by $\sim8$ \AA. The
redshifts of two other objects are confirmed by the presence of other
emission lines in their spectrum. In the spectrum of emitter
0943.LAE25 \nv\ $\lambda 1240$ is detected, confirming that the
object is a \lya\ emitter at $z \sim 2.90$. The redshift of emitter
0943.LAE18 ($z \sim 2.93$) is confirmed by the detection of
\civ\ $\lambda 1549$ (and a very weak detection of \heii\
$\lambda 1640$). The relatively small {\em FWHM} of the \lya\ line
({\em FWHM} $= 700$ \kms) suggests that this object is a type-II AGN
\citep[e.g.,][]{nor02}.

The remaining 27 sources merely show a single emission line in their
spectra, which typically cover a wavelength range of over 5500 \AA. The
most common and brightest low redshift emission line galaxies 
contaminating searches of $z \sim 3$ \lya\ emitters are [\oii]
$\lambda\lambda 3726,3729$ emitters \citep[e.g.,][]{fyn03}. To
distinguish high redshift \lya\ emitters from low redshift emission
line galaxies, various tests can be applied \citep[see][ for a
detailed review]{ster00}. Below we will briefly describe and apply
three tests (the asymmetry of the emission line, the presence of a
continuum break and the limits on accompanying emission lines) that
are frequently used in the literature to discriminate \lya\ emitters
from low redshift galaxies
\citep[e.g.,][]{ster00,fyn01,rho03,daw04,ven04}.

\noindent
~$\bullet$ {\em Emission line asymmetry.} This test, which makes use
of the characteristic blue wing absorption profile of the \lya\
emission line, has been successfully applied to confirm $z > 4.5$
\lya\ emitters \citep[e.g.,][]{kod03,rho03,daw04,hu04}. What makes
this test extra useful is that a blended [\oii] doublet would appear
as a single line with {\em excess} emission on the blue side, because
the redder line of the doublet is stronger in star forming galaxies
\citep{rho03}. To quantify the asymmetry of an emission line,
\citet{rho03} introduced the two asymmetry parameters $a_\lambda$
(``wavelength ratio'') and $a_f$ (``flux ratio''). These parameters
depend on the wavelength where the flux peaks ($\lambda_p$), and where
the flux is at 10\% of the peak value on the red side
($\lambda_{10,r}$) and the blue side ($\lambda_{10,b}$) of the
emission line. Using these wavelengths, the ``wavelength ratio'' is
defined as $a_\lambda =
(\lambda_{10,r}-\lambda_p)/(\lambda_p-\lambda_{10,b})$ and the ``flux
ratio'' as $a_f = \int_{\lambda_p}^{\lambda_{10,r}} f_\lambda
d\lambda/\int_{\lambda_{10,b}}^{\lambda_p} f_\lambda d\lambda$
\citep{rho03,rho04,daw04}. \citet{daw04} found that $z = 4.5$ \lya\
emitters have $a_\lambda > 1.0$ and/or $a_f > 1.0$, while [\oii]
emitters at $z \sim 1$ typically have $a_\lambda \approx 0.8$ and $a_f
\approx 0.8$.

\begin{figure*}
\includegraphics[width=17cm]{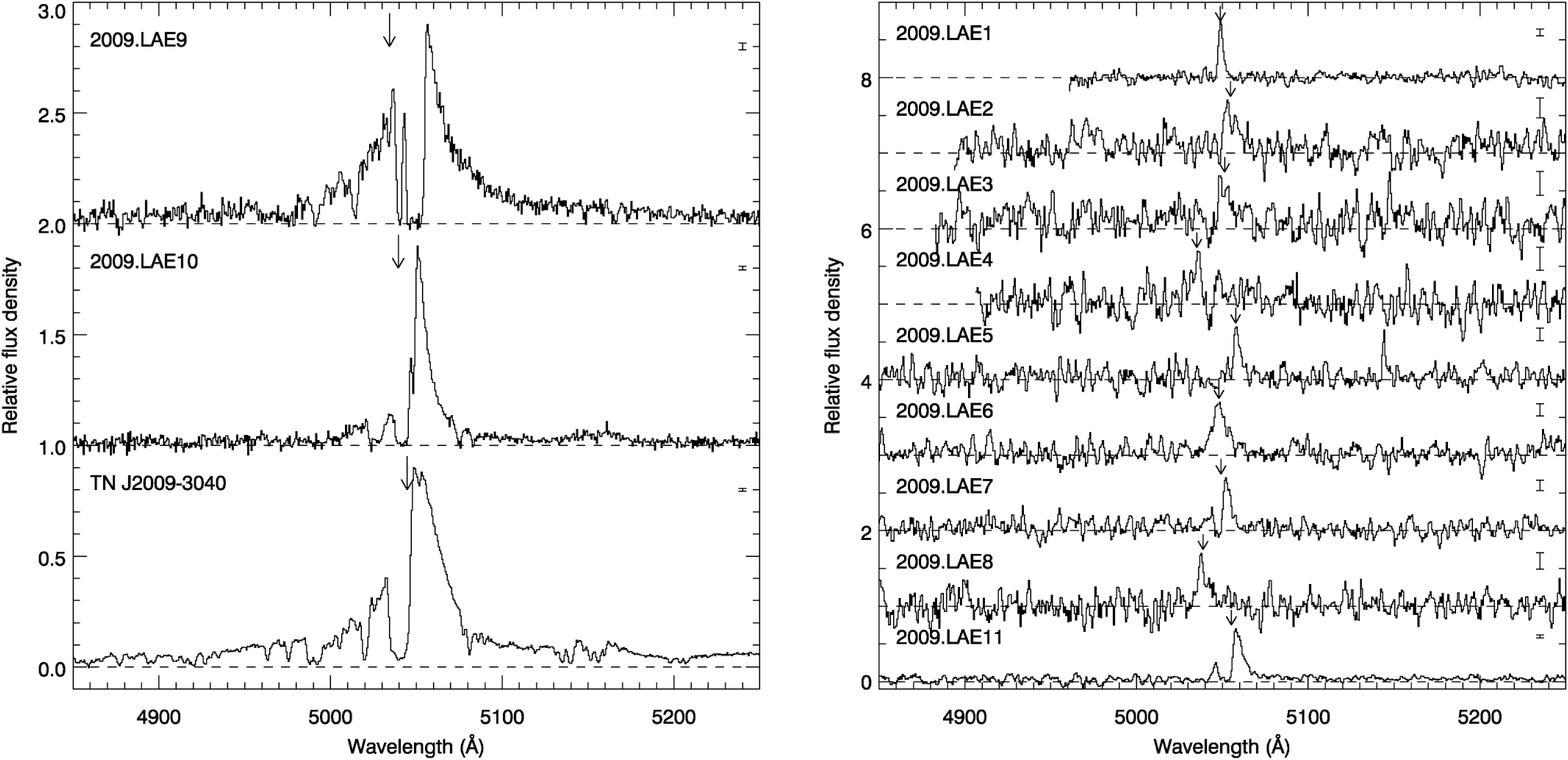}
\caption{\label{2009spectra} Spectra of the confirmed emitters near
  the radio source TN J2009--3040. The spectra of the radio source and
  two quasars companions are shown on the left, and on the right the confirmed
  \lya\ emitters are plotted. The spectra in the right panel are
  boxcar average over three pixels. The uncertainty per pixel is
  shown as error bar on the right of each spectrum. The arrow
  indicates the central wavelength of the emission line. The redshift,
  flux and width of the emission lines can be found in Table
  \ref{2009table}.}
\end{figure*}

However, this line asymmetry test critically depends on the assumption
that a large fraction of the blue part of the \lya\ emission line is
absorbed by neutral Hydrogen. Although absorption is often present,
there can still be a significant amount of flux on the blue wing
\citep[see the emission lines in e.g.\ Fig.\
\ref{0052spectra},][V05]{tap04}. Also, due to the absorption the peak
of the \lya\ line could move to the red, reducing the flux that
appears to be redward of the peak (see e.g.\ the spectra on the left
side of Fig.\ \ref{2009spectra} and the spectra in V05). As a
consequence, the asymmetry parameters for a \lya\ line can be smaller
than 1.0. For example, if the spectrum of the radio galaxy MRC
0943--242 would be convolved with a Gaussian with a {\em FWHM}
$\approx 250$ \kms\ (or would be observed with slightly lower
resolution), then the asymmetry parameters would give $a_l=0.67$ and
$a_f=0.92$. Also, the asymmetry parameters of emitter 0052.LAE1, which
is a QSO at $z = 2.86$ as confirmed by \nv, are $a_l=0.48$ and
$a_f=0.38$ and those of emitter 0052.LAE31 measure $a_\lambda = 1.0$
and $a_f = 0.41$. Thus, while lines with $a_\lambda > 1.0$ or $a_f >
1.0$ are most likely \lya\ lines, a profile with low values of
$a_\lambda$ and $a_f$ could be either an [\oii] or a \lya\ line.

With this caveat in mind, we measured the asymmetry of our single
emission line sources. 21 have $a_\lambda > 1.0$ and/or $a_f > 1.0$,
and these emission lines are most likely \lya\ lines at $z \simeq
2.9$. The identification of the remaining six lines remains problematic
and we have to rely on others tests to secure the redshift.

\noindent
~$\bullet$ {\em Continuum break}. A prominent feature of high redshift
galaxies is the absorption of the continuum blueward of the \lya\
line, caused by neutral Hydrogen located in the galaxy between the
galaxy and the observer \citep[e.g.,][]{mad95}. To measure whether the
continuum of the line emitters has a discontinuity, we looked at the
$u'$- and $V$-band images. For a galaxy at $z=2.9$ the central
wavelength of the $u'$ filter lies below the 912 \AA\ Lyman limit,
while the $V$-band begins just redward of the \lya\ line. Based on a
large imaging and spectroscopic survey of high redshift galaxies,
\citet{coo05} estimate that the vast majority of galaxies at $z \sim
3$ have a break of $u'-V > 1.2-1.6$. On the other hand, low redshift objects
with the Balmer/4000 \AA\ break that falls between the $u'$- and
$V$-band have a color of $u'-V \ll 1.5$ \citep[e.g.,][]{ham97}.

Of the 27 single emission line sources, 26 were inside the field of
view of the $u'$-band image.  Two objects were too close to a bright
foreground object to reliably measure the magnitude. Of the remaining
24 sources two are detected in the $u'$-band, the others are not
detected. The undetected emitters have typically $2 \sigma$ limits of
$m_{u'} > 27.6$ within an aperture radius of 1\farcs2, which is twice
the radius of the seeing disc. One of the two detected objects in the
$u'$-band has a signal-to-noise of $\sim 2.0$ and a $u'-V \simeq 2.5$
is measured for this object. The second detection, of emitter \# 2463
in Fig.\ \ref{0943spectra}, has a signal-to-noise of 8 and a $u'-V =
1.0$.  Combined with a measured line asymmetry of $a_\lambda = 0.7$
and $a_f = 0.7$, we conclude that this emitter is a foreground
galaxy. The emission line can be most likely identified with [\oii] at
a redshift of $z \sim 0.28$. It should be noted that this source had a
predicted equivalent width of 9 \AA, and was therefore not in our list
of priority candidate \lya\ emitters.

Five out of 24 emission line objects have large breaks with a
$2\sigma$ limit of $u'-V > 2.0$. These objects can securely be
identified with high redshift \lya\ emitting galaxies. For the
remaining 18 emission line galaxies, either the continuum was too faint
to detect a large $u'-V$ break in our images or the photometry in the
$V$-band was unreliable due to the presence of a nearby bright object.

\noindent
~$\bullet$ {\em Emission line intensity ratios.} Although no other
emission lines were found in the spectra of the 27 emission line sources,
we can derive upper limits on various line ratios. This is very useful
because several other emission lines are expected in the typical spectrum of an
[\oii] emission line galaxy, such as [\neiii]$\lambda 3869$, H$\beta$ and
[\oiii]$\lambda 5007$. 

\begin{figure}
\includegraphics[width=8.5cm]{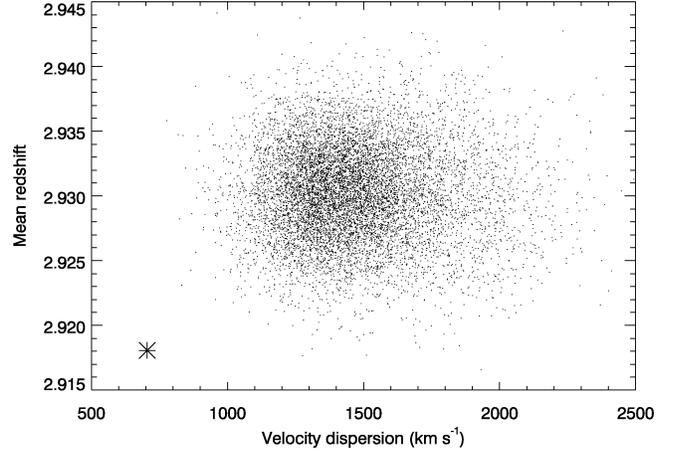}
\caption{\label{zdist} Velocity dispersion ($\sigma_v$) and mean
  redshift ($\bar{z}$) of 10\,000 simulated redshift distributions of
  28 emitters. In these simulations, the narrowband filter curve was
    used as redshift probability function. The simulated distributions
    have a mean redshift of $\bar{z} = 2.930 \pm 0.004$ and a velocity
    dispersion of $\sigma_v = 1475 \pm 235$ \kms\ (the errors
    represent the standard deviations between the simulations). These
    simulated values are significantly different (at the 4.7$\sigma$
    level) compared with the observed velocity dispersion and mean
    redshift of the emitters near MRC 0943--242 of $\sigma_v=715$
    \kms\ and $\bar{z} = 2.918$ (indicated with a cross).}
\end{figure}

What makes this test very useful to us, is the large wavelength range
of the spectra of the emission line sources. 
If the emission line in the spectrum is [\oii], then we are able to
place limits on the flux of the [\neiii]$\lambda 3869$, H$\beta$ and
[\oiii]$\lambda 5007$ lines. Although H$\alpha$ is also within the
probed wavelength range, it falls in a region where bright skylines
dominate the spectrum, and the H$\alpha$ upper limits are not
useful. Instead, we measured the limits on the expected positions of
[\neiii], H$\beta$ and [\oiii]. Two sigma limits were computed in a region
which is twice the {\em FWHM} of the detected emission line. The limits on
various emission line ratios are presented in Fig.\
\ref{0943nooii}. To compare our limits with low redshift [\oii]
emitters, we also plotted line ratios as found by \citet{ter91} for
local \hii\ galaxies. In all cases, our observed ratios lie in a
region outside the values found for local galaxies with [\oii] in
emission. We therefore conclude that the emission line objects are
\lya\ emitters.

To summarize, after applying three tests that can discriminate \lya\
emitters from low redshift emission line galaxies, 
we have identified one [\oii] emitter (object
\#2463 in Fig.\ \ref{0943spectra}) and confirm that the remaining 26
\lya\ emitters are at $z \sim 2.9$. Combined with the two emitters with
\civ\ in their spectrum, the total number of \lya\ emitters near the
radio galaxy MRC 0943--242 is 28. The properties of the \lya\ emission
lines of these $z = 2.9$ galaxies are printed in Table
\ref{0943table}. \\

\centerline{\em A large scale structure of Ly$\alpha$ emitters}

\noindent
The volume densities of emitters near MRC 0943--242 and near the radio
galaxy MRC 0316--257 (V05) are very similar. Down to a narrow-band
magnitude of $m_\mathrm{nb} = 26.1$ there are 63 emitters in the 0943
field compared to 59 in the 0316 field. Correcting for the difference
in observed comoving volume, the density ratio is $n_{0943}/n_{0316} =
1.0^{+0.2}_{-0.2}$. This translates to a density of $n_{0943} =
3.2^{+0.9}_{-0.7}$ times the field density. Comparing directly to
blank field surveys for $z\sim3$ \lya\ emitters gives similar results
(see also Sect.\ \ref{0052res}). For example, the field density of
\lya\ emitters at $z\simeq3.1$ was measured by \citet{hay04} in the
0.17 deg$^2$ Subaru Deep Field. They estimate a space density of
$n_{\mathrm{field}} = 3.5 \times 10^{-4}$ Mpc$^{-3}$ for \lya\
emitters with a rest frame equivalent width $> 37$ \AA\ down to a
narrow-band magnitude of 25.3. Applying the same selection criteria
(taking into account the difference in luminosity distance) gives a
space density of emitters in the field of MRC 0943--242 of $n_{0943} =
9.6 \times 10^{-4}$ Mpc$^{-3}$. The density ratio is
$n_{0943}/n_{\mathrm{field}} = 2.7^{+1.3}_{-0.9}$. When compared to
the space density of \lya\ emitters found by \citet{cia02}, the
overdensity in the 0943 field is $n_{0943}/n_{\mathrm{field}} =
5.2^{+2.7}_{-1.8}$. All three estimates of the density presented here
are consistent with each other. The weighted average of the estimates
is $3.4_{-0.6}^{+0.8}$.

In Fig.\ \ref{0943velskydist} the velocity and spatial distribution of
the emitters is shown. Although the emitters have no preferred
position on the sky, they are clustered in velocity space. We can
estimate the significance of the clustering by performing Monte Carlo
simulations of the redshift distribution. We reproduced 10\,000
realizations of 28 emitters using the narrow-band filter curve as
redshift probability function for each emitter. We find that both the
mean velocity and the dispersion of the emitters are significantly
smaller than that of the simulated redshift distributions at a level
of $\sim3.3 \sigma$ (Fig.\ \ref{zdist})\footnote{Recently,
\citet{mon05} investigated the effect of peculiar velocities on the
redshift distribution of \lya\ emitters using dark matter
simulations. They found that the velocity dispersion of \lya\ emitters
is smaller than based on the Monte Carlo simulations described above,
lowering the significance of the redshift clumping.}. This is a
strong indication that the suggested grouping of \lya\ emitters is not
a projection effect. The velocity dispersion of the confirmed emitters
is $715 \pm 105$ \kms. This dispersion could be a lower limit, since
no clear edge is visible in the distribution on the negative
velocities side.

\subsection{MRC 0316--257, $z = 3.13$}
\label{0316res}

This 1.5 Jy radio source from the Molonglo Reference Catalogue
\citep{lar81} lies at a redshift of 3.13 by \citet{mcc90}. This target
was especially interesting, because \citet{lef96} found two bright
\lya\ emitters close to the radio galaxy, which indicated that this
radio galaxy is in an overdense environment. Narrow- and broad-band
imaging of this field with the VLT resulted in the discovery of 77
candidate \lya\ emitters. Follow-up spectroscopy revealed 33 emission
line galaxies of which 31 are \lya\ emitters near the radio galaxy,
while the remaining two are foreground galaxies. By comparing the
number density of \lya\ emitters near the radio galaxy to that of the
field, the overdensity of emitters near 0316 is estimated to be a
factor $3.3^{+0.5}_{-0.4}$ times the field density (V05). The velocity
distribution of the emitters has a width of {\em FWHM} = 1510 \kms,
which is smaller than the width of the narrow-band filter ({\em FWHM}
$\sim 3500$ \kms).  The peak of the distribution is located within 200
\kms\ of the redshift of the radio galaxy. In V05 we have shown that
the confirmed emitters are members of a protocluster at $z \sim 3.13$
with an estimated mass of $> 3 \times 10^{14}$ \msun.

\subsection{TN J2009--3040, $z = 3.15$}

\begin{figure*}
\includegraphics[width=17cm]{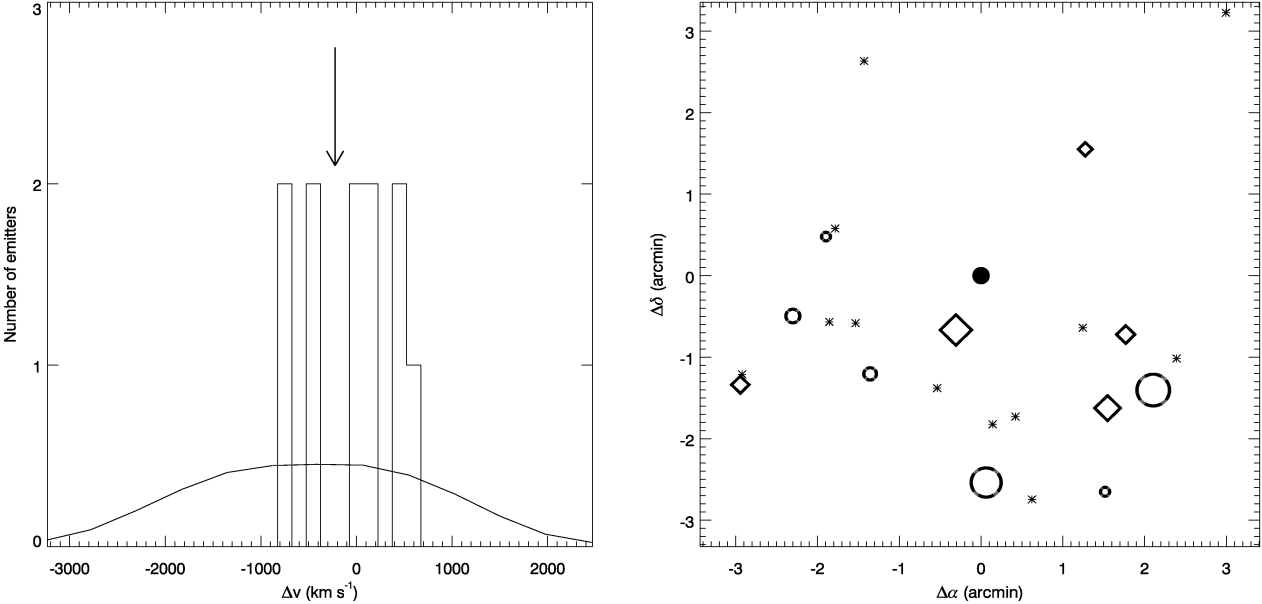}
\caption{\label{2009velskydist} Same as Fig.\ \ref{0052velskydist},
  but for the velocity distribution ({\em left}, the median
  redshift of the emitters ($z=3.1528$) is used as zeropoint) and spatial
  distribution ({\em right}) of the emitters near TN J2009--3040.}
\end{figure*}

This radio source was selected by \citet{deb00} as an ultra steep
spectrum radio source ($\alpha = -1.36$, with
$f_\nu\propto\nu^\alpha$) and was identified with a $m_K = 18.05 \pm
0.05$ object which has a strong unresolved component
\citep{deb02}. Spectroscopy of this object revealed three emission
lines, including \lya\ with a FWHM of $\sim2000$ \kms\ at a redshift
of 3.156. Based on the width of the emission lines and the morphology
of the source, this radio source is identified as a quasar (C. De
Breuck, private communications).  \\

\centerline{\em Imaging observations}

\noindent
Obtaining deep images of this field was challenging for two
reasons. First, the
field is located close to the galactic centre ($l = 11^\circ$ and $b =
-30^\circ$) and the extinction towards the radio source is high, with
an $E(B-V) = 0.181$ \citep{sch98}. Secondly, due to emission from
inside our Galaxy the background towards the quasar is not flat, but
varies up to 10\% over scales of roughly 10\arcsec. We removed this
variable background by subtracting the median pixel value calculated
in 10\arcsec$\times$10\arcsec\ boxes surrounding each pixel. \\ 

\centerline{\em Spectroscopic observations and results}

\noindent
Analysis of the narrow-band and $V$-band images resulted in a list of
21 candidate \lya\ emitters. In 2002 September spectra were taken of
nine of these candidates for 330 min with FORS2. The use of the 1400V
grism (with a resolution of $R = 2100$) allowed us to identify low
redshift interlopers like [\oii] emitters (see Fig.\ \ref{oiicivem}),
similar to the 0052 field. We confirmed that all nine candidates were
\lya\ emitters at $z \sim 3.15$ (Fig.\ \ref{2009spectra}). Two objects
with predicted {\em EW}$_0 \simeq 10$ \AA\ and $\Sigma \simeq 2.5$
(2009.LAE2 and 2009.LAE5 in Fig.\ \ref{2009spectra} and Table
\ref{2009table}) were also confirmed to be \lya\ emitters near the
radio source. Two galaxies with a predicted {\em EW}$_0 \simeq 6$ and
9 \AA\ were confirmed to be [\oii] emitters at $z \sim 0.35$ (bottom
two spectra in Fig.\ \ref{oiicivem}). In total, 11 \lya\ emitters are
found near the radio source. Two of these emitters are broad-line QSOs
with a {\em FWHM} of 1600 and 3200 \kms\ (see left panel of Fig.\
\ref{2009spectra}). 

The detection of (broad-line) QSOs near radio galaxies is quite
common. In each of the radio galaxy fields described in this paper, at
least one of the \lya\ emitters is confirmed to have a line {\em FWHM}
of $> 1000$ \kms\ (see Tables \ref{2048table}--\ref{1338table}), an
indication that these objects harbour an active galactic nucleus
(AGN). In the protocluster near the radio galaxy MRC 1138--262 at $z =
2.16$ (Sect.\ \ref{1138res}) at least four QSOs are confirmed
\citep{pen02,cro05}.

In Fig.\ \ref{2009velskydist} the velocity histogram and the spatial
distribution of the confirmed \lya\ emitters are shown. The median
redshift of the emitters lies 170 \kms\ away from the radio
source. The velocity dispersion of the emitters as calculated with the
Gapper scale estimator \citep[which is preferred for low ($n\simeq10$)
number statistics,][]{bee90} is $515 \pm 85$ \kms. This is a factor
$\sim$3 smaller than that of the narrow-band filter, which has a
$\sigma = 1490$ \kms. 
Monte Carlo simulations of the distribution of 11 emitters through the
narrow-band filter, we found that there is an 8\% chance that the
velocities of the emitters are drawn from a random distribution. On
the sky, the emitters appear to be concentrated on the southern half
of the field. The position of the radio source is on the edge of the
distribution of \lya\ emitters, very similar to the situation in the
field of TN J1338--1942 at $z = 4.1$ \citep[][ Sect.\
\ref{1338res}]{ven02}. \\

\centerline{\em Volume density}

\noindent
We can compare the number density of \lya\ emitters in this field
directly to that of the 0316 field, because the same narrow-band
filter was used. The number of (candidate)
emitters down to the same luminosity limit is in the 2009 field a
factor $2.0^{+0.7}_{-0.5}$ smaller than that in the 0316 field. The 0316
field is overdense by a factor 3.3$^{+0.5}_{-0.4}$ (V05), which
implies that the number density of \lya\ emitters near TN J2009--3040
is a factor $1.7^{+0.8}_{-0.6}$ times the field density, with is consistent
with no overdensity. 
However, nine out of 11 confirmed emitters lie south of the radio
galaxy, as do nine out of 12 unconfirmed candidates (Fig.\
\ref{2009velskydist}). This could indicate that (locally) the volume
density of \lya\ emitters near the radio source TN J2009--3040 is much
higher than the field density. Although the volume density of \lya\
emitters near TN J2009--3040 is consistent with the field density at
that redshift, the clustering both on the sky and in velocity space of
the emitters could points to a structure of galaxies. More observations
are needed to determine the reality of this clustering.

\subsection{TN J1338--1942, $z = 4.11$}
\label{1338res}

This radio galaxy has a redshift of 4.1 \citep{deb99,deb01} and is one
of the brightest known in \lya\ \citep{deb99,deb01}. 
Because no narrow-band filter is available at
the VLT that is centred on the wavelength of a \lya\ line at $z=4.1$,
we used a custom narrow-band filter with an effective wavelength of
6199 \AA\ and a FWHM of 59 \AA. Narrow-band and $R$-band imaging and
follow-up spectroscopy with the VLT of the field of TN
J1338--1942 revealed 20 \lya\ emitters within a projected distance of
1.3 Mpc and 600 \kms\ of the radio galaxy \citep{ven02}. The structure
is overdense in \lya\ emitters by a factor of 4--15 and could be the
ancestor of a rich cluster of galaxies. Multi-color imaging with the
Advanced Camera for Surveys (ACS) on board the {\em Hubble Space
Telescope (HST)} revealed an anomalously large number of 
LBGs near the radio galaxy, confirming the presence of a
protocluster at $z = 4.1$ \citep{mil04,ove06a}. \\

\centerline{\em New observations}

\begin{figure}
\includegraphics[width=8.5cm]{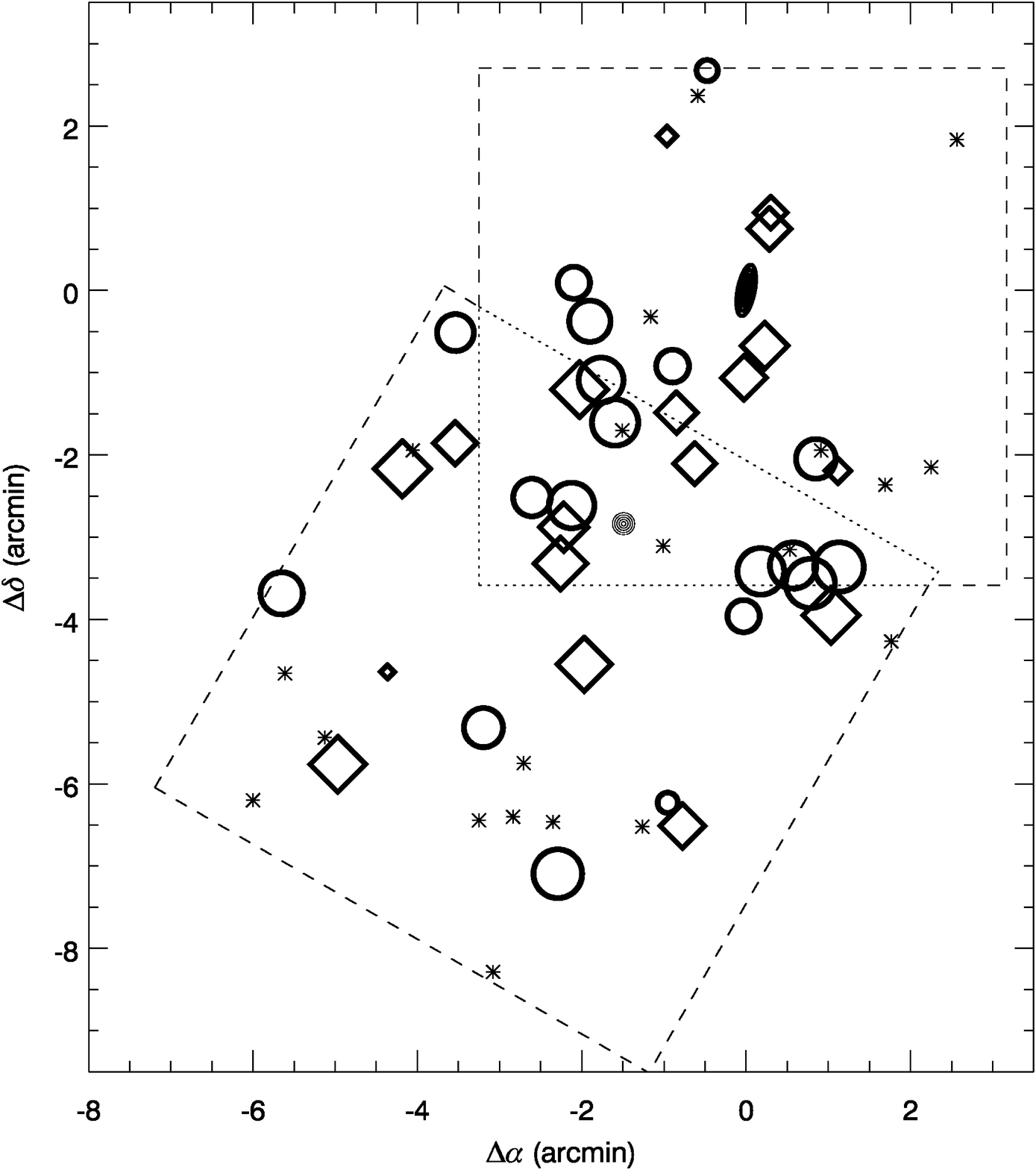}
\caption{\label{1338skydist} Spatial distribution of the
  confirmed and candidate \lya\ emitters near TN J1338--1942. Symbols
  and their sizes are the same as in Fig.\ \ref{0052velskydist}. The
  filled circle at (-1\farcm5,-2\farcm8) represents the centre of all the
  emitters. The dashed lines outline the area covered by the VLT images.}
\end{figure}

\noindent
The VLT observations presented in \citet{ven02} showed that the
emitters are not distributed homogeneously over the field, but appear
to have a boundary in the north-west. To further determine the extent
and shape of the protocluster region, the field near TN J1338--1942
was imaged at a second position. The second pointing is located
towards the south-east of the radio galaxy and overlaps the first
field (hereafter the 1338-1 field) at the position where the
concentration of \lya\ emitters seemed to be the highest (see Fig.\
\ref{1338skydist} for the outline of the imaging areas). The second
field (hereafter 1338-2) was observed for 420 min in the narrow-band
and for 75 min in the $R$-band. Analysis of data in the second field
resulted in the discovery of 35 candidate emitters. Ten candidate
emitters were also selected as candidates in the 1338-1 field and
eight of them were confirmed by \citet{ven02}. One candidate emitter
in the second field catalogue has an {\em EW}$_0 = 19^{+14}_{-5}$ \AA,
but a computed equivalent width of $9^{+6}_{-3}$ \AA\ in the first
field catalogue. The (weighted) average of these two measurements is
{\em EW}$_0 \simeq 14.5$ \AA. Because this is below our selection
criterion of {\em EW}$_0 = 15$ \AA, we removed the object from our
candidate list. Follow-up spectroscopy showed that the object is an
[\oii] emitter at $z \sim 0.66$. 
The total number of candidate emitters in the two fields combined is 54.

Spectroscopic observations of candidate emitters in the second field
were carried out in 2003 February with Keck/LRIS (150 min) and in 2003
March with VLT/FORS2 (310 min, see Table \ref{mosobs}). In the two
observing sessions combined, spectra were taken of 15 good candidate
emitters. In addition, objects were observed that were lower on our
priority list. These additional objects included four emitters that
were already confirmed by \citet{ven02}, six candidate emitters with
$9$ \AA$ <$ {\em EW}$_0 < 15$ \AA\ and five galaxies with an {\em
EW}$_0 > 15$ \AA, but with a signal-to-noise in the narrow-band image
of $\sim4$.

\begin{figure*}
\includegraphics[width=17cm]{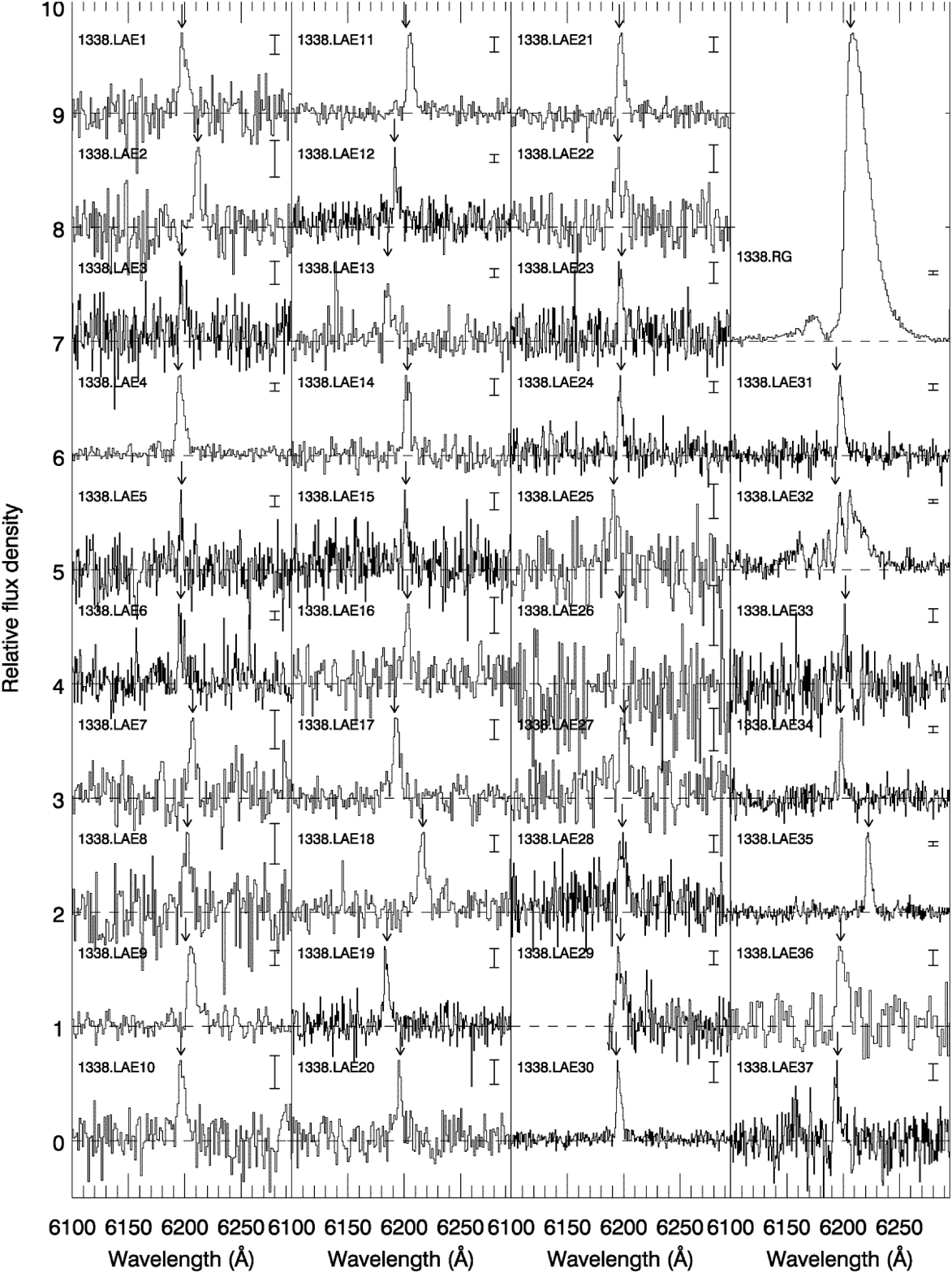}
\caption{\label{1338spectra} Spectra of the confirmed emitters near
  the radio galaxy TN J1338--1942. Error bars (right in each panel)
  represent the uncertainties in the flux density average over all
  pixels in the plotted wavelength range. All spectra are normalized
  to the peak of the \lya\ line, which has a relative flux density of
  0.7. The offset between the spectra is 1.0. The arrow indicates the
  central wavelength of the \lya\ emission line. The redshift, flux
  and width of the emission lines can be found in Table
  \ref{1338table}.}
\end{figure*}

Of the 15 good candidate emitters, 13 were confirmed to be \lya\
emitters at $z \sim 4.1$. The two unconfirmed emitters were among the
faintest candidates that were observed in that mask. From the
additional objects that were observed, the four emitters that were
confirmed by \citet{ven02} were all reconfirmed. Among the
low equivalent width sources was one \lya\ emitter at $z = 4.1$
and five [\oii] emitters at a redshift $z \simeq 0.66$. The low signal-to-noise
sources added another two \lya\ emitters at $z=4.1$ to the list of
confirmed emitters. Combined with the data described in \citet{ven02},
the total number of confirmed \lya\ emitters at $z \sim 4.1$ is 37. In
Fig.\ \ref{1338spectra} the spectra of the confirmed emitters and of
the radio galaxy are shown. In Table \ref{1338table} the properties of
the emission lines are listed. \\

\centerline{\em Volume density}

Several blank field surveys for \lya\ emitters at redshifts between $z
= 4$ and $z=5$ have been conducted in recent years. The two surveys
covering the largest area are the LALA survey, aimed at finding \lya\
emitters at $z = 4.5$ \citep{rho00,daw04}, and the Subaru Deep Field
(SDF) survey, in which was searched for \lya\ emitters at $z = 4.8$
\citep{ouc03,shi03,shi04}. 

\begin{figure}
\includegraphics[width=8.5cm]{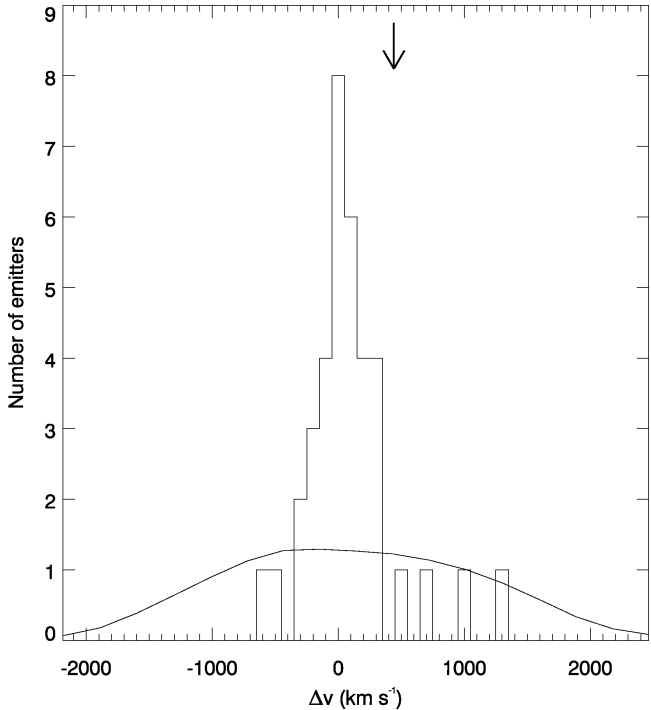}
\caption{\label{1338veldist} Velocity distribution of the confirmed
  emitters near TN J1338--1942. The median of the emitters
  ($z=4.0977$) is taken as zero-point. The velocity of the radio
  galaxy ($\Delta v = 440$ \kms) is indicated by an arrow. The solid
  line represents the selection function of the narrow-band filter,
  normalized to the total number of confirmed emitters.}
\end{figure}

To estimate the (over)density of \lya\ emitters near TN J1338--1942,
we compare our density to that found in the LALA and SDF surveys.
The LALA survey probed a volume of $1.5 \times 10^6$ comoving
Mpc$^3$ and $\sim350$ candidate \lya\ emitters were discovered in this
volume with an observed equivalent width $>80$ \AA\ and a $5 \sigma$
line flux $>2 \times 10^{-17}$ \ergscm\
\citep{rho00,daw04}. Follow-up spectroscopy confirmed 72\% of the
targeted sources \citep{daw04}, but we will conservatively assume
that all 350 LALA candidates are genuine \lya\ emitters. Applying the
same luminosity and equivalent width criteria used in the LALA survey
to our data selects 14 candidate \lya\ emitters in our field of which
13 are confirmed (one has not been observed spectroscopically
yet). Compared to the LALA survey, the density of \lya\ emitters near
TN J1338--1942 is a factor $5.2^{+1.8}_{-1.4}$ higher. The large
errors are due to small number statistics.

The SDF survey covers an area of 25\arcmin$\times$45\arcmin, which is
observed in two 1\% narrow-band filters, one centred on \lya\ at $z
\approx 4.79$ and one at $z \approx 4.86$. The volume density of
emitters with magnitudes $m_\mathrm{nb} < 25.5$ at these two redshifts
is very similar \citep[$2.7 \pm 0.4 \times 10^{-4}$ Mpc$^{-3}$ at $z
\sim 4.79$ and $3.1 \pm 0.5 \times 10^{-4}$ Mpc$^{-3}$ at $z \sim
4.86$,][]{shi04}. Selecting \lya\ emitters in our field in the same
way as was done in the SDF by \citet{shi03}, and comparing our numbers
to those at $z = 4.79$ and $z = 4.86$ gives an overdensity of
$4.5^{+1.3}_{-0.9}$. The weighted average of these density
measurements is $n_{1338}/n_{\mathrm{field}} = 4.8^{+1.1}_{-0.8}$,
consistent with the overdensity found by \citet{ven02}.  \\

\centerline{\em Structure properties}

\noindent
As mentioned in \citet{ven02}, the spatial distribution of emitters
near 1338 is not homogeneous over the field. In Fig.\
\ref{1338skydist} the spatial distribution is plotted of all the
confirmed and candidate emitters in the two fields. The average
position of the emitters (filled circle in Fig.\ \ref{1338skydist})
lies approximately 1\farcm5 east and 2\farcm8 south of the radio
galaxy. The radio galaxy appears to be at the (northern) edge of the
galaxy distribution. In contrast to the position on the sky of the
\lya\ emitters, \citet{deb04} found an overdensity of mm galaxies just
{\em north} of the radio galaxy in this field. Although there is no
spectroscopic confirmed whether the mm galaxies belong to a structure
at the redshift of the radio galaxy, it is possible that different
galaxy populations in the protocluster have different spatial
distributions (see also Sect.\ \ref{size}).

The redshift distribution of the emitters is shown in
Fig. \ref{1338veldist}. The velocity distribution is very narrow, with
a dispersion of only $260 \pm 65$ \kms\ ({\em FWHM} $= 625 \pm 150$
\kms). This is much narrower than the width of the narrow-band filter,
which has a {\em FWHM} of 2860 \kms. Monte-Carlo simulations indicate
that the chance that such a narrow redshift distribution is drawn from
a random sample is only 0.05\%. Although the number of confirmed
emitters has almost doubled since the discovery of this protocluster,
the velocity dispersion has not changed. This indicates that the
confirmed emitters in both the southern and northern field are members
of one single structure. The small value of the velocity dispersion
might suggest that we are looking at a flattened structure like a
filament perpendicular to the line of sight. However, as we will show
in Sect.\ \ref{veldisp}, the velocity dispersion of the emitters in
this field is in line with predictions of numerical models of the
formation of massive structures. The properties of this protocluster
at $z = 4.1$ will be described in more detail in Sect.\
\ref{protoclusters}.

\subsection{TN J0924--2201, $z = 5.20$}
\label{0924}

\begin{table}
  \caption{\label{halotable} Luminosity, size and position angle (PA) of the \lya\ halos surrounding the radio galaxies observed in our VLT program. The position angles of the halos are measured from the \lya\ images (Figs.\ \ref{halo16022048}--\ref{halo13380924}) and are accurate to $\sim$10 degrees.}
\begin{center}
\begin{tabular}{lcccc}
\hline
\hline
Name & $L_{\mathrm{Ly}\alpha}$ & Size & PA halo & PA radio\\
{} & erg\,s$^{-1}$ & kpc$\times$kpc & deg$^a$ & deg$^a$ \\
\hline
BRL 1602--174 & $7.5 \times 10^{44}$ & 90$\times$55 & 60 & 56$^b$ \\ 
MRC 2048--272 & $6.5 \times 10^{43}$ & 70$\times$40 & 25 & 42$^b$ \\ 
MRC 1138--262 & $2.5 \times 10^{45}$ & 250$\times$125 & 74 & 98$^b$ \\ 
MRC 0052--241 & $7.5 \times 10^{43}$ & 35$\times$30 & 5 & 15$^b$ \\  
MRC 0943--242 & $2.5 \times 10^{44}$ & 50$\times$40 & 55 & 74$^b$ \\ 
MRC 0316--257 & $7.0 \times 10^{43}$ & 35$\times$25 & 55 & 53$^b$\\ 
TN J2009--3040 & $3.0 \times 10^{44}$ & 40$\times$40 & --$^c$ & 144$^b$\\ 
TN J1338--1942 & $4.5 \times 10^{44}$ & 130$\times$45 & 170 & 152$^b$\\ 
TN J0924--2201 & $1.5 \times 10^{43}$ & 10$\times$10 & 90 & 74$^b$ \\ 
\hline
\end{tabular}
\end{center}
$^a$ Position Angle in degrees, measured east of north. \\
$^b$ References for the radio position angles: \citet{bes99} (1602),
\citet{pen00b} (2048), \citet{pen97} (1138), \citet{kap98} (0052),
\citet{pen99} (0943), \citet{car97} (0316), \citet{deb00} (2009, 1338
and 0924). \\ 
$^c$ The halo of TN J2009--3040 is circular. \\
\end{table}

\begin{figure*}
\includegraphics[width=17cm]{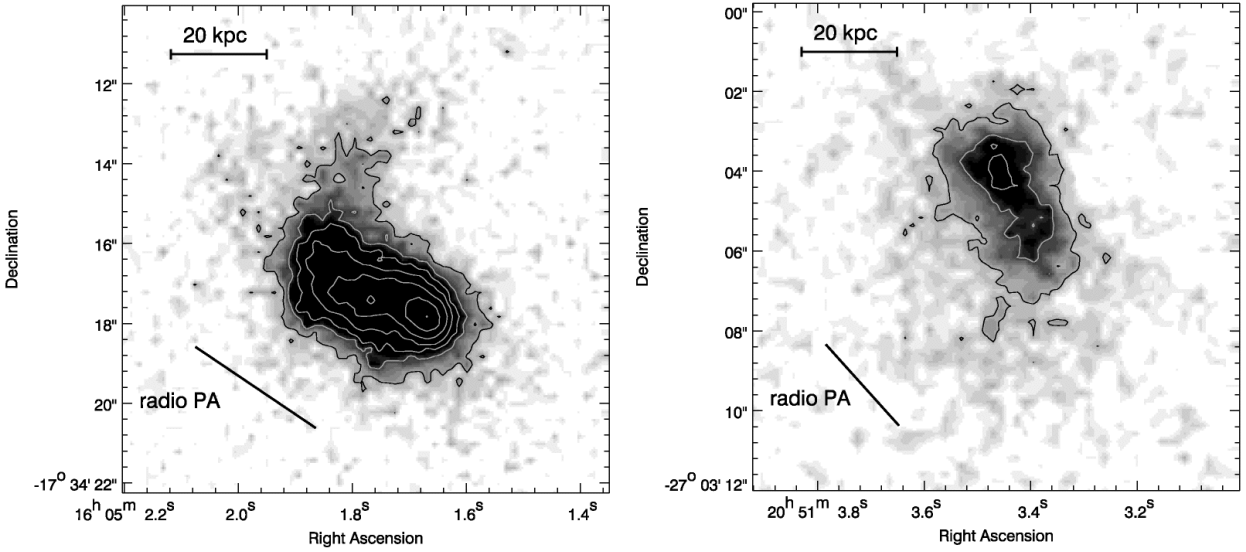}
\caption{\label{halo16022048} Continuum subtracted \lya\ halos of the
  radio galaxies BRL 1602--174 (left) and MRC 2048--272 (right). The
  depth of the images is given in Table \ref{imgobs}. The
  position angle of the radio emission is indicated on the left of
  each plot.}
\end{figure*}

\begin{figure*}
\includegraphics[width=17cm]{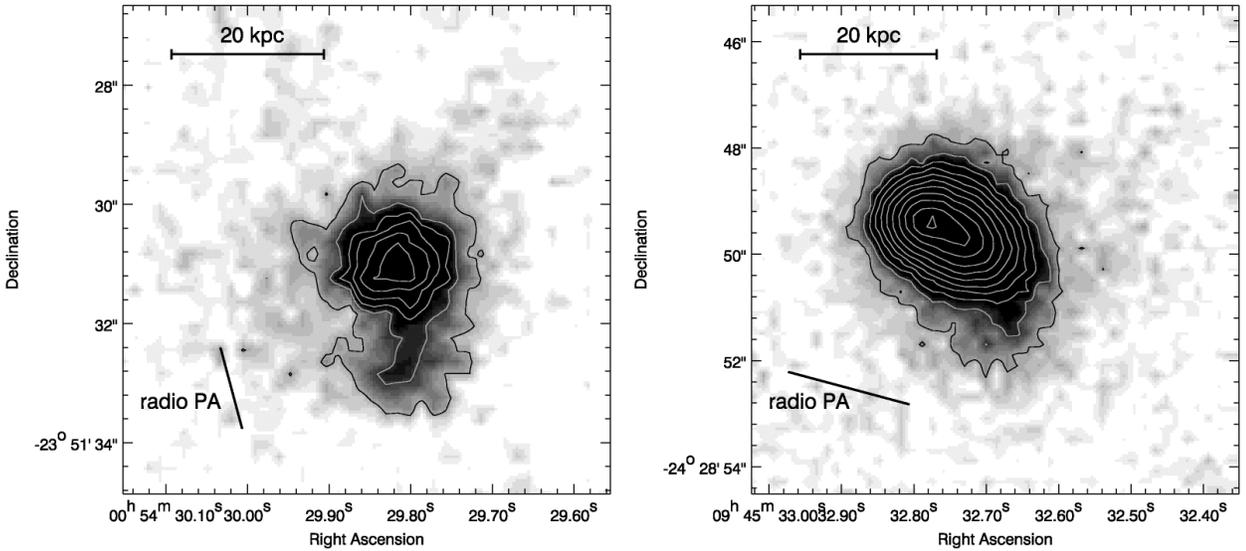}
\caption{\label{halo00520943} Same as Fig.\ \ref{halo16022048}, but
  showing the halos of the radio galaxies MRC 0052--241 (left) and MRC
  0943--242 (right).}
\end{figure*}

TN J0924--2201 is currently the highest redshift radio galaxy known
with a redshift of $z = 5.2$ \citep{bre99}. To select \lya\ emitters
in this field, a custom narrow-band filter encompassing the wavelength
of the \lya\ line of the radio galaxy at 7537 \AA\ was purchased. The
details of the observations and candidate selection in this field are
presented in \citet{ven04}. Follow-up spectroscopy of eight of 14
candidate \lya\ emitters resulted in the discovery of six \lya\
emitters near the radio galaxy \citep{ven04}. The field is overdense
in \lya\ emitters by a factor of 1.5--6.2 times the field density at
that redshift and comparable in density to the radio galaxy fields
1138, 0316 and 1338 at $z = 2.2, 3.1$ and 4.1 respectively.

\subsection{Radio galaxy \lya\ halos}
\label{halos}

\begin{figure*}
\includegraphics[width=17cm]{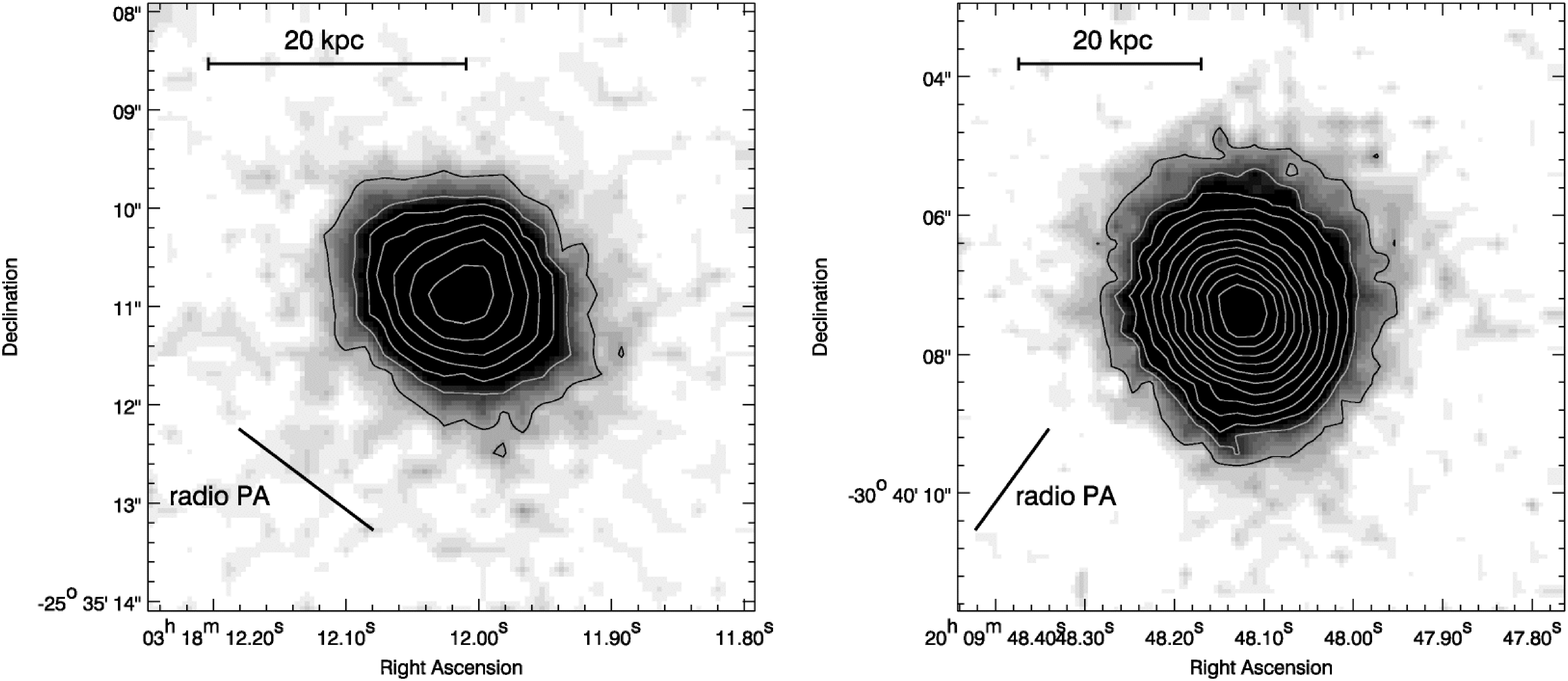}
\caption{\label{halo03162009} Same as Fig.\ \ref{halo16022048}, but
  showing the halos of the radio galaxies MRC 0316--257 (left) and TN
  J2009--3040 (right).}
\end{figure*}

\begin{figure*}
\includegraphics[width=17cm]{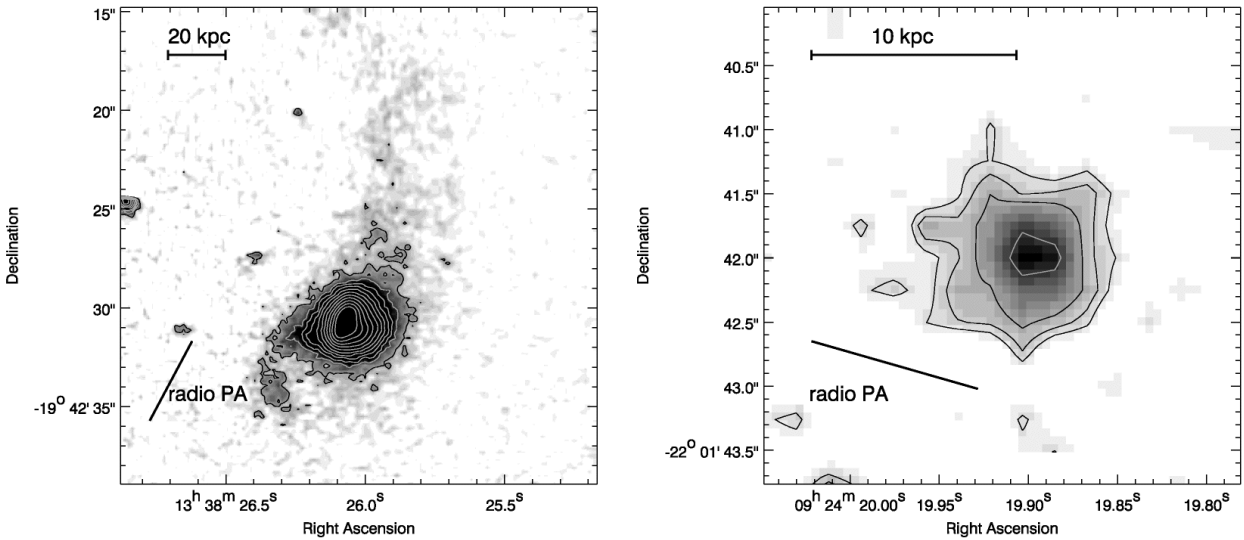}
\caption{\label{halo13380924} Same as Fig.\ \ref{halo16022048}, but
  showing the halos of the radio galaxies TN J1338--1942 (left) and TN
  J0924--2201 (right).}
\end{figure*}

Besides allowing the detection of candidate companion galaxies, the
deep narrowband \lya\ images of the radio galaxy fields also provide
deep maps of the diffuse emission line halos that surround the radio
galaxies. In each of the observed fields, the radio galaxy is the
brightest emitter of \lya\ emission, with line luminosities ranging
between $10^{43}$ and a few times $10^{45}$ erg\,s$^{-1}$. The
continuum subtracted \lya\ images of the radio galaxies observed in
our VLT program are shown in Figs. \ref{halo16022048}--\ref{halo13380924}.
The region from which the \lya\ emission originates is spatially
resolved and the emission line halos can be seen over 150 kpc in some
cases. In Table \ref{halotable}, we summarize the luminosity, size and
position angles of the halos.

The position angles of the emission line halos and the radio emission
are aligned, with the difference in the angle generally less than
$\sim$20$^{\circ}$ (see Table \ref{halotable}). This alignment between
the emission line halo and the radio jet is a common feature in radio
galaxies \citep[e.g.][]{mcc93,oji96,pen97,reu03a}, and suggests that a
least part of the \lya\ emission is caused by the interaction of the
gas with the radio jet or by ionisation by the photon beam that is
aligned with the radio jet. See for a detailed analysis of the halos
of the radio galaxies 1138 and 1338 the papers of \citet{pen97} and
\citet{zir05} respectively. Because an extensive study of the gas
halos surrounding the other radio galaxies is beyond the scope of this
article, we will present those results in a separate paper (Humphrey
et al.\ in prep).

It is interesting to mention that we do not find (angular) clustering
of \lya\ emitters along the axes of the radio sources. This so-called
``companion alignment'' has been predicted by models
\citep[e.g.][]{wes94} and statistical detections of this effect using
number counts have been reported by various authors
\citep[e.g.][]{rot96,bor06}. In contrast, the spatial distribution of
the \lya\ emitters near the radio galaxies studied in our program is
rather homogeneous throughout the field (e.g.\ Figs.\
\ref{0052velskydist}, \ref{0943velskydist}, and Fig.\ 19 in
V05). Recently, Croft et al.\ (2005) reported that near 1138 at least
three spectroscopically confirmed, X-ray selected AGN are, together
with the radio galaxy, members of a filamentary structure of at least
a few (proper) Mpc, in alignment with the radio jet axis. The \lya\
emitters in this protocluster do not show such a clear alignment
\citep{kur04b}. It is therefore possible that only certain types of
galaxies, e.g.\ the more massive galaxies, are preferentially aligned
with the radio galaxy axes, while the young and presumably less
massive \lya\ emitters (V05) do not show this angular
clustering. Future observations, aimed detecting the different galaxy
populations in the protoclusters, could test this picture.

\section{The environment of high redshift radio galaxies}
\label{resres}

\begin{table*}
\caption{\label{lptable} Summary of the results of the imaging and
  spectroscopic observations of the radio galaxy targets. For
  completeness, the target of the pilot project, MRC 1138--262 is
  included.}
\begin{center}
\begin{tabular}{rrrrrrrrrccc}
\hline
\hline
Field & $z$ & $N_\mathrm{img}^a$ & $N_\mathrm{spec}^b$ &
$N_\mathrm{conf}^c$ & $N_\mathrm{none}^d$ &
$N_{\mathrm{low}\,z}^e$ & $N_{\mathrm{extra}}^f$ & $N_\mathrm{tot}^g$ &
$n_\mathrm{rg}/n_\mathrm{field}^h$ & $\sigma_v^i$ &
$M_\mathrm{pcl}^j$ \\
{} & {} & {} & {} & {} & {} & {} & {} & {} & {} & \kms\ & 
$10^{14}$ \msun \\
\hline
1602 & 2.04 & ~2 & -- & -- & -- & -- & -- & -- & -- & -- & -- \\
2048 & 2.06 & 10 &  3 &  2 & 1 & 0 & 1 &  3 & $1.2^{+0.8}_{-0.7}$ & -- 
& -- \\
1138 & 2.16 & 37 & 11 & 11 & 0 & 0 & 4 & 15 & $4 \pm 2$ & $900 \pm
240$ & 3--4 \\
0052 & 2.86 & 57 & 36 & 35 & 1 & 0 & 2 & 37 & $3.0^{+0.5}_{-0.4}$ & 
$980 \pm 120$ & 3--4 \\
0943 & 2.92 & 65 & 30 & 25 & 4 & 1 & 3 & 28 & $3.2^{+0.9}_{-0.7}$ & 
$715 \pm 105$ & 4--5 \\
0316 & 3.13 & 77 & 30 & 28 & 1 & 1 & 3 & 31 & $3.3^{+0.5}_{-0.4}$ & 
$640 \pm 195$ & 3--5 \\
2009 & 3.16 & 21 &  9 &  9 & 0 & 0 & 2 & 11 & $1.7^{+0.8}_{-0.6}$ & 
$515 \pm 90$  & -- \\
1338 & 4.11 & 54 & 36 & 34 & 2 & 0 & 3 & 37 & $4.8^{+1.1}_{-0.8}$ & 
$265 \pm 65$  & 6--9 \\
0924 & 5.20 & 14 &  8 &  6 & 0 & 2 & 0 &  6 & $2.5^{+1.6}_{-1.0}$ & 
$305 \pm 110$ & 4--9 \\
\hline
\end{tabular}
\end{center}
$^a$ Number of candidate \lya\ emitters with {\em EW}$_0 > 15$ \AA\
and {\em EW}$_0/\Delta${\em EW}$_0> 3$. \\
$^b$ Number of spectroscopically observed candidates. \\ 
$^c$ Number of spectroscopically confirmed \lya\ emitters. \\
$^d$ Number of candidates not detected spectroscopically. \\
$^e$ Number of low redshift emission line galaxies among the observed
candidates. \\
$^f$ Number of confirmed \lya\ emitters not satisfying the imaging selection
criteria. \\ 
$^g$ Total number of confirmed \lya\ emitters near the radio galaxy. \\
$^h$ Density of the emitters as compared to the field density on the
basis of the imaging candidates. Note that the redshift range of
the confirmed emitters is generally smaller than the width of the
filter (see Sects.\ \ref{results} and \ref{mass}). \\
$^i$ Velocity dispersion of the confirmed emitters. \\
$^j$ Estimated mass of the protocluster (Sect.\ \ref{mass}). \\
\end{table*}

Besides the target of our pilot project, MRC 1138--262 at $z = 2.16$,
we imaged areas surrounding eight radio galaxies with $2.0 < z <
5.2$. A total of 300 candidate \lya\ emitters was selected fulfilling
the selection criteria {\em EW}$_0 > 15$ \AA\ and $\Sigma \equiv$ {\em
EW}$_0/\Delta${\em EW}$_0 > 3$ (Tables \ref{imgres} and
\ref{lptable}). One field (BRL 1602--174 at $z=2.04$) suffered from
high galactic extinction and a low response of the CCD. We will
discard this field in the following discussion. In the remaining seven
radio galaxy fields spectra were taken of 152 candidates. These
observations confirmed 139 \lya\ emitters. The success rate of our
selection procedure is $\sim91$\%. Only nine candidates could not be
confirmed spectroscopically. These candidates had a faint predicted
line flux ($F_{\mathrm{Ly}\alpha} < 10^{-17}$ \ergscm) and the
non-detections are most likely caused by a lack of signal-to-noise. An
overview of the number of candidates, spectroscopically observed and
confirmed \lya\ emitters per field is given in Table \ref{lptable}.

Four candidate emitters were identified with low redshift objects. We
estimate that the contamination of our (candidate) sample is roughly
$3$\%, although this percentage varies strongly from field to field
and with redshift.  For example, the fraction of contaminants in the
field surrounding TN J0924--2201 at $z = 5.2$ is 25\% \citep{ven04},
while in four fields near $z\sim3$, the contamination fraction is
$\sim2$\% (see Table \ref{lptable}).

Fourteen objects that fell outside our selection criteria were also
confirmed to be \lya\ emitters, increasing the number of confirmed \lya\
emitters to 153. Adding the 15 confirmed emitters in the 1138 field
\citep{pen00a}, a total of 168 \lya\ emitters are
confirmed near eight high redshift radio galaxies. 

Based on the (volume) overdensity and the clustering of emitters in
redshift space, we argue that at least five of the radio galaxies (MRC
1138--262, MRC 0052--241, MRC 0943--242, MRC 0316--257 and TN
J1338--1942) are associated with a forming cluster of galaxies
(protocluster). We are not able to make a definite statement about the
environment of TN J0924--2201 at $z = 5.2$, due to the small number of
confirmed galaxies \citep{ven04}. Recent multi-color observations with
the {\em HST} indicate that this field is overdense (at the 99.6\%
level) in $V$-band dropouts \citep{ove06b}, and these dropouts could
represent an additional galaxy population in the
protocluster. Although follow-up spectroscopy is needed to confirm
that the dropouts are at $z=5.2$, the richness of Lyman break galaxies
in the field strengthens our hypothesis that a large structure is
present at $z=5.2$. The two remaining radio galaxy fields, the 2048
and 2009 fields, have \lya\ volume densities consistent with the field
density. In the 2009 field, clustering can be seen in the position of
the emitters on the sky and in velocity space. More observations in
this field are needed to confirm the clustering and to identify a
possible structure of galaxies. Based on the results in the eight
radio galaxy fields, 75\% of luminous ($L_\mathrm{2.7\,GHz} > 10^{33}$
erg\,s$^{-1}$\,Hz$^{-1}$\,sr$^{-1}$) radio galaxies at $z > 2$ are
associated with a forming cluster of galaxies. 

In the next section, we will describe the properties of the radio
galaxy protoclusters as a whole. The properties of the high redshift \lya\
emitters are or will be discussed elsewhere (Kurk et al.\ 2004b; V05;
Venemans et al.\ in prep.).

\section{Properties of high redshift protoclusters}
\label{protoclusters}

\subsection{Structure size}
\label{size}

\begin{figure}
\includegraphics[width=8.5cm]{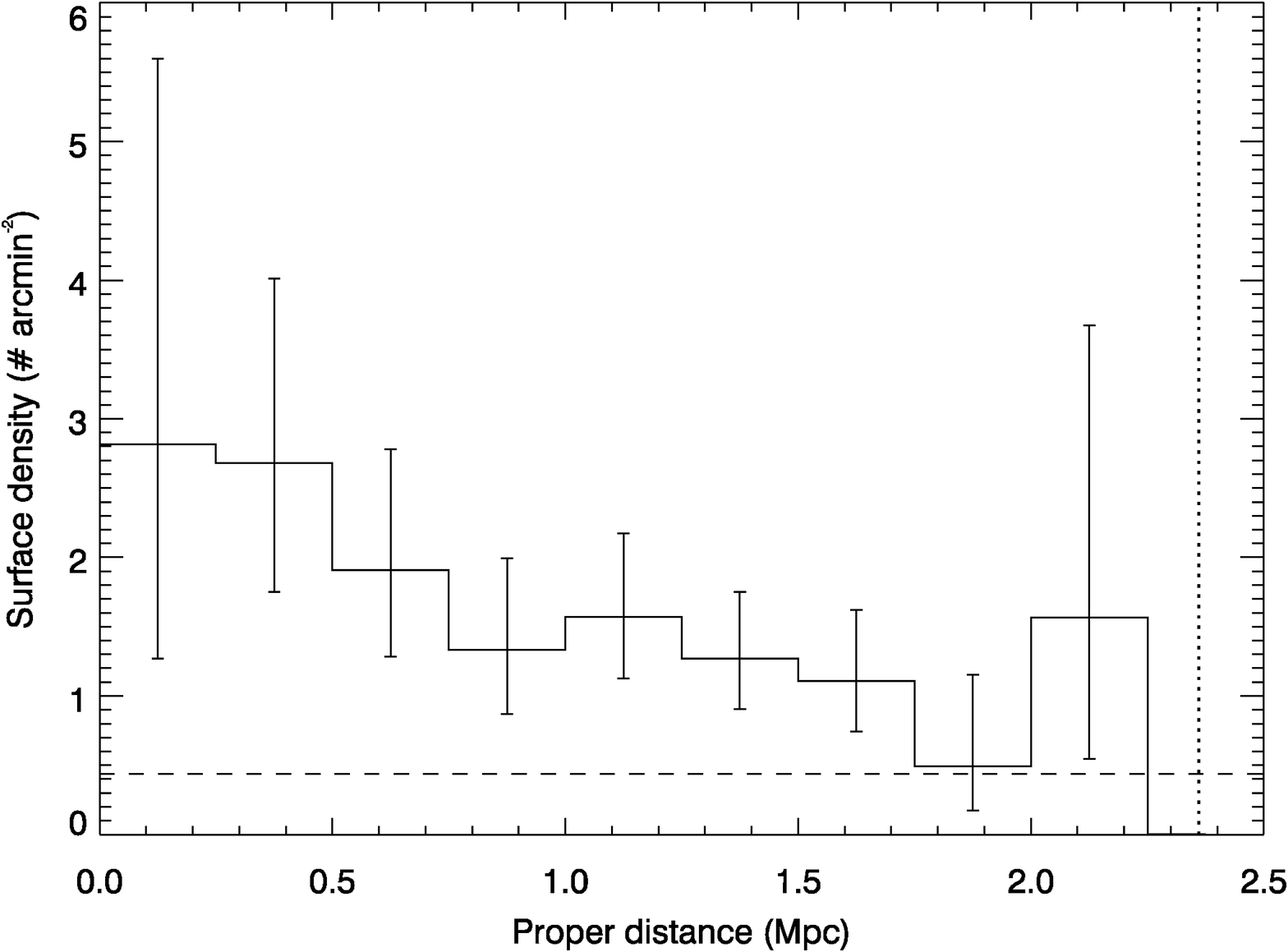}
\caption{\label{0943radial} Surface density of emitters as a function
  of proper distance from the radio galaxy MRC 0943--242. The error bars
  represent Poissonian errors. The horizontal dashed line is the
  surface density of \lya\ emitters in blank fields.  The vertical
  dotted line represents the maximum distance measurable in the
  image.}
\end{figure}

\begin{figure}
\includegraphics[width=8.5cm]{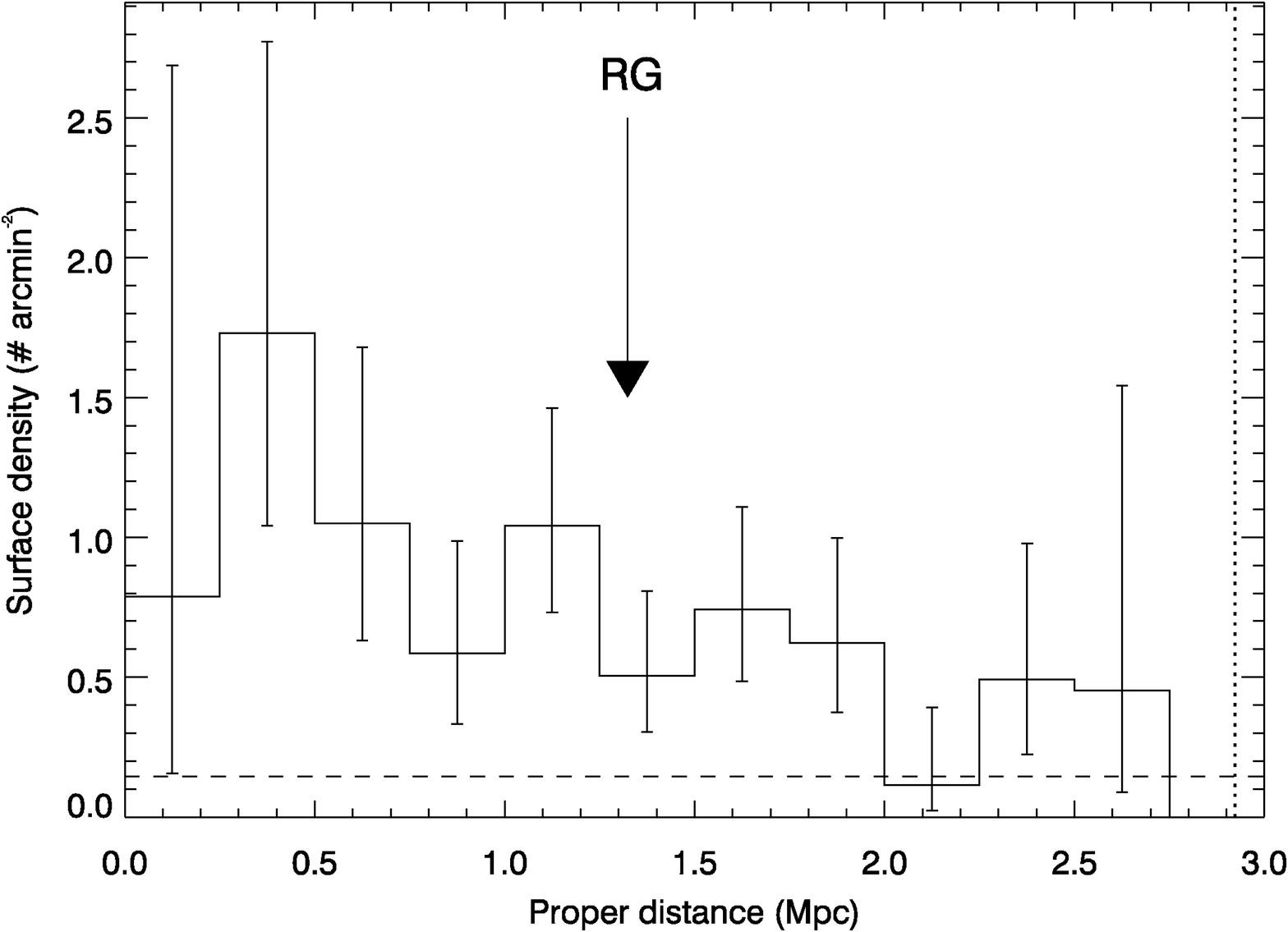}
\caption{\label{1338radial} Surface density of emitters as a function
  of proper distance from the average position of the emitters
  in the field near TN J1338--1942 at $z=4.1$. The position of the
  radio galaxy is indicated by the arrow. The error bars represent
  Poissonian errors. The dashed line is the surface density of field
  \lya\ emitters \citep[][ see Sect.\ \ref{1338res}]{daw04,shi04}. The
  dotted line is the maximum distance measurable in our imaging.}
\end{figure}

As can be seen in Figs.\ \ref{0052velskydist}, \ref{0943velskydist},
\ref{2009velskydist} and \ref{1338skydist}, the sizes of the
protoclusters are generally larger than the field of view of the FORS2
camera ($>$3.2$\times$3.2 Mpc$^2$). Only in the 1338 field a 
boundary can be seen to the north-west of the radio galaxy (but see
below). \citet{kur04b} found that the surface density of \lya\
emitters decreased with increasing distance to the radio galaxy in
their field. Within the limitations of the relatively small number of
objects, we tested each of our fields to see whether there is a
concentration of emitters near the radio galaxy. In two fields, the
0052 and 0316 fields, we found that density did not change with
distance to the radio galaxy, and that the distribution of emitters
was roughly homogeneous over the imaged area. Most likely, the extent
of the protocluster in these fields is larger than the area covered by
our imaging. In the 0943 field at $z = 2.9$, the surface density of
\lya\ emitters declines further away from the radio galaxy, as is
shown in Fig.\ \ref{0943radial}. At a proper distance of $\sim$1.75
Mpc from the radio galaxy, the surface density of \lya\ emitters is
consistent with the field density.
We regard this radius as the size of the protocluster. A similar
distribution was found in the 1338 field, but only if the average
position of the emitters was taken as centre (see Fig.\
\ref{1338radial}). As mentioned in Sect.\ \ref{1338res}, the radio
galaxy is located at a distance of 1.3 Mpc from the centre of the
emitters. We estimate that the size of this protocluster is 2.0 Mpc
(Fig.\ \ref{1338radial}). 

However, these size estimates of the protoclusters are based only on
the position of \lya\ emitters. It is possible that other galaxy
populations in the protoclusters are distributed differently. As
already mentioned in Sect.\ \ref{1338res}, an overdensity of mm
galaxies is located to the north of the radio galaxy TN J1338--1942
\citep{deb04}, whereas the highest concentrations of \lya\ emitters in
this field lies south of the radio source. Also, \citet{kur04b} found
that in the 1138 protocluster at $z=2.16$ the population of H$\alpha$
emitters and EROs are more concentrated towards the radio galaxy than
the \lya\ emitters. Despite this caveat, our size (radius) estimates
of $>$1.75--2.0 Mpc are consistent with the determination by other
groups. For example, \citet{kee99} estimate that the large structure
around the radio galaxy 53W002 at $z = 2.4$ has a diameter of
$\sim3.3$ Mpc. \citet{hay04} imaged a protocluster at $z\sim3.1$, that
was serendipitously discovered by \citet{ste98}. They found that the
region that is overdense in \lya\ emitters has a size of approximately
5$\times$5 Mpc$^2$. At a higher redshift, an overdensity of \lya\
emitters at $z = 4.9$ was discovered by \citet{shi03} in the SDF. This
overdensity has a roughly circular shape with a radius of $\sim$2.0
Mpc \citep{shi03}, which is very similar to the size of the 1338
protocluster.

To summarize, we find that the sizes of the protoclusters, as inferred
by the distribution of \lya\ emitters at $z=2.9$ and $z=4.1$, are
roughly $\sim2.0$ Mpc. The other protoclusters do not show a boundary
and could be larger. Wide field imaging is needed for these
protoclusters to see a boundary of the galaxy distribution.

\subsection{Mass}
\label{mass}

The age of the Universe at $z = 3$ is only $\sim 15$\% of its current
age, which corresponds to $\sim2$ Gyr in our adopted cosmology. A
protocluster galaxy with a relative velocity of 500 \kms\ would take
at least 4 Gyr to cross a structure with a size of 2 Mpc. Therefore,
the protoclusters are most likely still in the process of
formation. They are still far from virialization and we cannot use the
virial theorem to calculate the mass of the galaxy structures.

An alternative approach to calculate the mass of the structures is to
assume that the mass that is located within the volume that is
occupied by the protocluster galaxies will collapse into a single
structure \citep[see e.g.,][]{ste98}. To use this method we need to
know the mass overdensity $\delta_m$ within the volume. The mass
overdensity is related to the galaxy overdensity $\delta_\mathrm{gal}$
by the bias parameter $b \equiv \delta_\mathrm{gal}/\delta_m$.

To compute the volume of the protocluster, we estimate the redshift
range of the confirmed \lya\ emitters after removing outliers
(galaxies more than $\sim$1500 \kms\ away from the median
redshift). Because redshift space distortions modify the occupied
volume, we corrected for peculiar velocities by assuming the structure
is just breaking away from the Hubble flow (see Steidel et al.\ 1998;
V05 for more details). \nocite{ste98} If outliers are included, the
volume occupied by the protoclusters increases and the estimated
masses are higher.

Using this method, V05 estimate that the mass of the 0316 protocluster
is $\sim5 \times 10^{14}$ \msun. Below we will estimate the mass of
the protoclusters in the 1138, 0052, 0943, 1338 and 0924 fields. A
bias parameter of $b = 3-6$ will be assumed, as suggested by e.g.\
\citet{ste98} and \citet{shi03}.

~$\bullet$ {\em 1138:} near the radio galaxy MRC 1138--262, the
overdensity of emitters in a comoving volume of 4490 Mpc$^3$ is
$\delta_{\mathrm{gal}} > 3.4$ \citep{pen00a,kur04b}. The computed mass
in this volume is at least $3-4 \times 10^{14}$ \msun.

Interestingly, the redshift distribution of the \lya\ emitters in this
field appears to be bimodal \citep{pen00a}. The emitters are located
in two subgroups with velocity dispersions 520 and 280 \kms\
\citep{pen00a}. Calculating the mass of the individual groups gives
$\sim2 \times 10^{14}$ and $1-2 \times 10^{14}$ \msun. If the galaxies
in these two groups are close to virialization, we can apply the
virial theorem. \citet{kur04b} estimated that the virial radii of the
two groups is $\sim1$ Mpc. Using the velocity dispersions as given by
\citet{pen00a}, the virial masses of the groups are $4.0 \times 10^{14}$
and $1.0 \times 10^{14}$ \msun, comparable to the other estimates.

~$\bullet$ {\em 0052:} to estimate the volume that most likely will
collapse into a single structure, we removed the three lowest and the two
highest redshift galaxies. These galaxies are on the edges of the
redshift distribution (Fig.\ \ref{0052velskydist}), and we regard them
as outliers. The remaining 32 emitters have redshifts in the range
$2.8437 \leq z \leq 2.8673$. The range of $\Delta z = 0.0236$ is a
factor 2.3 smaller than the redshift range probed by the narrow-band filter
($\Delta z = 0.0541$). The volume occupied by the emitters is 3510 comoving
Mpc$^3$ (uncorrected for peculiar velocities). The overdensity within
the volume is $0.0541/0.0236 \times 32/37 \times 3.0 = 5.95$. Using
these values, the estimated mass is $3-4 \times 10^{14}$
\msun. Because the field of view of our imaging is not large enough to
show a boundary of the structure (Fig.\ \ref{0052velskydist}), this
mass is a lower limit. 

As in the 1138 field, the emitters are
concentrated in two velocity groups (see Fig.\ \ref{0052velskydist}). The
groups consist of 12 and 19 members and have velocity dispersions of
185 and 230 \kms\ respectively. Assuming that the virial theorem can
be applied for the two groups, this gives masses 
of $1.1 \times 10^{14}$ and $7 \times 10^{13}$ \msun\ for the
groups. Most likely, the mass of the protocluster lies in the range
$2-4 \times 10^{14}$ \msun.

~$\bullet$ {\em 0943:} the redshift range of the confirmed emitters is
0.034, and, as described in Sect.\ \ref{size}, the size of the
structure is roughly 1.75 Mpc in radius. This gives a comoving volume
of 4570 Mpc$^3$. Within this volume 61 of the 65 emitters (94\%) are
located. The overdensity of emitters in this volume is 5.5, giving a
protocluster mass of $4-5 \times 10^{14}$ \msun. Because the
protocluster might extend towards lower redshifts (see Fig.\
\ref{0943velskydist}), this mass must be regarded as a lower limit.

~$\bullet$ {\em 1338:} the $z=4.1$ protocluster has a size of 2 Mpc in
radius (Sect.\ \ref{size}), giving an area that is 78\% of the total
area observed (Fig.\ \ref{1338skydist}). The fraction of (candidate)
emitters in this area is 91\%. The volume occupied by these emitters
is 5815 Mpc$^3$. The overdensity of emitters in this volume is $\sim8.5$,
resulting in a mass of $6-9 \times 10^{14}$ \msun.

~$\bullet$ {\em 0924:} the redshifts of the confirmed emitters near TN
J0924--2201 only span a relatively narrow redshift range of $\Delta z = 0.0151$
\citep{ven04}. The corresponding volume is 1870 Mpc$^3$ and the
density of emitters as compared to the field density in this volume is
20.3. The structure mass is estimated to be $4-6 \times 10^{14}$
\msun, although the small number of confirmed emitters makes this mass
quite uncertain.

The mass estimates of the protoclusters are in the range
$2-9 \times 10^{14}$ \msun. Locally, such masses correspond to
clusters with Abell richness class 0 and higher \citep[e.g.,][]{bah93}. This
indicates that the protoclusters are the progenitors of present-day
massive clusters.

\subsection{Velocity dispersions}
\label{veldisp}

\begin{figure}
\includegraphics[width=8.5cm]{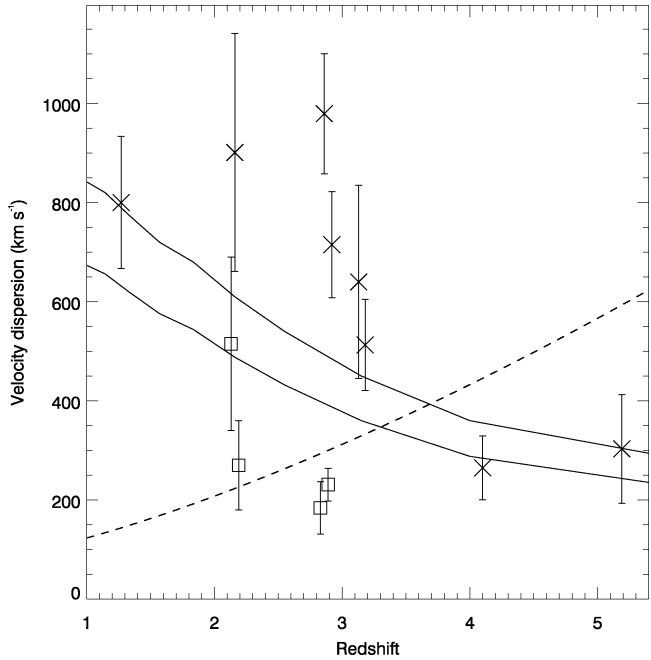}
\caption{\label{allveldists} Velocity dispersion of \lya\ emitters in
  our radio galaxy fields ({\em crosses}) as a function of
  redshift. The squares represent the velocity dispersions of the
  subgroups in the 1138 and 0052 fields (see Sect.\ \ref{mass}). The
  data point near $z = 1$ represents the velocity dispersion in the
  X-ray cluster RDCS 1252.9-2927 at $z=1.24$ \citep{ros04}. The
  dashed line is the Hubble flow $H(z)$ over 1 Mpc. The solid lines
  show the evolution of the dark matter velocity dispersions
  $\sigma_{\mathrm{dm}}$ of simulated massive clusters
  \citep{eke98}. The top line is scaled to 1000 \kms\ at $z=0$, the
  bottom line to 800 \kms\ at $z=0$.}
\end{figure}

In Fig.\ \ref{allveldists}, the velocity dispersion of the \lya\
emitters in the radio galaxy fields are plotted as a function of
redshift. Despite the small number of data points, the dispersion
appears to increase with decreasing redshift. While at $z > 4$ the
velocity dispersion is $\sim300$ \kms, the dispersion increases to
500--700 \kms\ at $z\sim3$. In our two lowest redshift fields
($z=2.86$ and $z=2.16$), the velocity dispersion of the emitters is
1000 \kms. However, it should be noted that the velocity distribution
of emitters at $z=2.86$ and $z=2.16$ is bimodal (Sect.\ \ref{mass}),
and the velocity dispersion of the individual subgroups is much lower
(200--500 \kms). Numerical simulations of the evolution of massive
clusters predict that the velocity dispersion of the dark matter
increases with cosmic time \citep[e.g.,][]{eke98}. Although based on
only one population of galaxies in the protoclusters, our observations
are consistent with these predictions. Our data approximately follow
the evolution of a massive cluster of galaxies, with a velocity
dispersion of 1000 \kms\ at $z=0$ (Fig.\
\ref{allveldists}). Additional observations of high redshift
protoclusters are needed to confirm this.

\section{Discussion}
\label{discussion}

A question is whether the radio galaxy selected protoclusters
presented in this paper are biased. To answer this question, we have
to assess the number density of powerful radio galaxies at $z > 2$. 

The radio galaxies in our program have luminosities
$L_\mathrm{2.7\,GHz} > 10^{33}$ erg\,s$^{-1}$\,Hz$^{-1}$\,sr$^{-1}$ 
(Table \ref{fields}). The number density of radio sources with this
luminosity limit in the redshift range $2 < z < 5$ is $4 \times
10^{-8}$ Mpc$^{-3}$ \citep{dun90}. Using the more recent radio
luminosity function at 151 MHz as derived by \citet{wil01} and
assuming a spectral slope of $\alpha \simeq -1.0$ gives a number
density that is similar within a factor of 2. It should be noted
that the number density depends strongly on the lower luminosity limit
that is assumed, since the density varies approximately with $n
\propto L^{-2}$ \citep{dun90}.

The association of 75\% of $z > 2$ radio galaxies with a protocluster
(Sect.\ \ref{resres}) implies that there are roughly $3 \times
10^{-8}$ forming clusters at $2.0 < z < 5.2$ per comoving Mpc$^3$ with
an active radio source. However, it is likely that there will be many
more forming clusters in this redshift range without an active radio
source. \citet{blu99} argue that high redshift radio galaxies are
visible only when they are younger than $10^7$ yr. Between $z = 5.2$
and $z = 2$, the age of the Universe increases with roughly 2 Gyr.
This means that between $z = 5.2$ and $z = 2$, there must be a factor
$\sim2\times10^2$ more forming clusters {\em without} a bright radio
source. Therefore, most high redshift protoclusters will not have a
powerful radio galaxy.

Based on these calculations, we estimate that the number density of
protoclusters such as the ones presented in this paper is $\sim6
\times 10^{-6}$ Mpc$^{-3}$. It is instructive to compare this number
with the density of protoclusters as derived by other
groups. \citet{ste98} estimate that they find one galaxy overdensity
in a 9\arcmin$\times$18\arcmin\ field in a redshift range $2.7 < z <
3.4$ in their Lyman break galaxy survey. In our cosmology, this
translates to a number density of protoclusters of $3 \times 10^{-6}$
Mpc$^{-3}$. High redshift blank field \lya\ searches also found
several galaxy overdensities. \citet{shi03} reports the discovery of a
structure of \lya\ emitters at $z = 4.9$ in the Subaru Deep Field with
an estimated mass of $3 \times 10^{14}$ \msun. The implied volume
density of such structures is $7 \times 10^{-6}$ Mpc$^{-3}$. At even
higher redshifts, \citet{ouc05} find two overdensities of \lya\
emitters at $z = 5.7$ in a 1.0 deg$^2$ region of the sky. They
estimate a density of $> 2 \times 10^{-6}$ Mpc$^{-3}$ for such
structures. These blank field density estimates are very
similar to our calculated density of $\sim6 \times 10^{-6}$
Mpc$^{-3}$. This indicates that the overdensities discovered in blank
field surveys and our protoclusters near radio galaxies could very
well be the same type of object. It is likely that we see the
protocluster at the time that the brightest cluster galaxy undergoes a
burst of nuclear activity. 

Interestingly, in the local Universe objects with a similar volume
density as the protoclusters are clusters with masses $\gtrsim$ $2
\times 10^{14}$ \msun\ \citep{gir98,rei02}, which compares favourably
with the mass estimates presented in Sect.\ \ref{mass}. This further
supports the idea that the protoclusters are the progenitors of (rich)
clusters of galaxies.

\section{Summary}
\label{summary}

In this paper we have presented the results of a large observational
program to search for galaxy overdensities near luminous radio
galaxies at $2 < z < 5.2$. The targets for our project were selected from a
list of roughly $150$ $z>2$ radio sources, and were required to be
luminous in the radio with $L_\mathrm{2.7\,GHz} > 10^{33}$
erg\,s$^{-1}$\,Hz$^{-1}$\,sr$^{-1}$ and to have redshifts and
positions optimum for deep narrow-band \lya\ imaging with the FORS2
camera on the VLT. We selected three radio galaxies at $z\simeq2.1$
(including the target of our pilot project MRC 1138--262), two
at $z\simeq2.9$, two at $z\simeq3.1$, one at $z=4.1$ and one at $z=5.2$.

To search for \lya\ emitting galaxies at the redshift of the radio
galaxies, we obtained deep narrow-band \lya\ and broad-band images
with the VLT and the Keck telescope of the fields surrounding the
radio galaxies. The size of the images was roughly
6\farcm8$\times$6\farcm8, which corresponds at a redshift of 3 to a
proper size of 3.1$\times$3.1 Mpc$^2$. The (typical) sensitivity
reached in the imaging is a \lya\ luminosity of $L_{\mathrm{Ly}\alpha}
\sim 10^{42}$ erg\,s$^{-1}$ for an emitter with no continuum detected
at the $5\sigma$ level in the narrow-band image. An exception is the
field of BRL 1602--174 at $z=2.04$, the lowest redshift radio galaxy
in our program. Due to the high galactic extinction towards this radio
source and the low quantum efficiency of the FORS2 detector at the
wavelengths below 4000 \AA, the imaging of this field was significant
shallower than of the other fields and therefore only two candidate
emitters were detected. No follow-up spectroscopy was conducted in
this field. Using the narrow-band and broad-band images we found 300
candidate \lya\ emitters with a rest-frame equivalent width of {\em
EW}$_0>15$ \AA\ and a significance $\Sigma \equiv$ {\em
EW}$_0/\Delta${\em EW}$_0 > 3$.

To confirm whether these candidate emitters have redshifts similar to
that of the radio galaxies, we conducted follow-up spectroscopy with
the VLT and Keck. Candidate emitters were
confirmed to be \lya\ emitters based on an observed (asymmetric) line
profile, emission line ratios and/or continuum break. Of the 152
candidate \lya\ emitters observed spectroscopically, 139 ($91$\%) were
confirmed to be \lya\ emitters, four ($3$\%) were low redshift
interlopers and nine (6\%) did not show a line (or continuum) in their
spectra, which is most likely due to a lack of
signal-to-noise. Furthermore, 14 sources that were not selected as
emitters were also confirmed to be high redshift \lya\
emitters. Combined with the 15 emitters found near MRC 1138--262, the
total number of confirmed \lya\ emitters is 168 in eight radio
galaxies fields. We will describe the properties of these \lya\
emitters in a forthcoming paper (Venemans et al.\ in prep). 

Six of the fields are overdense in \lya\ emitters by a factor of 3--5
as compared to the volume density of (field) \lya\ emitters at similar
redshifts. The remaining two radio galaxy fields have volume densities
that are within the errors consistent with the field density. The
significance of the volume overdensity near our highest redshift radio
galaxy, at $z=5.2$, is only marginal due to the low number (six) of
confirmed emitters. Recently, an overdensity of LBGs was found in this
field using multi-color {\em HST} observations. This strengthens our
hypothesis that the radio galaxy is located in an overdense region.

The confirmed emitters show clustering in redshift space, and the
width of the redshift distribution of the confirmed emitters is a
factor 2--5 smaller than the width of the narrow-band filters. Taken
together with the volume overdensity of \lya\ emitters, we argue that
six of the radio galaxies reside in a forming cluster of galaxies or
protocluster. We estimate that of all powerful high redshift radio
galaxies at least 6/8 = 75\% are located in a protocluster, making our
program very efficient in finding high redshift forming clusters.

The size of the protoclusters is, based on the distribution of \lya\
emitters, at least 1.75 Mpc. This is consistent with the size
measurements by other groups. The mass of the protoclusters can be
calculated assuming the galaxies in the protoclusters are just
breaking away from the Hubble flow and that the mass inside the volume
occupied by the protocluster galaxies will collapse into a single
structure. This method yields masses in the range $2-9 \times 10^{14}$
\msun, suggesting that the protoclusters are the progenitors of
clusters with Abell richness class 0 or higher.  The results of our
program allows us to follow the evolution of the velocity dispersion
of the \lya\ emitters in the overdense regions as a function of
redshift. Although based on only a small number of data points, we find
that the velocity dispersion increases with decreasing redshift, in
agreement with the predictions of numerical simulations of forming
clusters. In a future paper, we will present a more detailed
comparison of our observations with numerical simulations of forming
massive clusters (Venemans et al.\ in prep).

Based on the results of our program and the density of luminous radio
sources at $z > 2$ and assuming that radio galaxies at high redshift
are visible only for $10^7$ yr, we estimate that the number density of
protoclusters at $2 < z < 5.2$ is $6\times10^{-6}$
Mpc$^{-3}$. Locally, objects that have such number densities are
clusters of galaxies with masses in excess of $2\times10^{14}$
\msun. This further supports the idea that the protoclusters are the
progenitors of (rich) clusters of galaxies. Using the protoclusters
presented in this paper, we are now able to systematically study
the evolution of clusters and their galaxies out to redshifts $z > 5$.

\begin{acknowledgements} 
  We are grateful to the staff of Paranal, Chile, for their excellent
  support. We thank Gero Rupprecht at ESO and William Grenier of
  Andover Corporation for their help in our purchase of the custom
  narrow-band filters. BPV thanks Michiel Reuland, Hyron Spinrad,
  Steve Dawson and Curtis Manning for useful discussions, and Philip
  Best for carefully reading an early draft of this manuscript. GKM
  acknowledges funding by an Academy Professorship of the Royal
  Netherlands Academy of Arts and Sciences (KNAW). The work by WvB, SC
  and SAS at LLNL was performed under the auspices of the U.S.\
  Department of Energy, National Nuclear Security Administration, by
  the University of California, Lawrence Livermore National
  Laboratory, under contract No.\ W-7405-Eng-48. The authors wish to
  recognize and acknowledge the very significant cultural role and
  reverence that the summit of Mauna Kea has always had within the
  indigenous Hawaiian community. We are most fortunate to have the
  opportunity to conduct observations from this mountain. This
  research has made use of the NASA/IPAC Extragalactic Database (NED)
  which is operated by the Jet Propulsion Laboratory, California
  Institute of Technology, under contract with the National
  Aeronautics and Space Administration. This work was supported by the
  European Community Research and Training Network ``The physics of
  the Intergalactic Medium''. SC and WvB acknowledge support for radio
  galaxy studies at UC Merced, including the work reported here, with
  the Hubble, Spitzer and Chandra space telescopes via NASA grants HST
  10127, SST 3482, SST 3329 and CXO 06701011.
\end{acknowledgements}

\appendix
\section{Object lists}

\begin{table*}
\caption{\label{2048table} Position and properties of the \lya\ emission
  line of the confirmed \lya\ emitters and the radio galaxy in the
  2048 field.}
\begin{tabular}{ccccccc}
\hline \hline Object & \multicolumn{2}{c}{Position} & $z$ & Flux &
EW$_0$ & FWHM \\ 
{} & $\alpha_{\mathrm{J}2000}$ & $\delta_{\mathrm{J}2000}$ & {} &
\ergscm\ & \AA & \kms\ \\  
\hline
2048.LAE1 & 20 51 02.05 & $-$27 04 52.7 & 2.1058 $\pm$ 0.0004 & (1.5
$\pm$ 0.2)$\times 10^{-16}$ & $145_{-35}^{+56}$ & 720 $\pm$ 70 \\
2048.LAE2 & 20 51 07.53 & $-$27 04 46.1 & 2.0591 $\pm$ 0.0005 & (1.1
$\pm$ 0.1)$\times 10^{-16}$ & $37_{-3}^{+4}$ & 1015 $\pm$ 115 \\
2048.LAE3 & 20 51 13.26 & $-$27 00 57.8 & 2.0580 $\pm$ 0.0005 & (1.5
$\pm$ 0.8)$\times 10^{-17}$ & $23_{-5}^{+6}$ & $<$415 \\
\hline 2048.RG & 20 51 03.45 & $-$27 03 04.1 & 2.0590 $\pm$ 0.0004 &
(2.9 $\pm$ 0.3)$\times 10^{-16}$ & $309_{-27}^{+50}$ & 1580 $\pm$ 110
\\ 
\hline
\end{tabular}
\end{table*}

\begin{table*}
\caption{\label{0052table} Position and properties of the \lya\ emission
  line of the confirmed \lya\ emitters and the radio galaxy in the
  0052 field.}
\begin{tabular}{ccccccc}
\hline
\hline
Object & \multicolumn{2}{c}{Position} & $z$ & Flux & EW$_0$ & FWHM \\ 
{} & $\alpha_{\mathrm{J}2000}$ & $\delta_{\mathrm{J}2000}$ & {} &
\ergscm\ & \AA & \kms\ \\
\hline
0052.LAE1 & 00 54 14.83 & $-$23 51 27.4 & 2.8301 $\pm$ 0.0010 & (4.8 $\pm$
0.5)$\times 10^{-16}$ & 104$_{-5}^{+6}$ & 1075 $\pm$ 85 \\
0052.LAE2 & 00 54 15.01 & $-$23 49 42.9 & 2.8655 $\pm$ 0.0003 & (1.1 $\pm$
0.2)$\times 10^{-17}$ & 76$_{-21}^{+1000}$ & 305 $\pm$ 50 \\
0052.LAE3 & 00 54 19.05 & $-$23 51 41.7 & 2.8359 $\pm$ 0.0002 & (1.0 $\pm$
0.3)$\times 10^{-17}$ & 9$_{-3}^{+4}$ & 155 $\pm$ 40 \\
0052.LAE4 & 00 54 21.74 & $-$23 51 51.1 & 2.8691 $\pm$ 0.0004 & (4.4 $\pm$
1.6)$\times 10^{-17}$ & 123$_{-25}^{+488}$ & 430 $\pm$ 50 \\
0052.LAE5 & 00 54 21.95 & $-$23 54 08.9 & 2.8666 $\pm$ 0.0002 & (1.2 $\pm$
0.2)$\times 10^{-16}$ & 443$_{-62}^{+156}$ & 220 $\pm$ 30 \\
0052.LAE6 & 00 54 22.62 & $-$23 55 07.0 & 2.8489 $\pm$ 0.0002 & (1.8 $\pm$
0.5)$\times 10^{-17}$ & 52$_{-13}^{+492}$ & 255 $\pm$ 40 \\
0052.LAE7 & 00 54 22.75 & $-$23 53 06.7 & 2.8599 $\pm$ 0.0002 & (1.0 $\pm$
0.3)$\times 10^{-17}$ & 94$_{-18}^{+24}$ & 205 $\pm$ 50 \\
0052.LAE8 & 00 54 24.57 & $-$23 53 12.8 & 2.8449 $\pm$ 0.0002 & (2.0 $\pm$
0.4)$\times 10^{-17}$ & 218$_{-28}^{+42}$ & 220 $\pm$ 40 \\
0052.LAE9 & 00 54 24.58 & $-$23 54 54.7 & 2.8636 $\pm$ 0.0004 & (5.2 $\pm$
2.3)$\times 10^{-17}$ & 49$_{-9}^{+27}$ & 525 $\pm$ 85 \\
0052.LAE10 & 00 54 25.35 & $-$23 53 07.5 & 2.8628 $\pm$ 0.0003 & (4.2 $\pm$
1.0)$\times 10^{-18}$ & 54$_{-14}^{+495}$ & 305 $\pm$ 50 \\
0052.LAE11 & 00 54 25.64 & $-$23 53 40.6 & 2.8448 $\pm$ 0.0001 & (6.1 $\pm$
1.1)$\times 10^{-18}$ & 105$_{-18}^{+24}$ & 130 $\pm$ 25 \\
0052.LAE12 & 00 54 26.05 & $-$23 51 40.6 & 2.8656 $\pm$ 0.0004 & (2.7 $\pm$
0.9)$\times 10^{-18}$ & 68$_{-16}^{+365}$ & 280 $\pm$ 80 \\
0052.LAE13 & 00 54 27.31 & $-$23 51 52.9 & 2.8651 $\pm$ 0.0008 & (2.3 $\pm$
0.6)$\times 10^{-17}$ & 114$_{-24}^{+33}$ & 765 $\pm$ 150 \\
0052.LAE14 & 00 54 28.17 & $-$23 51 52.3 & 2.8444 $\pm$ 0.0012 & (1.0 $\pm$
0.3)$\times 10^{-17}$ & 56$_{-12}^{+199}$ & 950 $\pm$ 235 \\
0052.LAE15 & 00 54 28.59 & $-$23 53 48.6 & 2.8550 $\pm$ 0.0002 & (1.2 $\pm$
0.1)$\times 10^{-16}$ & 108$_{-11}^{+26}$ & 585 $\pm$ 25 \\
0052.LAE16 & 00 54 28.65 & $-$23 51.55.3 & 2.8479 $\pm$ 0.0003 & (2.1 $\pm$
0.4)$\times 10^{-17}$ & 45$_{-8}^{+26}$ & 325 $\pm$ 50 \\
0052.LAE17 & 00 54 28.89 & $-$23 52 26.3 & 2.8673 $\pm$ 0.0004 & (1.0 $\pm$
0.2)$\times 10^{-17}$ & 26$_{-5}^{+9}$ & 395 $\pm$ 70 \\
0052.LAE18 & 00 54 29.31 & $-$23 53 18.1 & 2.8669 $\pm$ 0.0002 & (5.2 $\pm$
0.5)$\times 10^{-17}$ & 132$_{-19}^{+78}$ & 425 $\pm$ 30 \\
0052.LAE19 & 00 54 29.84 & $-$23 54 42.8 & 2.8618 $\pm$ 0.0001 & (2.3 $\pm$
0.3)$\times 10^{-17}$ & 168$_{-39}^{+1000}$ & 285 $\pm$ 30 \\
0052.LAE20 & 00 54 30.53 & $-$23 53 34.1 & 2.8592 $\pm$ 0.0003 & (6.9 $\pm$
2.6)$\times 10^{-18}$ & 78$_{-17}^{+22}$ & 150 $\pm$ 60 \\
0052.LAE21 & 00 54 30.61 & $-$23 53 36.2 & 2.8600 $\pm$ 0.0003 & (7.2 $\pm$
1.6)$\times 10^{-18}$ & 27$_{-5}^{+11}$ & 260 $\pm$ 50 \\
0052.LAE22 & 00 54 31.99 & $-$23 49 09.3 & 2.8503 $\pm$ 0.0002 & (4.8 $\pm$
0.8)$\times 10^{-18}$ & 118$_{-20}^{+26}$ & 225 $\pm$ 30 \\
0052.LAE23 & 00 54 32.86 & $-$23 52 14.1 & 2.8660 $\pm$ 0.0003 & (1.1 $\pm$
0.3)$\times 10^{-17}$ & 74$_{-16}^{+211}$ & 365 $\pm$ 65 \\
0052.LAE24 & 00 54 34.64 & $-$23 53 27.7 & 2.8460 $\pm$ 0.0004 & (1.0 $\pm$
0.3)$\times 10^{-17}$ & 76$_{-16}^{+21}$ & 275 $\pm$ 75 \\
0052.LAE25 & 00 54 35.61 & $-$23 53 27.9 & 2.8441 $\pm$ 0.0003 & (1.7 $\pm$
0.3)$\times 10^{-16}$ & 468$_{-65}^{+165}$ & 430 $\pm$ 50 \\
0052.LAE26 & 00 54 35.66 & $-$23 54 16.7 & 2.8602 $\pm$ 0.0002 & (6.6 $\pm$
2.3)$\times 10^{-18}$ & 34$_{-8}^{+34}$ & 90 $\pm$ 55 \\
0052.LAE27 & 00 54 36.28 & $-$23 50 01.6 & 2.8437 $\pm$ 0.0003 & (2.0 $\pm$
0.2)$\times 10^{-17}$ & 27$_{-6}^{+14}$ & 670 $\pm$ 60 \\
0052.LAE28 & 00 54 36.30 & $-$23 53 39.1 & 2.8451 $\pm$ 0.0010 & (6.2 $\pm$
2.6)$\times 10^{-18}$ & 115$_{-19}^{+46}$ & 610 $\pm$ 200 \\
0052.LAE29 & 00 54 36.55 & $-$23 53 40.2 & 2.8447 $\pm$ 0.0002 & (4.8 $\pm$
1.3)$\times 10^{-18}$ & 146$_{-23}^{+33}$ & 120 $\pm$ 40 \\
0052.LAE30 & 00 54 37.13 & $-$23 51 39.2 & 2.8647 $\pm$ 0.0003 & (1.5 $\pm$
0.6)$\times 10^{-17}$ & 102$_{-22}^{+31}$ & 145 $\pm$ 55 \\
0052.LAE31 & 00 54 38.54 & $-$23 48 39.8 & 2.8647 $\pm$ 0.0003 & (6.4 $\pm$
0.9)$\times 10^{-17}$ & 105$_{-10}^{+21}$ & 860 $\pm$ 50 \\
0052.LAE32 & 00 54 39.91 & $-$23 52 06.6 & 2.8311 $\pm$ 0.0003 & (1.3 $\pm$
0.3)$\times 10^{-17}$ & 410$_{-82}^{+130}$ & 495 $\pm$ 70 \\
0052.LAE33 & 00 54 40.36 & $-$23 49 13.1 & 2.8499 $\pm$ 0.0001 & (4.8 $\pm$
1.4)$\times 10^{-18}$ & 136$_{-21}^{+28}$ & 65 $\pm$ 40 \\
0052.LAE34 & 00 54 42.46 & $-$23 50 50.5 & 2.8663 $\pm$ 0.0004 & (5.4 $\pm$
1.3)$\times 10^{-18}$ & 101$_{-25}^{+1000}$ & 410 $\pm$ 75 \\
0052.LAE35 & 00 54 42.90 & $-$23 52 55.9 & 2.8756 $\pm$ 0.0003 & (4.2 $\pm$
0.5)$\times 10^{-17}$ & 165$_{-39}^{+1000}$ & 440 $\pm$ 35 \\
0052.LAE36 & 00 54 43.06 & $-$23 53 14.1 & 2.8764 $\pm$ 0.0004 & (1.7 $\pm$
0.3)$\times 10^{-16}$ & 28$_{-2}^{+3}$ & 590 $\pm$ 30 \\
0052.LAE37 & 00 54 44.36 & $-$23 51 52.4 & 2.8619 $\pm$ 0.0002 & (1.1 $\pm$
0.2)$\times 10^{-17}$ & 38$_{-8}^{+24}$ & 265 $\pm$ 45 \\
\hline
0052.RG & 00 54 29.83 & $-$23 51 31.1 & 2.8609 $\pm$ 0.0011 & (1.11
$\pm$ 0.09)$\times 10^{-15}$ & 133$_{-5}^{+6}$ & 2040 $\pm$ 85 \\
\hline
\end{tabular}
\end{table*}

\begin{table*}
\caption{\label{0943table} Position and properties of the \lya\ emission
  line of the confirmed \lya\ emitters and the radio galaxy in the
  0943 field.}
\begin{tabular}{ccccccc}
\hline
\hline
Object & \multicolumn{2}{c}{Position} & $z$ & Flux & EW$_0$ & FWHM \\ 
{} & $\alpha_{\mathrm{J}2000}$ & $\delta_{\mathrm{J}2000}$ & {} &
\ergscm\ & \AA & \kms\ \\
\hline
0943.LAE1 & 09 45 20.70 & $-$24 28 12.5 & 2.8970 $\pm$ 0.0002 & (3.1
$\pm$ 0.2)$\times 10^{-17}$ & $131_{-44}^{+1000}$ & 600 $\pm$ 35 \\
0943.LAE2 & 09 45 21.99 & $-$24 29 55.1 & 2.9202 $\pm$ 0.0001 & (2.2
$\pm$ 0.2)$\times 10^{-17}$ & $88_{-16}^{+231}$ & 290 $\pm$ 25 \\
0943.LAE3 & 09 45 22.17 & $-$24 28 56.2 & 2.9259 $\pm$ 0.0002 & (2.2
$\pm$ 0.2)$\times 10^{-17}$ & $144_{-10}^{+11}$ & 425 $\pm$ 45 \\
0943.LAE4 & 09 45 23.77 & $-$24 28 48.0 & 2.9263 $\pm$ 0.0007 & (6.7
$\pm$ 1.0)$\times 10^{-17}$ & $38_{-6}^{+26}$ & 555 $\pm$ 75 \\
0943.LAE5 & 09 45 27.14 & $-$24 27 52.6 & 2.9167 $\pm$ 0.0005 & (1.1
$\pm$ 0.3)$\times 10^{-17}$ & $61_{-8}^{+9}$ & 360 $\pm$ 115 \\
0943.LAE6 & 09 45 29.09 & $-$24 26 49.7 & 2.9242 $\pm$ 0.0003 & (1.3
$\pm$ 0.2)$\times 10^{-17}$ & $70_{-8}^{+9}$ & 225 $\pm$ 75 \\
0943.LAE7 & 09 45 30.60 & $-$24 25 31.4 & 2.9265 $\pm$ 0.0005 & (7.4
$\pm$ 1.8)$\times 10^{-18}$ & $9_{-2}^{+6}$ & 455 $\pm$ 100 \\
0943.LAE8 & 09 45 30.66 & $-$24 28 06.2 & 2.9113 $\pm$ 0.0002 & (4.7
$\pm$ 0.4)$\times 10^{-17}$ & $35_{-5}^{+14}$ & 455 $\pm$ 35 \\
0943.LAE9 & 09 45 30.83 & $-$24 28 01.4 & 2.9214 $\pm$ 0.0006 & (6.1
$\pm$ 2.1)$\times 10^{-18}$ & $8_{-3}^{+14}$ & 410 $\pm$ 140 \\
0943.LAE10 & 09 45 31.13 & $-$24 27 34.4 & 2.9122 $\pm$ 0.0006 & (1.0
$\pm$ 0.2)$\times 10^{-17}$ & $71_{-10}^{+11}$ & 440 $\pm$ 65 \\
0943.LAE11 & 09 45 32.26 & $-$24 31 21.6 & 2.9259 $\pm$ 0.0003 & (9.8
$\pm$ 1.4)$\times 10^{-18}$ & $49_{-11}^{+319}$ & 445 $\pm$ 60 \\
0943.LAE12 & 09 45 32.68 & $-$24 29 46.5 & 2.9004 $\pm$ 0.0003 & (2.1
$\pm$ 0.2)$\times 10^{-16}$ & $67_{-10}^{+18}$ & 625 $\pm$ 30 \\
0943.LAE13 & 09 45 32.77 & $-$24 29 05.4 & 2.9189 $\pm$ 0.0003 & (7.7
$\pm$ 0.8)$\times 10^{-17}$ & $103_{-20}^{+569}$ & 630 $\pm$ 35 \\
0943.LAE14 & 09 45 32.86 & $-$24 31 06.9 & 2.9073 $\pm$ 0.0000 & (4.1
$\pm$ 0.1)$\times 10^{-17}$ & $177_{-19}^{+23}$ & 180 $\pm$ 10 \\
0943.LAE15 & 09 45 34.34 & $-$24 29 23.7 & 2.9094 $\pm$ 0.0001 & (3.2
$\pm$ 0.2)$\times 10^{-17}$ & $111_{-26}^{+1000}$ & 330 $\pm$ 20 \\
0943.LAE16 & 09 45 34.96 & $-$24 30 46.9 & 2.9242 $\pm$ 0.0002 & (1.7
$\pm$ 0.2)$\times 10^{-17}$ & $28_{-4}^{+9}$ & 275 $\pm$ 50 \\
0943.LAE17 & 09 45 36.18 & $-$24 30 00.5 & 2.9174 $\pm$ 0.0005 & (1.7
$\pm$ 0.5)$\times 10^{-17}$ & $20_{-5}^{+36}$ & 305 $\pm$ 90 \\
0943.LAE18 & 09 45 36.36 & $-$24 29 56.4 & 2.9313 $\pm$ 0.0002 & (5.4
$\pm$ 0.3)$\times 10^{-17}$ & $133_{-24}^{+418}$ & 695 $\pm$ 35 \\
0943.LAE19 & 09 45 37.19 & $-$24 26 27.2 & 2.9081 $\pm$ 0.0003 & (2.8
$\pm$ 0.6)$\times 10^{-17}$ & $103_{-14}^{+17}$ & 430 $\pm$ 45 \\
0943.LAE20 & 09 45 37.52 & $-$24 25 23.7 & 2.9266 $\pm$ 0.0005 & (5.1
$\pm$ 0.6)$\times 10^{-17}$ & $165_{-11}^{+12}$ & 720 $\pm$ 60 \\
0943.LAE21 & 09 45 38.64 & $-$24 29 54.3 & 2.9159 $\pm$ 0.0006 & (2.2
$\pm$ 0.8)$\times 10^{-17}$ & $136_{-12}^{+14}$ & 595 $\pm$ 80 \\
0943.LAE22 & 09 45 38.71 & $-$24 29 57.5 & 2.9201 $\pm$ 0.0003 & (1.6
$\pm$ 0.2)$\times 10^{-17}$ & $20_{-3}^{+6}$ & 510 $\pm$ 55 \\
0943.LAE23 & 09 45 39.08 & $-$24 30 10.2 & 2.9189 $\pm$ 0.0004 & (5.6
$\pm$ 1.3)$\times 10^{-17}$ & $33_{-6}^{+20}$ & 680 $\pm$ 65 \\
0943.LAE24 & 09 45 39.17 & $-$24 29 20.1 & 2.9089 $\pm$ 0.0002 & (2.1
$\pm$ 0.2)$\times 10^{-17}$ & $26_{-6}^{+19}$ & 515 $\pm$ 40 \\
0943.LAE25 & 09 45 39.46 & $-$24 26 15.9 & 2.9037 $\pm$ 0.0010 & (2.0
$\pm$ 0.4)$\times 10^{-17}$ & $221_{-30}^{+42}$ & 1185 $\pm$ 195 \\
0943.LAE26 & 09 45 41.88 & $-$24 28 35.1 & 2.9309 $\pm$ 0.0008 & (4.3
$\pm$ 1.9)$\times 10^{-18}$ & $7_{-2}^{+5}$ & 360 $\pm$ 165 \\
0943.LAE27 & 09 45 42.61 & $-$24 28 57.9 & 2.9302 $\pm$ 0.0001 & (2.1
$\pm$ 0.2)$\times 10^{-17}$ & $150_{-32}^{+1000}$ & 285 $\pm$ 25 \\
0943.LAE28 & 09 45 45.72 & $-$24 26 49.1 & 2.9233 $\pm$ 0.0007 & (3.1
$\pm$ 0.7)$\times 10^{-17}$ & $86_{-17}^{+477}$ & 630 $\pm$ 60 \\
\hline
0943.RG & 09 45 32.74 & $-$24 28 49.7 & 2.9209 $\pm$ 0.0003 & (3.6
$\pm$ 0.1)$\times 10^{-15}$ & $173_{-3}^{+3}$ & 1755 $\pm$ 20 \\
\hline
\end{tabular}
\end{table*}

\begin{table*}
\caption{\label{2009table} Position and properties of the \lya\ emission
  line of the confirmed \lya\ emitters and the radio galaxy in the
  2009 field.}
\begin{tabular}{ccccccc}
\hline
\hline
Object & \multicolumn{2}{c}{Position} & $z$ & Flux & EW$_0$ & FWHM \\ 
{} & $\alpha_{\mathrm{J}2000}$ & $\delta_{\mathrm{J}2000}$ & {} &
\ergscm\ & \AA & \kms\ \\
\hline
2009.LAE1 & 20 09 38.32 & $-$30 41 31.6 & 3.1528 $\pm$ 0.0001 & (7.2 $\pm$
0.7)$\times 10^{-18}$ & $75_{-8}^{+10}$ & 175 $\pm$ 20 \\
2009.LAE2 & 20 09 39.89 & $-$30 40 50.7 & 3.1577 $\pm$ 0.0007 & (2.6 $\pm$
0.8)$\times 10^{-18}$ & $12_{-4}^{+20}$ & 530 $\pm$ 135 \\
2009.LAE3 & 20 09 40.92 & $-$30 41 44.9 & 3.1548 $\pm$ 0.0006 & (1.6 $\pm$
0.6)$\times 10^{-18}$ & $25_{-6}^{+206}$ & 385 $\pm$ 115 \\
2009.LAE4 & 20 09 41.06 & $-$30 42 46.4 & 3.1415 $\pm$ 0.0005 & (1.9 $\pm$
0.7)$\times 10^{-18}$ & $73_{-9}^{+11}$ & 320 $\pm$ 95 \\
2009.LAE5 & 20 09 42.19 & $-$30 38 34.3 & 3.1601 $\pm$ 0.0003 & (4.2 $\pm$
0.7)$\times 10^{-18}$ & $16_{-6}^{+52}$ & 370 $\pm$ 50 \\
2009.LAE6 & 20 09 47.83 & $-$30 42 39.6 & 3.1521 $\pm$ 0.0004 & (1.4 $\pm$
0.2)$\times 10^{-17}$ & $27_{-6}^{+50}$ & 720 $\pm$ 75 \\
2009.LAE7 & 20 09 49.55 & $-$30 40 47.4 & 3.1531 $\pm$ 0.0003 & (1.7 $\pm$
0.4)$\times 10^{-17}$ & $23_{-5}^{+37}$ & 485 $\pm$ 55 \\
2009.LAE8 & 20 09 54.43 & $-$30 41 19.7 & 3.1445 $\pm$ 0.0006 & (1.4 $\pm$
0.4)$\times 10^{-16}$ & $38_{-10}^{+699}$ & 505 $\pm$ 120 \\
2009.LAE9 & 20 09 56.94 & $-$30 39 38.6 & 3.1412 $\pm$ 0.0006 & (6.5 $\pm$
0.3)$\times 10^{-16}$ & $49_{-3}^{+4}$ & 3170 $\pm$ 50 \\
2009.LAE10 & 20 09 58.82 & $-$30 40 37.1 & 3.1454 $\pm$ 0.0004 & (7.3 $\pm$
0.3)$\times 10^{-16}$ & $84_{-8}^{+14}$ & 1565 $\pm$ 25 \\
2009.LAE11 & 20 10 01.81 & $-$30 41 27.7 & 3.1581 $\pm$ 0.0003 & (5.8 $\pm$
0.5)$\times 10^{-17}$ & $23_{-3}^{+6}$ & 750 $\pm$ 30 \\
\hline
2009.RG & 20 09 48.12 & $-$30 40 07.4 & 3.1497 $\pm$ 0.0002 & (2.41
$\pm$ 0.04)$\times 10^{-15}$ & $68_{-1}^{+1}$ & 2300 $\pm$ 25 \\
\hline
\end{tabular}
\end{table*}

\begin{table*}
\caption{\label{1338table} Position and properties of the \lya\ emission
  line of the confirmed \lya\ emitters and the radio galaxy in the two
  1338 fields.}
\begin{tabular}{ccccccc}
\hline
\hline
Object & \multicolumn{2}{c}{Position} & $z$ & Flux & EW$_0$ & FWHM \\ 
{} & $\alpha_{\mathrm{J}2000}$ & $\delta_{\mathrm{J}2000}$ & {} &
\ergscm\ & \AA & \kms\ \\
\hline
1338.LAE1 & 13 38 21.27 & $-$19 45 52.7 & 4.0977 $\pm$ 0.0010 & (1.5 $\pm$
0.4)$\times 10^{-17}$ & $55_{-7}^{+38}$ & 495 $\pm$ 105 \\
1338.LAE2 & 13 38 21.32 & $-$19 44 42.5 & 4.1092 $\pm$ 0.0003 & (8.1 $\pm$
2.2)$\times 10^{-18}$ & $33_{-6}^{+17}$ & $<$255 \\
1338.LAE3 & 13 38 21.68 & $-$19 46 27.9 & 4.0978 $\pm$ 0.0003 & (1.0 $\pm$
0.4)$\times 10^{-18}$ & $14_{-3}^{+7}$ & 75 $\pm$ 75 \\
1338.LAE4 & 13 38 22.47 & $-$19 44 33.8 & 4.0950 $\pm$ 0.0002 & (2.7 $\pm$
0.3)$\times 10^{-17}$ & $51_{-4}^{+7}$ & 230 $\pm$ 40 \\
1338.LAE5 & 13 38 22.78 & $-$19 46 04.8 & 4.0976 $\pm$ 0.0003 & (3.0 $\pm$
1.1)$\times 10^{-18}$ & $49_{-13}^{+16}$ & 120 $\pm$ 65 \\
1338.LAE6 & 13 38 23.65 & $-$19 45 51.7 & 4.0969 $\pm$ 0.0003 & (2.2 $\pm$
0.7)$\times 10^{-18}$ & $24_{-3}^{+5}$ & 95 $\pm$ 55 \\
1338.LAE7 & 13 38 24.79 & $-$19 41 34.2 & 4.1055 $\pm$ 0.0004 & (5.9 $\pm$
1.4)$\times 10^{-18}$ & $29_{-7}^{+51}$ & $<$260 \\
1338.LAE8 & 13 38 24.87 & $-$19 41 46.0 & 4.1017 $\pm$ 0.0007 & (9.2 $\pm$
2.9)$\times 10^{-18}$ & $31_{-6}^{+23}$ & 270 $\pm$ 140 \\
1338.LAE9 & 13 38 25.11 & $-$19 43 11.2 & 4.1004 $\pm$ 0.0007 & (7.0 $\pm$
1.3)$\times 10^{-17}$ & $63_{-7}^{+15}$ & 720 $\pm$ 55 \\
1338.LAE10 & 13 38 25.32 & $-$19 45 55.9 & 4.0970 $\pm$ 0.0002 & (1.6 $\pm$
0.3)$\times 10^{-17}$ & $167_{-17}^{+25}$ & 420 $\pm$ 90 \\
1338.LAE11 & 13 38 26.18 & $-$19 43 34.7 & 4.1010 $\pm$ 0.0006 & (6.0 $\pm$
1.5)$\times 10^{-17}$ & $70_{-9}^{+23}$ & 445 $\pm$ 45 \\
1338.LAE12 & 13 38 26.20 & $-$19 46 28.5 & 4.0925 $\pm$ 0.0005 & (9.0 $\pm$
1.7)$\times 10^{-18}$ & $152_{-22}^{+32}$ & 460 $\pm$ 55 \\
1338.LAE13 & 13 38 28.08 & $-$19 39 50.5 & 4.0877 $\pm$ 0.0004 & (1.4 $\pm$
0.4)$\times 10^{-17}$ & $278_{-62}^{+1000}$ & $<$170 \\
1338.LAE14 & 13 38 28.73 & $-$19 44 37.2 & 4.1020 $\pm$ 0.0002 & (4.5 $\pm$
0.6)$\times 10^{-17}$ & $123_{-18}^{+1000}$ & $<$200 \\
1338.LAE15 & 13 38 29.41 & $-$19 49 01.7 & 4.1010 $\pm$ 0.0005 & (2.5 $\pm$
0.6)$\times 10^{-18}$ & $80_{-15}^{+22}$ & 405 $\pm$ 75 \\
1338.LAE16 & 13 38 29.68 & $-$19 44 00.0 & 4.1021 $\pm$ 0.0003 & (4.6 $\pm$
1.2)$\times 10^{-18}$ & $31_{-4}^{+9}$ & $<$165 \\
1338.LAE17 & 13 38 29.88 & $-$19 43 26.1 & 4.0927 $\pm$ 0.0004 & (7.0 $\pm$
1.3)$\times 10^{-18}$ & $267_{-71}^{+1000}$ & 350 $\pm$ 80 \\
1338.LAE18 & 13 38 30.16 & $-$19 40 38.3 & 4.1132 $\pm$ 0.0005 & (1.6 $\pm$
0.4)$\times 10^{-17}$ & $98_{-19}^{+391}$ & 275 $\pm$ 105 \\
1338.LAE19 & 13 38 30.17 & $-$19 48 44.9 & 4.0872 $\pm$ 0.0003 & (7.1 $\pm$
1.2)$\times 10^{-18}$ & $144_{-37}^{+1000}$ & 240 $\pm$ 35 \\
1338.LAE20 & 13 38 32.85 & $-$19 44 07.2 & 4.0969 $\pm$ 0.0003 & (7.6 $\pm$
1.8)$\times 10^{-18}$ & $45_{-6}^{+30}$ & $<$170 \\
1338.LAE21 & 13 38 33.58 & $-$19 43 36.3 & 4.0965 $\pm$ 0.0009 & (4.6 $\pm$
0.9)$\times 10^{-17}$ & $31_{-2}^{+3}$ & 380 $\pm$ 80 \\
1338.LAE22 & 13 38 34.15 & $-$19 42 53.4 & 4.0955 $\pm$ 0.0009 & (2.6 $\pm$
0.8)$\times 10^{-17}$ & $42_{-9}^{+105}$ & 505 $\pm$ 150 \\
1338.LAE23 & 13 38 34.44 & $-$19 47 03.7 & 4.0983 $\pm$ 0.0003 & (3.7 $\pm$
0.8)$\times 10^{-18}$ & $66_{-13}^{+348}$ & 215 $\pm$ 45 \\
1338.LAE24 & 13 38 34.69 & $-$19 43 43.2 & 4.0979 $\pm$ 0.0002 & (3.3 $\pm$
0.6)$\times 10^{-18}$ & $35_{-6}^{+301}$ & 135 $\pm$ 30 \\
1338.LAE25 & 13 38 34.98 & $-$19 42 25.4 & 4.0925 $\pm$ 0.0008 & (7.4 $\pm$
2.6)$\times 10^{-18}$ & $19_{-4}^{+7}$ & 275 $\pm$ 160 \\
1338.LAE26 & 13 38 35.10 & $-$19 45 07.8 & 4.0967 $\pm$ 0.0005 & (5.4 $\pm$
2.3)$\times 10^{-18}$ & $53_{-13}^{+18}$ & $<$150 \\
1338.LAE27 & 13 38 35.52 & $-$19 45 23.7 & 4.0999 $\pm$ 0.0005 & (9.4 $\pm$
2.3)$\times 10^{-18}$ & $90_{-15}^{+420}$ & 230 $\pm$ 110 \\
1338.LAE28 & 13 38 35.67 & $-$19 45 50.1 & 4.0987 $\pm$ 0.0003 & (3.1 $\pm$
0.7)$\times 10^{-18}$ & $54_{-9}^{+14}$ & 395 $\pm$ 70 \\
1338.LAE29 & 13 38 35.82 & $-$19 49 36.5 & 4.0975 $\pm$ 0.0004 & (3.5 $\pm$
0.9)$\times 10^{-18}$ & $27_{-3}^{+5}$ & 335 $\pm$ 70 \\
1338.LAE30 & 13 38 37.15 & $-$19 45 02.0 & 4.0943 $\pm$ 0.0003 & (6.0 $\pm$
1.0)$\times 10^{-17}$ & $103_{-12}^{+141}$ & 440 $\pm$ 35 \\
1338.LAE31 & 13 38 39.69 & $-$19 47 50.1 & 4.0948 $\pm$ 0.0007 & (3.6 $\pm$
1.0)$\times 10^{-17}$ & $72_{-10}^{+32}$ & 575 $\pm$ 45 \\
1338.LAE32 & 13 38 41.10 & $-$19 43 01.7 & 4.0941 $\pm$ 0.0007 & (2.3 $\pm$
0.2)$\times 10^{-16}$ & $16_{-1}^{+1}$ & 2110 $\pm$ 85 \\
1338.LAE33 & 13 38 41.11 & $-$19 44 22.4 & 4.1015 $\pm$ 0.0005 & (1.9 $\pm$
0.6)$\times 10^{-17}$ & $49_{-13}^{+17}$ & 235 $\pm$ 75 \\
1338.LAE34 & 13 38 43.85 & $-$19 44 41.0 & 4.0978 $\pm$ 0.0003 & (2.4 $\pm$
0.6)$\times 10^{-17}$ & $101_{-26}^{+1000}$ & 220 $\pm$ 30 \\
1338.LAE35 & 13 38 44.65 & $-$19 47 09.4 & 4.1185 $\pm$ 0.0002 & (3.2 $\pm$
0.3)$\times 10^{-17}$ & $313_{-53}^{+1000}$ & 300 $\pm$ 40 \\
1338.LAE36 & 13 38 47.19 & $-$19 48 16.7 & 4.0979 $\pm$ 0.0008 & (1.8 $\pm$
0.4)$\times 10^{-17}$ & $119_{-18}^{+25}$ & 655 $\pm$ 125 \\
1338.LAE37 & 13 38 50.12 & $-$19 46 12.2 & 4.0958 $\pm$ 0.0004 & (1.2 $\pm$
0.4)$\times 10^{-18}$ & $37_{-8}^{+52}$ & 165 $\pm$ 55 \\
\hline
1338.RG & 13 38 26.06 & $-$19 42 30.8 & 4.1052 $\pm$ 0.0006 & (4.4 $\pm$
0.2)$\times 10^{-15}$ & $578_{-14}^{+16}$ & 1810 $\pm$ 40 \\
\hline
\end{tabular}
\end{table*}


\begin{thebibliography}{127}
\expandafter\ifx\csname natexlab\endcsname\relax\def\natexlab#1{#1}\fi

\bibitem[{{Appenzeller} \& {Rupprecht}(1992)}]{app92}
{Appenzeller}, I. \& {Rupprecht}, G. 1992, The Messenger, 67, 18

\bibitem[{{Archibald} {et~al.}(2001){Archibald}, {Dunlop}, {Hughes},
  {Rawlings}, {Eales}, \& {Ivison}}]{arc01}
{Archibald}, E.~N., {Dunlop}, J.~S., {Hughes}, D.~H., {et~al.} 2001, \mnras,
  323, 417

\bibitem[{{Athreya} {et~al.}(1998){Athreya}, {Kapahi}, {McCarthy}, \& {van
  Breugel}}]{ath98}
{Athreya}, R.~M., {Kapahi}, V.~K., {McCarthy}, P.~J., \& {van Breugel},
  W.~J.~M. 1998, \aap, 329, 809

\bibitem[{{Bahcall} \& {Cen}(1993)}]{bah93}
{Bahcall}, N.~A. \& {Cen}, R. 1993, \apjl, 407, L49

\bibitem[{{Bahcall} \& {Fan}(1998)}]{bah98}
{Bahcall}, N.~A. \& {Fan}, X. 1998, \apj, 504, 1

\bibitem[{{Bahcall} {et~al.}(1997){Bahcall}, {Fan}, \& {Cen}}]{bah97}
{Bahcall}, N.~A., {Fan}, X., \& {Cen}, R. 1997, \apjl, 485, L53

\bibitem[{{Baldwin} \& {Stone}(1984)}]{bal84}
{Baldwin}, J.~A. \& {Stone}, R.~P.~S. 1984, \mnras, 206, 241

\bibitem[{{Barr} {et~al.}(2004){Barr}, {Baker}, {Bremer}, {Hunstead}, \&
  {Bland-Hawthorn}}]{bar04}
{Barr}, J.~M., {Baker}, J.~C., {Bremer}, M.~N., {Hunstead}, R.~W., \&
  {Bland-Hawthorn}, J. 2004, \aj, 128, 2660

\bibitem[{{Beers} {et~al.}(1990){Beers}, {Flynn}, \& {Gebhardt}}]{bee90}
{Beers}, T.~C., {Flynn}, K., \& {Gebhardt}, K. 1990, \aj, 100, 32

\bibitem[{{Bertin} \& {Arnouts}(1996)}]{ber96}
{Bertin}, E. \& {Arnouts}, S. 1996, \aaps, 117, 393

\bibitem[{{Bessell}(1979)}]{bes79}
{Bessell}, M.~S. 1979, \pasp, 91, 589

\bibitem[{{Best}(2000)}]{bes00}
{Best}, P.~N. 2000, \mnras, 317, 720

\bibitem[{{Best} {et~al.}(1999){Best}, {R{\" o}ttgering}, \& {Lehnert}}]{bes99}
{Best}, P.~N., {R{\" o}ttgering}, H.~J.~A., \& {Lehnert}, M.~D. 1999, \mnras,
  310, 223

\bibitem[{{Best} {et~al.}(2003){Best}, {Lehnert}, {Miley}, \& {R{\"
  o}ttgering}}]{bes03}
{Best}, P.~N., {Lehnert}, M.~D., {Miley}, G.~K., \& {R{\" o}ttgering}, H.~J.~A.
  2003, \mnras, 343, 1

\bibitem[{{Blakeslee} {et~al.}(2003){Blakeslee}, {Franx}, {Postman}, {Rosati},
  {Holden}, {Illingworth}, {Ford}, {Cross}, {Gronwall}, {Ben{\'{\i}}tez},
  {Bouwens}, {Broadhurst}, {Clampin}, {Demarco}, {Golimowski}, {Hartig},
  {Infante}, {Martel}, {Miley}, {Menanteau}, {Meurer}, {Sirianni}, \&
  {White}}]{bla03a}
{Blakeslee}, J.~P., {Franx}, M., {Postman}, M., {et~al.} 2003, \apjl, 596, L143

\bibitem[{{Blundell} \& {Rawlings}(1999)}]{blu99}
{Blundell}, K.~M. \& {Rawlings}, S. 1999, \nat, 399, 330

\bibitem[{{Bornancini} {et~al.}(2006){Bornancini}, {Lambas}, \& {De
  Breuck}}]{bor06}
{Bornancini}, C.~G., {Lambas}, D.~G., \& {De Breuck}, C. 2006, \mnras, 366,
  1067

\bibitem[{{Carilli} {et~al.}(1997){Carilli}, {R\"{o}ttgering}, {van Ojik},
  {Miley}, \& {van Breugel}}]{car97}
{Carilli}, C.~L., {R\"{o}ttgering}, H.~J.~A., {van Ojik}, R., {Miley}, G.~K.,
  \& {van Breugel}, W.~J.~M. 1997, \apjs, 109, 1

\bibitem[{{Carilli} {et~al.}(2001){Carilli}, {Miley}, {R{\" o}ttgering},
  {Kurk}, {Pentericci}, {Harris}, {Bertoldi}, {Menten}, \& {van
  Breugel}}]{car01}
{Carilli}, C.~L., {Miley}, G., {R{\" o}ttgering}, H.~J.~A., {et~al.} 2001, in
  Astronomical Society of the Pacific Conference Series, 101

\bibitem[{{Carlstrom} {et~al.}(2002){Carlstrom}, {Holder}, \& {Reese}}]{carl02}
{Carlstrom}, J.~E., {Holder}, G.~P., \& {Reese}, E.~D. 2002, \araa, 40, 643

\bibitem[{{Ciardullo} {et~al.}(2002){Ciardullo}, {Feldmeier}, {Krelove},
  {Jacoby}, \& {Gronwall}}]{cia02}
{Ciardullo}, R., {Feldmeier}, J.~J., {Krelove}, K., {Jacoby}, G.~H., \&
  {Gronwall}, C. 2002, \apj, 566, 784

\bibitem[{{Cooke} {et~al.}(2005){Cooke}, {Wolfe}, {Prochaska}, \&
  {Gawiser}}]{coo05}
{Cooke}, J., {Wolfe}, A.~M., {Prochaska}, J.~X., \& {Gawiser}, E. 2005, \apj,
  621, 596

\bibitem[{{Croft} {et~al.}(2005){Croft}, {Kurk}, {van Breugel}, {Stanford}, {de
  Vries}, {Pentericci}, \& {Rottgering}}]{cro05}{
Croft}, S., {Kurk}, J., {van Breugel}, W., {et~al.} 2005, \aj, 130, 867

\bibitem[{{Dawson} {et~al.}(2004){Dawson}, {Rhoads}, {Malhotra}, {Stern},
  {Dey}, {Spinrad}, {Jannuzi}, {Wang}, \& {Landes}}]{daw04}
{Dawson}, S., {Rhoads}, J.~E., {Malhotra}, S., {et~al.} 2004, \apj, 617, 707

\bibitem[{{De Breuck} {et~al.}(1999){De Breuck}, {van Breugel}, {Minniti},
  {Miley}, {R{\" o}ttgering}, {Stanford}, \& {Carilli}}]{deb99}
{De Breuck}, C., {van Breugel}, W.~J.~M., {Minniti}, D., {et~al.} 1999, \aap,
  352, L51

\bibitem[{{De Breuck} {et~al.}(2000){De Breuck}, {van Breugel}, {R{\"
  o}ttgering}, \& {Miley}}]{deb00}
{De Breuck}, C., {van Breugel}, W., {R{\" o}ttgering}, H.~J.~A., \& {Miley}, G.
  2000, \aaps, 143, 303

\bibitem[{{De Breuck} {et~al.}(2001){De Breuck}, {van Breugel}, {R{\"
  o}ttgering}, {Stern}, {Miley}, {de Vries}, {Stanford}, {Kurk}, \&
  {Overzier}}]{deb01}
{De Breuck}, C., {van Breugel}, W.~J.~M., {R{\" o}ttgering}, H., {et~al.} 2001,
  \aj, 121, 1241

\bibitem[{{De Breuck} {et~al.}(2002){De Breuck}, {van Breugel}, {Stanford},
  {R{\" o}ttgering}, {Miley}, \& {Stern}}]{deb02}
{De Breuck}, C., {van Breugel}, W.~J.~M., {Stanford}, S.~A., {et~al.} 2002,
  \aj, 123, 637

\bibitem[{{De Breuck} {et~al.}(2003{\natexlab{a}}){De Breuck}, {Neri},
  {Morganti}, {Omont}, {Rocca-Volmerange}, {Stern}, {Reuland}, {van Breugel},
  {R{\" o}ttgering}, {Stanford}, {Spinrad}, {Vigotti}, \& {Wright}}]{deb03a}
{De Breuck}, C., {Neri}, R., {Morganti}, R., {et~al.} 2003{\natexlab{a}}, \aap,
  401, 911

\bibitem[{{De Breuck} {et~al.}(2003{\natexlab{b}}){De Breuck}, {Neri}, \&
  {Omont}}]{deb03b}
{De Breuck}, C., {Neri}, R., \& {Omont}, A. 2003{\natexlab{b}}, New Astronomy
  Review, 47, 285

\bibitem[{{De Breuck} {et~al.}(2004){De Breuck}, {Bertoldi}, {Carilli},
  {Omont}, {Venemans}, {R{\" o}ttgering}, {Overzier}, {Reuland}, {Miley},
  {Ivison}, \& {van Breugel}}]{deb04}
{De Breuck}, C., {Bertoldi}, F., {Carilli}, C., {et~al.} 2004, \aap, 424, 1

\bibitem[{{Deutsch}(1999)}]{deu99}
{Deutsch}, E.~W. 1999, \aj, 118, 1882

\bibitem[{{Dey} {et~al.}(1997){Dey}, {van Breugel}, {Vacca}, \&
  {Antonucci}}]{dey97}
{Dey}, A., {van Breugel}, W.~J.~M., {Vacca}, W.~D., \& {Antonucci}, R. 1997,
  \apj, 490, 698

\bibitem[{{Dunlop} \& {Peacock}(1990)}]{dun90}
{Dunlop}, J.~S. \& {Peacock}, J.~A. 1990, \mnras, 247, 19

\bibitem[{{Eggen} {et~al.}(1962){Eggen}, {Lynden-Bell}, \& {Sandage}}]{egg62}
{Eggen}, O.~J., {Lynden-Bell}, D., \& {Sandage}, A.~R. 1962, \apj, 136, 748

\bibitem[{{Eke} {et~al.}(1996){Eke}, {Cole}, \& {Frenk}}]{eke96}
{Eke}, V.~R., {Cole}, S., \& {Frenk}, C.~S. 1996, \mnras, 282, 263

\bibitem[{{Eke} {et~al.}(1998){Eke}, {Navarro}, \& {Frenk}}]{eke98}
{Eke}, V.~R., {Navarro}, J.~F., \& {Frenk}, C.~S. 1998, \apj, 503, 569

\bibitem[{{Ellis} {et~al.}(1997){Ellis}, {Smail}, {Dressler}, {Couch},
  {Oemler}, {Butcher}, \& {Sharples}}]{ell97}
{Ellis}, R.~S., {Smail}, I., {Dressler}, A., {et~al.} 1997, \apj, 483, 582

\bibitem[{{Ettori} {et~al.}(2003){Ettori}, {Tozzi}, \& {Rosati}}]{ett03}
{Ettori}, S., {Tozzi}, P., \& {Rosati}, P. 2003, \aap, 398, 879

\bibitem[{{Fynbo} {et~al.}(2001){Fynbo}, {M{\o}ller}, \& {Thomsen}}]{fyn01}
{Fynbo}, J.~P.~U., {M{\o}ller}, P., \& {Thomsen}, B. 2001, \aap, 374, 443

\bibitem[{{Fynbo} {et~al.}(2003){Fynbo}, {Ledoux}, {M{\o}ller}, {Thomsen}, \&
  {Burud}}]{fyn03}
{Fynbo}, J.~P.~U., {Ledoux}, C., {M{\o}ller}, P., {Thomsen}, B., \& {Burud}, I.
  2003, \aap, 407, 147

\bibitem[{{Gehrels}(1986)}]{geh86}
{Gehrels}, N. 1986, \apj, 303, 336

\bibitem[{{Girardi} {et~al.}(1998){Girardi}, {Borgani}, {Giuricin},
  {Mardirossian}, \& {Mezzetti}}]{gir98}
{Girardi}, M., {Borgani}, S., {Giuricin}, G., {Mardirossian}, F., \&
  {Mezzetti}, M. 1998, \apj, 506, 45

\bibitem[{{Gladders}(2002)}]{gla02}
{Gladders}, M.~D. 2002, Ph.D.~Thesis

\bibitem[{{Goto}(2005)}]{got05}
{Goto}, T. 2005, \mnras, 356, L6

\bibitem[{{Hall} {et~al.}(2001){Hall}, {Sawicki}, {Martini}, {Finn},
  {Pritchet}, {Osmer}, {McCarthy}, {Evans}, {Lin}, \& {Hartwick}}]{hal01}
{Hall}, P.~B., {Sawicki}, M., {Martini}, P., {et~al.} 2001, \aj, 121, 1840

\bibitem[{{Hammer} {et~al.}(1997){Hammer}, {Flores}, {Lilly}, {Crampton}, {Le
  Fevre}, {Rola}, {Mallen-Ornelas}, {Schade}, \& {Tresse}}]{ham97}
{Hammer}, F., {Flores}, H., {Lilly}, S.~J., {et~al.} 1997, \apj, 481, 49

\bibitem[{{Hashimoto} {et~al.}(2004){Hashimoto}, {Barcons}, {B{\" o}hringer},
  {Fabian}, {Hasinger}, {Mainieri}, \& {Brunner}}]{has04}
{Hashimoto}, Y., {Barcons}, X., {B{\" o}hringer}, H., {et~al.} 2004, \aap, 417,
  819

\bibitem[{{Hayashino} {et~al.}(2004){Hayashino}, {Matsuda}, {Tamura},
  {Yamauchi}, {Yamada}, {Ajiki}, {Fujita}, {Murayama}, {Nagao}, {Ohta},
  {Okamura}, {Ouchi}, {Shimasaku}, {Shioya}, \& {Taniguchi}}]{hay04}
{Hayashino}, T., {Matsuda}, Y., {Tamura}, H., {et~al.} 2004, \aj, 128, 2073

\bibitem[{{Holden} {et~al.}(2005){Holden}, {van der Wel}, {Franx},
  {Illingworth}, {Blakeslee}, {van Dokkum}, {Ford}, {Magee}, {Postman}, {Rix},
  \& {Rosati}}]{hol05}
{Holden}, B.~P., {van der Wel}, A., {Franx}, M., {et~al.} 2005, \apjl, 620, L83

\bibitem[{{Hu} {et~al.}(2004){Hu}, {Cowie}, {Capak}, {McMahon}, {Hayashino}, \&
  {Komiyama}}]{hu04}
{Hu}, E.~M., {Cowie}, L.~L., {Capak}, P., {et~al.} 2004, \aj, 127, 563

\bibitem[{{Jarvis} {et~al.}(2001){Jarvis}, {Rawlings}, {Eales}, {Blundell},
  {Bunker}, {Croft}, {McLure}, \& {Willott}}]{jar01}
{Jarvis}, M.~J., {Rawlings}, S., {Eales}, S., {et~al.} 2001, \mnras, 326, 1585

\bibitem[{{Jarvis} {et~al.}(2003){Jarvis}, {Wilman}, {R{\" o}ttgering}, \&
  {Binette}}]{jar03}
{Jarvis}, M.~J., {Wilman}, R.~J., {R{\" o}ttgering}, H.~J.~A., \& {Binette}, L.
  2003, \mnras, 338, 263

\bibitem[{{Kapahi} {et~al.}(1998){Kapahi}, {Athreya}, {van Breugel},
  {McCarthy}, \& {Subrahmanya}}]{kap98}
{Kapahi}, V.~K., {Athreya}, R.~M., {van Breugel}, W., {McCarthy}, P.~J., \&
  {Subrahmanya}, C.~R. 1998, \apjs, 118, 275

\bibitem[{{Keel} {et~al.}(1999){Keel}, {Cohen}, {Windhorst}, \&
  {Waddington}}]{kee99}
{Keel}, W.~C., {Cohen}, S.~H., {Windhorst}, R.~A., \& {Waddington}, I. 1999,
  \aj, 118, 2547

\bibitem[{{Kodaira} {et~al.}(2003){Kodaira}, {Taniguchi}, {Kashikawa}, {Kaifu},
  {Ando}, {Karoji}, {Ajiki}, {Akiyama}, {Aoki}, {Doi}, {Fujita}, {Furusawa},
  {Hayashino}, {Imanishi}, {Iwamuro}, {Iye}, {Kawabata}, {Kobayashi}, {Kodama},
  {Komiyama}, {Kosugi}, {Matsuda}, {Miyazaki}, {Mizumoto}, {Motohara},
  {Murayama}, {Nagao}, {Nariai}, {Ohta}, {Ohyama}, {Okamura}, {Ouchi},
  {Sasaki}, {Sekiguchi}, {Shimasaku}, {Shioya}, {Takata}, {Tamura}, {Terada},
  {Umemura}, {Usuda}, {Yagi}, {Yamada}, {Yasuda}, \& {Yoshida}}]{kod03}
{Kodaira}, K., {Taniguchi}, Y., {Kashikawa}, N., {et~al.} 2003, \pasj, 55, L17

\bibitem[{{Kurk} {et~al.}(2000){Kurk}, {R{\" o}ttgering}, {Pentericci},
  {Miley}, {van Breugel}, {Carilli}, {Ford}, {Heckman}, {McCarthy}, \&
  {Moorwood}}]{kur00}
{Kurk}, J.~D., {R{\" o}ttgering}, H.~J.~A., {Pentericci}, L., {et~al.} 2000,
  \aap, 358, L1

\bibitem[{{Kurk} {et~al.}(2004{\natexlab{a}}){Kurk}, {Pentericci}, {Overzier},
  {R{\" o}ttgering}, \& {Miley}}]{kur04a}
{Kurk}, J.~D., {Pentericci}, L., {Overzier}, R.~A., {R{\" o}ttgering},
  H.~J.~A., \& {Miley}, G.~K. 2004{\natexlab{a}}, \aap, 428, 817

\bibitem[{{Kurk} {et~al.}(2004{\natexlab{b}}){Kurk}, {Pentericci}, {R{\"
  o}ttgering}, \& {Miley}}]{kur04b}
{Kurk}, J.~D., {Pentericci}, L., {R{\" o}ttgering}, H.~J.~A., \& {Miley}, G.~K.
  2004{\natexlab{b}}, \aap, 428, 793

\bibitem[{{Landolt}(1992)}]{lan92}
{Landolt}, A.~U. 1992, \aj, 104, 340

\bibitem[{{Larson}(1974)}]{lar74}
{Larson}, R.~B. 1974, \mnras, 166, 585

\bibitem[{{Large} {et~al.}(1981){Large}, {Mills}, {Little}, {Crawford}, \&
  {Sutton}}]{lar81}
{Large}, M.~I., {Mills}, B.~Y., {Little}, A.~G., {Crawford}, D.~F., \&
  {Sutton}, J.~M. 1981, \mnras, 194, 693

\bibitem[{{Le F\`evre} {et~al.}(1996){Le F\`evre}, {Deltorn}, {Crampton}, \&
  {Dickinson}}]{lef96}
{Le F\`evre}, O., {Deltorn}, J.~M., {Crampton}, D., \& {Dickinson}, M. 1996,
  \apjl, 471, L11

\bibitem[{{Ma} \& {Feissel}(1998)}]{ma98}
{Ma}, C. \& {Feissel}, M. 1998, VizieR Online Data Catalog, 1251, 0

\bibitem[{{Madau}(1995)}]{mad95}
{Madau}, P. 1995, \apj, 441, 18

\bibitem[{{Maughan} {et~al.}(2004){Maughan}, {Jones}, {Ebeling}, \&
  {Scharf}}]{mau04}
{Maughan}, B.~J., {Jones}, L.~R., {Ebeling}, H., \& {Scharf}, C. 2004, \mnras,
  351, 1193

\bibitem[{{McCarthy}(1993)}]{mcc93}
{McCarthy}, P.~J. 1993, \araa, 31, 639

\bibitem[{{McCarthy} {et~al.}(1990){McCarthy}, {Kapahi}, {van Breugel}, \&
  {Subrahmanya}}]{mcc90}
{McCarthy}, P.~J., {Kapahi}, V.~K., {van Breugel}, W.~J.~M., \& {Subrahmanya},
  C.~R. 1990, \aj, 100, 1014

\bibitem[{{McCarthy} {et~al.}(1996){McCarthy}, {Kapahi}, {van Breugel},
  {Persson}, {Athreya}, \& {Subrahmanya}}]{mcc96}
{McCarthy}, P.~J., {Kapahi}, V.~K., {van Breugel}, W., {et~al.} 1996, \apjs,
  107, 19

\bibitem[{{McCarthy} {et~al.}(1998){McCarthy}, {Cohen}, {Butcher}, {Cromer},
  {Croner}, {Douglas}, {Goeden}, {Grewal}, {Lu}, {Petrie}, {Weng}, {Weber},
  {Koch}, \& {Rodgers}}]{mcc98b}
{McCarthy}, J.~K., {Cohen}, J.~G., {Butcher}, B., {et~al.} 1998, in Proc. SPIE
  Vol. 3355, p. 81-92, Optical Astronomical Instrumentation, Sandro D'Odorico;
  Ed., 81

\bibitem[{{Miles} {et~al.}(2004){Miles}, {Raychaudhury}, {Forbes},
  {Goudfrooij}, {Ponman}, \& {Kozhurina-Platais}}]{miles04}
{Miles}, T.~A., {Raychaudhury}, S., {Forbes}, D.~A., {et~al.} 2004, \mnras,
  355, 785

\bibitem[{{Miley} {et~al.}(2004){Miley}, {Overzier}, {Tsvetanov}, {Bouwens},
  {Ben{\'{\i}}tez}, {Blakeslee}, {Ford}, {Illingworth}, {Postman}, {Rosati},
  {Clampin}, {Hartig}, {Zirm}, {R{\" o}ttgering}, {Venemans}, {Ardila},
  {Bartko}, {Broadhurst}, {Brown}, {Burrows}, {Cheng}, {Cross}, {De Breuck},
  {Feldman}, {Franx}, {Golimowski}, {Gronwall}, {Infante}, {Martel},
  {Menanteau}, {Meurer}, {Sirianni}, {Kimble}, {Krist}, {Sparks}, {Tran},
  {White}, \& {Zheng}}]{mil04}
{Miley}, G.~K., {Overzier}, R.~A., {Tsvetanov}, Z.~I., {et~al.} 2004, \nat,
  427, 47

\bibitem[{{Monaco} {et~al.}(2005){Monaco}, {Moller}, {Fynbo}, {Weidinger},
  {Ledoux}, \& {Theuns}}]{mon05}
{Monaco}, P., {Moller}, P., {Fynbo}, J., {et~al.} 2005, \aap, 440, 799

\bibitem[{{Monet}(1998)}]{mon98b}
{Monet}, D.~G. 1998, Bulletin of the American Astronomical Society, 30, 1427

\bibitem[{{Monet} {et~al.}(1998){Monet}, {Canzian}, {Dahn}, {Guetter},
  {Harris}, {Henden}, {Levine}, {Luginbuhl}, {Monet}, {Rhodes}, {Riepe},
  {Sell}, {Stone}, {Vrba}, \& {Walker}}]{mon98a}
{Monet}, D.~B.~A., {Canzian}, B., {Dahn}, C., {et~al.} 1998, VizieR Online Data
  Catalog, 1252, 0

\bibitem[{{Mullis} {et~al.}(2005){Mullis}, {Rosati}, {Lamer}, {B{\" o}hringer},
  {Schwope}, {Schuecker}, \& {Fassbender}}]{mul05}
{Mullis}, C.~R., {Rosati}, P., {Lamer}, G., {et~al.} 2005, \apjl, 623, L85

\bibitem[{{Nakata} {et~al.}(2001){Nakata}, {Kajisawa}, {Yamada}, {Kodama},
  {Shimasaku}, {Tanaka}, {Doi}, {Furusawa}, {Hamabe}, {Iye}, {Kimura},
  {Komiyama}, {Miyazaki}, {Okamura}, {Ouchi}, {Sasaki}, {Sekiguchi}, {Yagi}, \&
  {Yasuda}}]{nak01}
{Nakata}, F., {Kajisawa}, M., {Yamada}, T., {et~al.} 2001, \pasj, 53, 1139

\bibitem[{{Nakata} {et~al.}(2005){Nakata}, {Bower}, {Balogh}, \&
  {Wilman}}]{nak05}
{Nakata}, F., {Bower}, R.~G., {Balogh}, M.~L., \& {Wilman}, D.~J. 2005, \mnras,
  357, 679

\bibitem[{{Norman} {et~al.}(2002){Norman}, {Hasinger}, {Giacconi}, {Gilli},
  {Kewley}, {Nonino}, {Rosati}, {Szokoly}, {Tozzi}, {Wang}, {Zheng}, {Zirm},
  {Bergeron}, {Gilmozzi}, {Grogin}, {Koekemoer}, \& {Schreier}}]{nor02}
{Norman}, C., {Hasinger}, G., {Giacconi}, R., {et~al.} 2002, \apj, 571, 218

\bibitem[{{Oke}(1974)}]{oke74}
{Oke}, J.~B. 1974, \apjs, 27, 21

\bibitem[{{Oke}(1990)}]{oke90}
---. 1990, \aj, 99, 1621

\bibitem[{{Oke} {et~al.}(1995){Oke}, {Cohen}, {Carr}, {Cromer}, {Dingizian},
  {Harris}, {Labrecque}, {Lucinio}, {Schaal}, {Epps}, \& {Miller}}]{oke95}
{Oke}, J.~B., {Cohen}, J.~G., {Carr}, M., {et~al.} 1995, \pasp, 107, 375

\bibitem[{{Ouchi} {et~al.}(2003){Ouchi}, {Shimasaku}, {Furusawa}, {Miyazaki},
  {Doi}, {Hamabe}, {Hayashino}, {Kimura}, {Kodaira}, {Komiyama}, {Matsuda},
  {Miyazaki}, {Nakata}, {Okamura}, {Sekiguchi}, {Shioya}, {Tamura},
  {Taniguchi}, {Yagi}, \& {Yasuda}}]{ouc03}
{Ouchi}, M., {Shimasaku}, K., {Furusawa}, H., {et~al.} 2003, \apj, 582, 60

\bibitem[{{Ouchi} {et~al.}(2005){Ouchi}, {Shimasaku}, {Akiyama}, {Sekiguchi},
  {Furusawa}, {Okamura}, {Kashikawa}, {Iye}, {Kodama}, {Saito}, {Sasaki},
  {Simpson}, {Takata}, {Yamada}, {Yamanoi}, {Yoshida}, \& {Yoshida}}]{ouc05}
{Ouchi}, M., {Shimasaku}, K., {Akiyama}, M., {et~al.} 2005, \apjl, 620, L1

\bibitem[{{Overzier} {et~al.}(2006a){Overzier}, {Bouwens}, {Cross}, {Venemans},
  {Miley}, {Zirm}, {Benitez}, {Blakeslee}, {Coe}, {Demarco},
  {Ford}, {Homeier}, {Illingworth}, {Kurk}, {Martel}, {Mei},
  {Rottgering}, {Tsvetanov}, \& {Zheng}}]{ove06a}
{Overzier}, R.~A., {Bouwens}, R.~J., {Cross}, N.~J.~G., {et~al.} 2006a, ApJ submitted (astro-ph/0601223)

\bibitem[{{Overzier} {et~al.}(2006b){Overzier}, {Zirm}, {Miley}, {Tsvetanov},
  {Bouwens}, {Ben{\'{\i}}tez}, {Blakeslee}, {Ford}, {Illingworth}, {Postman},
  {Rosati}, {Clampin}, {Hartig}, {R{\" o}ttgering}, {Venemans}, {Ardila},
  {Bartko}, {Broadhurst}, {Brown}, {Burrows}, {Cheng}, {Cross}, {De Breuck},
  {Feldman}, {Franx}, {Golimowski}, {Gronwall}, {Infante}, {Martel},
  {Menanteau}, {Meurer}, {Sirianni}, {Kimble}, {Krist}, {Sparks}, {Tran},
  {White}, \& {Zheng}}]{ove06b}
{Overzier}, R.~A., {Zirm}, A.~W., {Miley}, G.~K., {et~al.} 2006b, \apj, 637, 58

\bibitem[{{Papadopoulos} {et~al.}(2000){Papadopoulos}, {R{\" o}ttgering}, {van
  der Werf}, {Guilloteau}, {Omont}, {van Breugel}, \& {Tilanus}}]{pap00}
{Papadopoulos}, P.~P., {R{\" o}ttgering}, H.~J.~A., {van der Werf}, P.~P.,
  {et~al.} 2000, \apj, 528, 626

\bibitem[{{Pascarelle} {et~al.}(1996){Pascarelle}, {Windhorst}, {Driver},
  {Ostrander}, \& {Keel}}]{pas96}
{Pascarelle}, S.~M., {Windhorst}, R.~A., {Driver}, S.~P., {Ostrander}, E.~J.,
  \& {Keel}, W.~C. 1996, \apjl, 456, L21

\bibitem[{{Pentericci} {et~al.}(1997){Pentericci}, {R\"{o}ttgering}, {Miley},
  {Carilli}, \& {McCarthy}}]{pen97}
{Pentericci}, L., {R\"{o}ttgering}, H.~J.~A., {Miley}, G.~K., {Carilli}, C.~L.,
  \& {McCarthy}, P. 1997, \aap, 326, 580

\bibitem[{{Pentericci} {et~al.}(1998){Pentericci}, {R\"ottgering}, {Miley},
  {Spinrad}, {McCarthy}, {van Breugel}, \& {Macchetto}}]{pen98}
{Pentericci}, L., {R\"ottgering}, H.~J.~A., {Miley}, G.~K., {et~al.} 1998,
  \apj, 504, 139

\bibitem[{{Pentericci} {et~al.}(1999){Pentericci}, {R{\" o}ttgering}, {Miley},
  {McCarthy}, {Spinrad}, {van Breugel}, \& {Macchetto}}]{pen99}
{Pentericci}, L., {R{\"o}ttgering}, H.~J.~A., {Miley}, G.~K., {et~al.}
  1999, \aap, 341, 329

\bibitem[{{Pentericci} {et~al.}(2000{\natexlab{a}}){Pentericci}, {Kurk}, {R{\"
  o}ttgering}, {Miley}, {van Breugel}, {Carilli}, {Ford}, {Heckman},
  {McCarthy}, \& {Moorwood}}]{pen00a}
{Pentericci}, L., {Kurk}, J.~D., {R{\" o}ttgering}, H.~J.~A., {et~al.}
  2000{\natexlab{a}}, \aap, 361, L25

\bibitem[{{Pentericci} {et~al.}(2000{\natexlab{b}}){Pentericci}, {Van Reeven},
  {Carilli}, {R{\" o}ttgering}, \& {Miley}}]{pen00b}
{Pentericci}, L., {Van Reeven}, W., {Carilli}, C.~L., {R{\" o}ttgering},
  H.~J.~A., \& {Miley}, G.~K. 2000{\natexlab{b}}, \aaps, 145, 121

\bibitem[{{Pentericci} {et~al.}(2001){Pentericci}, {McCarthy}, {R{\"
  o}ttgering}, {Miley}, {van Breugel}, \& {Fosbury}}]{pen01}
{Pentericci}, L., {McCarthy}, P.~J., {R{\" o}ttgering}, H.~J.~A., {et~al.}
  2001, \apjs, 135, 63

\bibitem[{{Pentericci} {et~al.}(2002){Pentericci}, {Kurk}, {Carilli}, {Harris},
  {Miley}, \& {R{\" o}ttgering}}]{pen02}
{Pentericci}, L., {Kurk}, J.~D., {Carilli}, C.~L., {et~al.} 2002, \aap, 396,
  109

\bibitem[{{Reiprich} \& {B{\" o}hringer}(2002)}]{rei02}
{Reiprich}, T.~H. \& {B{\" o}hringer}, H. 2002, \apj, 567, 716

\bibitem[{{Reuland} {et~al.}(2003){Reuland}, {van Breugel}, {R{\" o}ttgering},
  {de Vries}, {Stanford}, {Dey}, {Lacy}, {Bland-Hawthorn}, {Dopita}, \&
  {Miley}}]{reu03a}
{Reuland}, M., {van Breugel}, W.~J.~M., {R{\" o}ttgering}, H.~J.~A., {et~al.}
   2003, \apj, 592, 755

\bibitem[{{Reuland} {et~al.}(2004){Reuland}, {R{\" o}ttgering}, {van Breugel},
  \& {De Breuck}}]{reu04}
{Reuland}, M., {R{\" o}ttgering}, H., {van Breugel}, W., \& {De Breuck}, C.
  2004, \mnras, 353, 377

\bibitem[{{Rhoads} {et~al.}(2000){Rhoads}, {Malhotra}, {Dey}, {Stern},
  {Spinrad}, \& {Jannuzi}}]{rho00}
{Rhoads}, J.~E., {Malhotra}, S., {Dey}, A., {et~al.} 2000, \apjl, 545, L85

\bibitem[{{Rhoads} {et~al.}(2003){Rhoads}, {Dey}, {Malhotra}, {Stern},
  {Spinrad}, {Jannuzi}, {Dawson}, {Brown}, \& {Landes}}]{rho03}
{Rhoads}, J.~E., {Dey}, A., {Malhotra}, S., {et~al.} 2003, \aj, 125, 1006

\bibitem[{{Rhoads} {et~al.}(2004){Rhoads}, {Xu}, {Dawson}, {Dey}, {Malhotra},
  {Wang}, {Jannuzi}, {Spinrad}, \& {Stern}}]{rho04}
{Rhoads}, J.~E., {Xu}, C., {Dawson}, S., {et~al.} 2004, \apj, 611, 59

\bibitem[{{Rosati} {et~al.}(2004){Rosati}, {Tozzi}, {Ettori}, {Mainieri},
  {Demarco}, {Stanford}, {Lidman}, {Nonino}, {Borgani}, {Della Ceca},
  {Eisenhardt}, {Holden}, \& {Norman}}]{ros04}
{Rosati}, P., {Tozzi}, P., {Ettori}, S., {et~al.} 2004, \aj, 127, 230

\bibitem[{{R\"ottgering} {et~al.}(1995){R\"ottgering}, {Hunstead}, {Miley},
  {van Ojik}, \& {Wieringa}}]{rot95}
{R\"ottgering}, H.~J.~A., {Hunstead}, R.~W., {Miley}, G.~K., {van Ojik}, R., \&
  {Wieringa}, M.~H. 1995, \mnras, 277, 389

\bibitem[{{R{\"o}ttgering} {et~al.}(1997){R{\"o}ttgering}, {van Ojik}, {Miley},
  {Chambers}, {van Breugel}, \& {de Koff}}]{rot97}
{R{\"o}ttgering}, H.~J.~A., {van Ojik}, R., {Miley}, G.~K., {et~al.} 1997,
  \aap, 326, 505

\bibitem[{{R\"{o}ttgering} {et~al.}(1996){R\"{o}ttgering}, {West}, {Miley}, \&
  {Chambers}}]{rot96}
{R\"{o}ttgering}, H.~J.~A., {West}, M.~J., {Miley}, G.~K., \& {Chambers}, K.~C.
  1996, \aap, 307, 376

\bibitem[{{S{\' a}nchez} \& {Gonz{\' a}lez-Serrano}(1999)}]{san99}
{S{\' a}nchez}, S.~F. \& {Gonz{\' a}lez-Serrano}, J.~I. 1999, \aap, 352, 383

\bibitem[{{S{\' a}nchez} \& {Gonz{\' a}lez-Serrano}(2002)}]{san02}
---. 2002, \aap, 396, 773

\bibitem[{{Schlegel} {et~al.}(1998){Schlegel}, {Finkbeiner}, \&
  {Davis}}]{sch98}
{Schlegel}, D.~J., {Finkbeiner}, D.~P., \& {Davis}, M. 1998, \apj, 500, 525

\bibitem[{{Shimasaku} {et~al.}(2003){Shimasaku}, {Ouchi}, {Okamura},
  {Kashikawa}, {Doi}, {Furusawa}, {Hamabe}, {Hayashino}, {Kawabata}, {Kimura},
  {Kodaira}, {Komiyama}, {Matsuda}, {Miyazaki}, {Miyazaki}, {Nakata}, {Ohta},
  {Ohyama}, {Sekiguchi}, {Shioya}, {Tamura}, {Taniguchi}, {Yagi}, {Yamada}, \&
  {Yasuda}}]{shi03}
{Shimasaku}, K., {Ouchi}, M., {Okamura}, S., {et~al.} 2003, \apjl, 586, L111

\bibitem[{{Shimasaku} {et~al.}(2004){Shimasaku}, {Hayashino}, {Matsuda},
  {Ouchi}, {Ohta}, {Okamura}, {Tamura}, {Yamada}, \& {Yamauchi}}]{shi04}
{Shimasaku}, K., {Hayashino}, T., {Matsuda}, Y., {et~al.} 2004, \apjl, 605, L93

\bibitem[{{Smith} {et~al.}(2002){Smith}, {Tucker}, {Kent}, {Richmond},
  {Fukugita}, {Ichikawa}, {Ichikawa}, {Jorgensen}, {Uomoto}, {Gunn}, {Hamabe},
  {Watanabe}, {Tolea}, {Henden}, {Annis}, {Pier}, {McKay}, {Brinkmann}, {Chen},
  {Holtzman}, {Shimasaku}, \& {York}}]{smi02}
{Smith}, J.~A., {Tucker}, D.~L., {Kent}, S., {et~al.} 2002, \aj, 123, 2121

\bibitem[{{Somerville} {et~al.}(2004){Somerville}, {Lee}, {Ferguson},
  {Gardner}, {Moustakas}, \& {Giavalisco}}]{som04}
{Somerville}, R.~S., {Lee}, K., {Ferguson}, H.~C., {et~al.} 2004, \apjl, 600,
  L171

\bibitem[{{Stanford} {et~al.}(1998){Stanford}, {Eisenhardt}, \&
  {Dickinson}}]{sta98}
{Stanford}, S.~A., {Eisenhardt}, P.~R., \& {Dickinson}, M. 1998, \apj, 492, 461

\bibitem[{{Steidel} {et~al.}(1998){Steidel}, {Adelberger}, {Dickinson},
  {Giavalisco}, {Pettini}, \& {Kellogg}}]{ste98}
{Steidel}, C.~C., {Adelberger}, K.~L., {Dickinson}, M., {et~al.} 1998, \apj,
  492, 428

\bibitem[{{Steidel} {et~al.}(2004){Steidel}, {Shapley}, {Pettini},
  {Adelberger}, {Erb}, {Reddy}, \& {Hunt}}]{ste04}
{Steidel}, C.~C., {Shapley}, A.~E., {Pettini}, M., {et~al.} 2004, \apj, 604,
  534

\bibitem[{{Steidel} {et~al.}(2005){Steidel}, {Adelberger}, {Shapley}, {Erb},
  {Reddy}, \& {Pettini}}]{ste05}
{Steidel}, C.~C., {Adelberger}, K.~L., {Shapley}, A.~E., {et~al.} 2005, \apj,
  626, 44

\bibitem[{{Stern} {et~al.}(2000){Stern}, {Bunker}, {Spinrad}, \&
  {Dey}}]{ster00}
{Stern}, D., {Bunker}, A., {Spinrad}, H., \& {Dey}, A. 2000, \apj, 537, 73

\bibitem[{{Stevens} {et~al.}(2003){Stevens}, {Ivison}, {Dunlop}, {Smail},
  {Percival}, {Hughes}, {R{\" o}ttgering}, {van Breugel}, \&
  {Reuland}}]{stev03}
{Stevens}, J.~A., {Ivison}, R.~J., {Dunlop}, J.~S., {et~al.} 2003, \nat, 425,
  264

\bibitem[{{Stiavelli} {et~al.}(2001){Stiavelli}, {Scarlata}, {Panagia}, {Treu},
  {Bertin}, \& {Bertola}}]{sti01}
{Stiavelli}, M., {Scarlata}, C., {Panagia}, N., {et~al.} 2001, \apjl, 561, L37

\bibitem[{{Stone} \& {Baldwin}(1983)}]{sto83}
{Stone}, R.~P.~S. \& {Baldwin}, J.~A. 1983, \mnras, 204, 347

\bibitem[{{Tanaka} {et~al.}(2004){Tanaka}, {Goto}, {Okamura}, {Shimasaku}, \&
  {Brinkmann}}]{tan04}
{Tanaka}, M., {Goto}, T., {Okamura}, S., {Shimasaku}, K., \& {Brinkmann}, J.
  2004, \aj, 128, 2677

\bibitem[{{Tapken} {et~al.}(2004){Tapken}, {Appenzeller}, {Mehlert}, {Noll}, \&
  {Richling}}]{tap04}
{Tapken}, C., {Appenzeller}, I., {Mehlert}, D., {Noll}, S., \& {Richling}, S.
  2004, \aap, 416, L1

\bibitem[{{Terlevich} {et~al.}(1991){Terlevich}, {Melnick}, {Masegosa},
  {Moles}, \& {Copetti}}]{ter91}
{Terlevich}, R., {Melnick}, J., {Masegosa}, J., {Moles}, M., \& {Copetti},
  M.~V.~F. 1991, \aaps, 91, 285

\bibitem[{{Tozzi} {et~al.}(2003){Tozzi}, {Rosati}, {Ettori}, {Borgani},
  {Mainieri}, \& {Norman}}]{toz03}
{Tozzi}, P., {Rosati}, P., {Ettori}, S., {et~al.} 2003, \apj, 593, 705

\bibitem[{{Tran} {et~al.}(2005){Tran}, {van Dokkum}, {Illingworth}, {Kelson},
  {Gonzalez}, \& {Franx}}]{tra05}
{Tran}, K.~H., {van Dokkum}, P., {Illingworth}, G.~D., {et~al.} 2005, \apj,
  619, 134

\bibitem[{{van Breugel} {et~al.}(1999){van Breugel}, {De Breuck}, {Stanford},
  {Stern}, {R{\" o}ttgering}, \& {Miley}}]{bre99}
{van Breugel}, W., {De Breuck}, C., {Stanford}, S.~A., {et~al.} 1999, \apjl,
  518, L61

\bibitem[{{van Dokkum} \& {Stanford}(2003)}]{dok03}
{van Dokkum}, P.~G. \& {Stanford}, S.~A. 2003, \apj, 585, 78

\bibitem[{{van Ojik} {et~al.}(1996){van Ojik}, {R\"{o}ttgering}, {Carilli},
  {Miley}, {Bremer}, \& {Macchetto}}]{oji96}
{van Ojik}, R., {R\"{o}ttgering}, H.~J.~A., {Carilli}, C.~L., {Miley}, G.~K.,
  {Bremer}, M.~N., \& {Macchetto}, F. 1996, \aap, 313, 25

\bibitem[{{van Zee} {et~al.}(2004){van Zee}, {Barton}, \& {Skillman}}]{zee04}
{van Zee}, L., {Barton}, E.~J., \& {Skillman}, E.~D. 2004, \aj, 128, 2797

\bibitem[{{Venemans} {et~al.}(2002){Venemans}, {Kurk}, {Miley}, {R{\"
  o}ttgering}, {van Breugel}, {Carilli}, {De Breuck}, {Ford}, {Heckman},
  {McCarthy}, \& {Pentericci}}]{ven02}
{Venemans}, B.~P., {Kurk}, J.~D., {Miley}, G.~K., {et~al.} 2002, \apjl, 569,
  L11

\bibitem[{{Venemans} {et~al.}(2004){Venemans}, {R{\" o}ttgering}, {Overzier},
  {Miley}, {De Breuck}, {Kurk}, {van Breugel}, {Carilli}, {Ford}, {Heckman},
  {McCarthy}, \& {Pentericci}}]{ven04}
{Venemans}, B.~P., {R{\" o}ttgering}, H.~J.~A., {Overzier}, R.~A., {et~al.}
  2004, \aap, 424, L17

\bibitem[{{Venemans} {et~al.}(2005){Venemans}, {R{\" o}ttgering}, {Miley},
  {Kurk}, {de Breuck}, {Overzier}, {van Breugel}, {Carilli}, {Ford}, {Heckman},
  {Pentericci}, \& {McCarthy}}]{ven05}
{Venemans}, B.~P., {R{\" o}ttgering}, H.~J.~A., {Miley}, G.~K., {et~al.} 2005,
  \aap, 431, 793

\bibitem[{{Villar-Mart{\'{\i}}n} {et~al.}(2003){Villar-Mart{\'{\i}}n},
  {Vernet}, {di Serego Alighieri}, {Fosbury}, {Humphrey}, \&
  {Pentericci}}]{vil03}
{Villar-Mart{\'{\i}}n}, M., {Vernet}, J., {di Serego Alighieri}, S., {et~al.}
  2003, \mnras, 346, 273

\bibitem[{{West}(1994)}]{wes94}
{West}, M.~J. 1994, \mnras, 268, 79

\bibitem[{{Willott} {et~al.}(2001){Willott}, {Rawlings}, {Blundell}, {Lacy}, \&
  {Eales}}]{wil01}
{Willott}, C.~J., {Rawlings}, S., {Blundell}, K.~M., {Lacy}, M., \& {Eales},
  S.~A. 2001, \mnras, 322, 536

\bibitem[{{Wold} {et~al.}(2003){Wold}, {Armus}, {Neugebauer}, {Jarrett}, \&
  {Lehnert}}]{wol03}
{Wold}, M., {Armus}, L., {Neugebauer}, G., {Jarrett}, T.~H., \& {Lehnert},
  M.~D. 2003, \aj, 126, 1776

\bibitem[{{Zirm} {et~al.}(2003){Zirm}, {Dickinson}, \& {Dey}}]{zir03}
{Zirm}, A.~W., {Dickinson}, M., \& {Dey}, A. 2003, \apj, 585, 90

\bibitem[{{Zirm} {et~al.}(2005){Zirm}, {Overzier}, {Miley}, {Blakeslee},
  {Clampin}, {De Breuck}, {Demarco}, {Ford}, {Hartig}, {Homeier},
  {Illingworth}, {Martel}, {R{\"o}ttgering}, {Venemans}, {Ardila}, {Bartko},
  {Ben{\'{\i}}tez}, {Bouwens}, {Bradley}, {Broadhurst}, {Brown}, {Burrows},
  {Cheng}, {Cross}, {Feldman}, {Franx}, {Golimowski}, {Goto}, {Gronwall},
  {Holden}, {Infante}, {Kimble}, {Krist}, {Lesser}, {Mei}, {Menanteau},
  {Meurer}, {Motta}, {Postman}, {Rosati}, {Sirianni}, {Sparks}, {Tran},
  {Tsvetanov}, {White}, \& {Zheng}}]{zir05}
{Zirm}, A.~W., {Overzier}, R.~A., {Miley}, G.~K., {et~al.} 2005, \apj,
630, 68

\end{thebibliography}
\end{document}